\documentclass[]{jfm}
\usepackage{graphicx}
\usepackage{newtxtext}
\usepackage{newtxmath}
\usepackage{natbib}
\usepackage{hyperref}
\usepackage[section]{placeins}
\usepackage{float}
\usepackage{caption}
\usepackage{subcaption}
\usepackage{xcolor}
\usepackage[normalem]{ulem}
\usepackage{cleveref}
\usepackage{color,soul}
\usepackage{booktabs}
\hypersetup{
    colorlinks = true,
    urlcolor   = blue,
    citecolor  = black,
}

\newcommand{\RomanNumeralCaps}[1]
\linenumbers

\title{Addressing \textit{A Posteriori} Performance Degradation in Neural Network Subgrid Stress Models}

\author{Andy Wu\aff{1}
  \corresp{\email{awu1018@stanford.edu}} \and
  Sanjiva K. Lele\aff{1, 2}}

\affiliation{\aff{1}Department of Aeronautics and Astronautics, Stanford University, Stanford, California, USA
\aff{2}Department of Mechanical Engineering, Stanford University, Stanford, California, USA}

\begin{document}
\maketitle

\begin{abstract}
Neural network subgrid stress models often have \textit{a priori} performance that is far better than the \textit{a posteriori} performance, leading to neural network models that look very promising \textit{a priori} completely failing in \textit{a posteriori} Large Eddy Simulations (LES). This performance gap can be decreased by combining two different methods, training data augmentation and reducing input complexity to the neural network. Augmenting the training data with two different filter shapes before training the neural networks has no performance degradation \textit{a priori} as compared to a neural network trained with one filter. \textit{A posteriori}, neural networks trained with two different filters are more robust across two LES codes with different high order numerical schemes. In addition, by ablating away the higher order terms input into the neural network, the \textit{a priori} versus \textit{a posteriori} performance changes become less apparent. When combined, neural networks that use both training data augmentation and a less complex set of inputs have \textit{a posteriori} performance more reflective of their \textit{a priori} evaluation. 
\end{abstract}


\section{Introduction}
\label{sec:headings}
Large Eddy Simulation (LES) provides an economical paradigm for high fidelity flow prediction by resolving the larger, energy containing scales and modelling the spatial scales of turbulence that are smaller than the grid scale through a subgrid scale (SGS) closure~\citep{sagaut}. Spatially filtering the Navier-Stokes equations results in an unclosed term called the subgrid stress tensor, $\tau_{ij}$, which represents the effect of the subgrid spatial scales of turbulence on the resolved flow~\citep{pope}.

Recently, with the availability of Direct Numerical Simulation (DNS) data online and advances in deep learning theory, neural networks have been explored as subgrid stress models~\citep{sarg, beck, park, Choi_2024, xie, xie2, stoffer, Kang, kawai, cheng2022deep, Wu_PrF}. A persistent challenge is the discrepancy in neural network \textit{a priori} and \textit{a posteriori} performance~\citep{beck, beckreview, stoffer, park}, which can be interpreted as a distribution shift between the training data and the LES simulation~\citep{hu2025stable}. Due to this distribution shift, ad-hoc methods are sometimes used, such as adding an additional Smagorinsky term to help the neural network dissipate energy so that simulations remains stable~\citep{beck}. Two mechanisms may contribute to the distribution shift. First, the input fields provided to the neural network during training, which are generated assuming an explicit filter (oftentimes a box filter), can differ from the LES resolved fields, since the LES filter is implicit and is a product of both the grid and the numerical method~\citep{sagaut}. Second, the output SGS stress statistics also differ across filtering operations. For example, the omission or use of two-thirds dealiasing in spectral methods modifies the effective filter transfer function, producing different SGS statistics. This motivates the hypothesis that neural networks trained on one filter may struggle to generalize to LES solvers~\textit{a posteriori}. To address these two mechanisms of distribution shift, this work introduces a multi-filter data augmentation strategy that exposes the neural network to various plausible filtered inputs and SGS stress distributions. These sources of distribution shift are shifts in the spatial distributions. There are also potential distribution shifts as the neural network subgrid stress model is time advanced in the simulation itself, which can manifest as a source of error accumulation that causes a distribution shift~\citep{sirignano2020dpm}. This work focuses on addressing the spatial distribution shifts instead of those due to time advancement.

Reinforcement learning approaches~\citep{kim2022deep, bae2022scientific, kurz2023deep, guan2023learning} aim to bypass these issues by learning the SGS model (or wall model) directly in a LES solver. However, models learned in this manner are limited in their transferability to other solvers, as they learn the combined effect of solver-specific numerical effects and turbulence physics. Similarly, end-to-end learning approaches, where a differentiable solver is used alongside a neural network model~\citep{sirignano2023deep, kim2024generalizable, duraisamy2021perspectives} can suffer similar issues, as the neural network learned in this setting is also unlikely to transfer directly to different flow solvers, having been optimized along a specific flow solver and obtaining gradients that are impacted by the flow solver's numerical method. In addition, the training cost of the neural network is much higher, as it requires LES to be run during training. In contrast, the present supervised learning approach aims to not learn solver peculiarities, and incorporates known filtering operations into the training data. The filters used do not mimic the specific transfer function of any specific LES solver and instead are inspired by common LES numerical behavior. However, many works that do adopt a supervised approach where the filtered DNS data is the ground truth will also not learn solver peculiarities. Instead, these works only use one filter (oftentimes a box filter) when assembling the training data. Therefore, instead of learning solver peculiarities, the neural networks can now overfit to filter peculiarities, causing a distribution shift \textit{a priori} versus \textit{a posteriori}. 

Furthermore, another potential source of distribution shift can come from the use of increasingly complex neural network inputs, such as higher order powers of the velocity gradient tensor ($V_{ij}$). For example, the second invariant $Q = 0.5(P^2-\operatorname{trace}(V_{ij}^2)))$  where $P = \operatorname{trace}(V_{ij})$ and $V_{ij}$ is the velocity gradient tensor, and the third invariant $R = det(V_{ij})$ are more sensitive to aliasing and differences in numerical methods (where $det$ denotes the determinant). Therefore, a systematic ablation of neural network input complexity is performed to analyze the effect of inputs to neural network performance \textit{a posteriori}.

The contributions of this work are twofold. First, multi-filter (two and three filter) data augmentation is shown to improve the robustness of learned SGS models across small architecture perturbations and across LES solvers with different high-order numerical methods. Second, an ablation of input feature complexity reveals that more complex neural network inputs suffer more from \textit{a posteriori} performance degradation. Together, these results enable the training of neural networks that are more robust \textit{a posteriori}.  

\section{Numerical Details}
\label{sec:Nums}

\subsection {LES}
The LES governing equations involve the filtered Navier-Stokes equations. The filtering operation is defined below:

\begin{equation} \label{eqn1}
\Bar{\chi}(\mathbf{x}, t) = \int{G(\mathbf{r})\chi(\mathbf{x - r}, t) d\mathbf{r}},  \int{G(\mathbf{r}) d\mathbf{r}} = 1
\end{equation}

where $\chi$ is the flow variable to be filtered, $G(\mathbf{r})$ is the filtering kernel, and $\Bar{\chi}$ is the filtered flow variable. In all sections, the overbar operator is used to denote filtered quantities. For robust subgrid stress modeling, the filtering operation must not be oscillatory in physical space, as this would create additional small scales purely through the act of filtering~\citep{sagaut}. This means that the spectral cutoff filter, which is in the shape of a sinc function in physical space (while being sharp in spectral space) is not appropriate as a filter for subgrid stress modeling for general application. With the definition of the filter above, the LES governing equations are written below: (in Einstein's summation notation) 

 \begin{equation} \label{eqn3}
\begin{aligned}
\frac{\partial \Bar{u}_i}{\partial x_i} &= 0 \\
\frac{\partial \Bar{u}_i}{\partial t} + \Bar{u}_j \frac{\partial \Bar{u}_i}{\partial x_j} &= - \frac{1}{\rho} \frac{\partial \Bar{p}}{\partial x_i} + \nu \frac{\partial^2 \Bar{u}_i}{\partial x_j \partial x_j} - \frac{\partial \tau_{ij}}{\partial x_j} \\
\tau_{ij} &= \overline{u_i u_j} - \Bar{u}_i\Bar{u}_j
\end{aligned}
\end{equation}

where $\overline{p}$ is the filtered pressure, $\rho$ is the density, $\overline{u}$ is the filtered velocity, and $\nu$ is the kinematic viscosity. As seen, an unclosed term, $\tau_{ij}$, is obtained by filtering the momentum equation. In LES, only the filtered values of the dependent variables are available, thus, $\tau_{ij}$ can't be computed and has to be modelled. In the present work, the neural network will learn to approximate $\tau_{ij}$ through filtered DNS data.

\subsection {\textit{A Posteriori Flow Solvers}}
Since this work involves drawing conclusions about \textit{a posteriori} performance, two in-house LES flow solvers, PadeOps and PadeLibs (with different numerical methods) are used~\citep{ghate, SONG2024padelibs}. PadeOps is a Fourier-Spectral code with RK4-5 or RK3 time stepping~\citep{Bogacki}. Therefore, two-thirds dealiasing is applied due to the non-linear component in the convective acceleration term. Meanwhile, PadeLibs is a 6th order compact code with RK3 time stepping (with no 2/3 dealiasing). These two codes have different numerical methods, so the \textit{a posteriori} conclusions obtained are generalizable across these high order discretizations. As seen in a later section, these two flow solvers create two different environments for the neural network when considering subgrid stress modelling.

\section{Filtering and Neural Network Details}
Neural network SGS models trained outside of the LES solver and tested \textit{a posteriori} depend on explicitly filtered DNS data, making the choice of filters critical. Mismatches between the explicit filters chosen and the implicit filters encountered in \textit{a posteriori} LES can produce both input-side and output-side distribution shifts. To reduce such mismatches, the present work adopts multi-filter training, selecting explicit filters to diversify the training distribution so that the filters span representative LES filtering behaviors without approximating a specific LES solver's transfer function. The first explicit filter chosen is the box filter, as it is a common filter used in previous studies. The box filter is defined as follows:

\begin{equation} \label{eqn4}
\Bar{\chi}(\mathbf{x}, t) = \int_V \frac{\chi(\mathbf{x - r}, t)}{|V|} \, d\mathbf{r}
\end{equation}

where \(V\) is the volume (or region) of the box filter, \(|V|\) is its measure. The box filter is not oscillatory in physical space, but is oscillatory in spectral space. The 1-D kernel of the box filter, $G_{box}(r)$, is also given below:

\begin{equation} \label{eqn44}
G_{box}(r) = \begin{cases} \frac{1}{\Delta} & \text{for } |r| \le \frac{\Delta}{2} \\ 0 & \text{for } |r| > \frac{\Delta}{2} \end{cases}
\end{equation}

where $\Delta$ is the filter width. 

\subsubsection{DSCF: A Localized Low-Pass Filter}
To obtain a ``cleaner" grid-cutoff at the Nyquist wavenumber, a discrete spatial cosine filter (DSCF) is used. The DSCF attenuates frequencies faster than the box filter after the grid cutoff while avoiding the oscillatory kernel of an ideal spectral cutoff filter in physical space. The 1-D kernel for the localized low-pass filter is given in Equation~\ref{eqn45}\footnote{The authors are grateful to an anonymous referee for suggesting the physical space formulation of the DSCF adopted in the paper. The original treatment in spectral space was equivalent but cumbersome.}.

\begin{equation} \label{eqn45}
G_{DSCF}(r) = \begin{cases} \frac{1}{\Delta} \left( 1 + \cos\left( \frac{2\pi r}{\Delta} \right) \right) & \text{for } |r| \le \frac{\Delta}{2} \\ 0 & \text{for } |r| > \frac{\Delta}{2} \end{cases}
\end{equation}

\begin{figure}
\centering
\includegraphics[width=.99\textwidth]{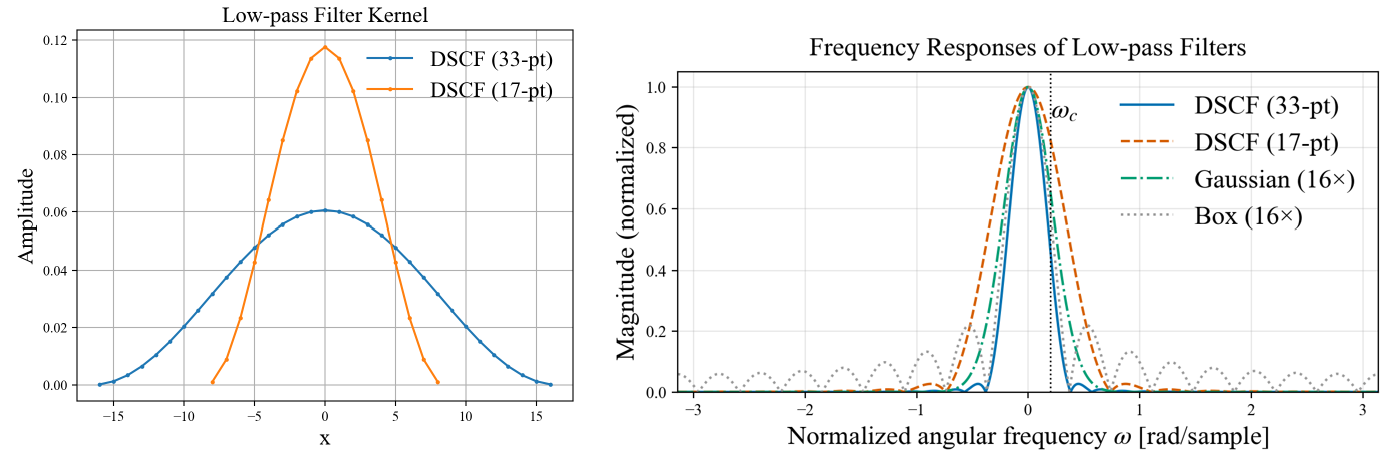}
\caption{Characterizations of the DSCF in both physical and spectral space. $\omega_c = 2 \pi f_c$ corresponds to the filter cutoff, here designated at a $16\Delta_{DNS}$ filter width, while $\omega = 2 \pi f_n, f_n \in [-0.5, 0.5]$ denotes the normalized angular frequency. Two different DSCF widths are shown, one with a 17 and the other with a 33 point support. The DSCF with the smaller filter width has twice the cutoff frequency.}
\label{fig:kernels}
\end{figure}

As seen, this filter is a single mode approximation to the spectral cutoff filter (plus the zero wavenumber component). For a specific filter width, the DSCF cutoff frequency $f_c$ can be found to be $f_c = \frac{1}{\Delta}$, with $\Delta$ being the filter width. In general, when preparing training data, the cutoff frequency is set to be the grid scale cutoff for LES. For example, if one wants to filter the DNS data at $\Delta_{LES} = 16 \Delta_{DNS}$, then the desired $f_{cd} = \frac{0.5}{16} = 0.03125$ (0.5 due to Nyquist), while when filtering at $\Delta_{LES} = 32 \Delta_{DNS}$, the desired $f_{cd} = \frac{0.5}{32} = 0.0156$ is set. This means that when filtering at $\Delta_{LES} = 16 \Delta_{DNS}$, the desired $f_{cd}$ implies that a 33 point DSCF is needed (which has a filter width of 16 points to the left and 16 points to the right to obtain $f_{cd} = f_c$), and when filtering at $\Delta_{LES} = 32 \Delta_{DNS}$, a 65 point DSCF is needed. Both the physical space and spectral space characterizations of the DSCF are shown in figure~\ref{fig:kernels}.

As seen in figure~\ref{fig:kernels}, the frequency response is not perfectly rectangular, instead exhibiting a smoother roll-off with mild high wavenumber oscillations when compared to a box filter. Meanwhile, in physical space, the filter is strictly positive, non-oscillatory, and has finite support. As seen, the 17-point DSCF has the same filter shape as the 33-point DSCF, but has around 2x less points in the spatial kernel, and thus has twice the cutoff frequency. The 33 point DSCF matches the initial decay of the box filter but significantly suppresses the oscillations (peak oscillation magnitude of 0.028 vs 0.22). In addition, it attenuates frequencies immediately beyond the grid cutoff more than the Gaussian filter. This is especially seen in the normalized angular frequency range of 0.2-0.4, where the 33 point DSCF is over 4-5 times lower than the Gaussian.

However, one advantage of the Gaussian filter kernel is that it decays higher wavenumbers exponentially, while the DSCF has mild oscillations. While a wider Gaussian filter kernel (and potentially tuning the standard deviation) can match the passband attenuation of the DSCF, near the grid cutoff, the DSCF will still attenuate frequencies faster. This is especially important as filters attenuating frequencies slightly larger than the grid cut-off more aggressively result in less grid-scale noise contaminating the filtered fields. This ensures that the neural network is fed a cleaner signal, which is more free from the grid-scale noise contamination that can hinder convergence in deep learning models.

\subsubsection{Two-Thirds Dealiasing as an Additional Filter}
PadeOps, the Fourier-Spectral code with two-thirds dealiasing, uses a filtering operation that modifies the effective transfer function due to the two-thirds dealiasing operation. To study how multi-filter training interacts with this procedure, a filter is derived that is consistent with two-thirds dealiasing. The total subgrid stress can be calculated by applying the two-thirds dealiasing operation on top of the already filtered Navier-Stokes equations, defined as the tilde operation:
\begin{equation} \label{eqn:doublefilt}
\begin{aligned}
\tau_{ij}^f &= \widetilde{\overline{u_i u_j}} - \Tilde{\Bar{u}}_i\Tilde{\Bar{u}}_j
\end{aligned}
\end{equation}
As seen in equation~\ref{eqn:doublefilt}, the subgrid stress tensor, after accounting for this two-thirds dealiasing operation, is denoted as $\tau_{ij}^f$. The two-thirds dealiasing operation is computed as an additional spectral cutoff filter on top of the box filtered data. This filtering operation will be called box and two-thirds (BTF), and is used alongside the box filter to form the augmented dataset for the Fourier-Spectral code. Note that the BTF is not intended to exactly replicate the transfer function of Fourier-Spectral codes, and it is unlikely that the transfer function of a Fourier-Spectral code with 2/3 dealiasing is identical to BTF. Rather, BTF provides the neural network with exposure to the type of behavior that two-thirds dealiasing qualitatively introduces, to reduce the training data distribution shift \textit{a posteriori}. 

\subsection{Neural Network details}
The neural network used is a hybrid tensor basis neural network (TBNN) from~\cite{ling} and graph neural network (GNN). In a classic TBNN, an Artificial Neural Network (ANN) maps inputs to scalar coefficients of the tensor basis expansion. Here, the ANN front-end architecture is substituted with a GNN, which allows for learned input stencils instead of prescribed input stencils~\citep{abekawa2023exploration}. To recap, the tensor basis expansion for LES, where the neural network predicts 8 scalar coefficients $c_1-c_8$, can be written as~\citep{stallcup, Wu_PrF}: 
\begin{equation}\label{eqn:tblc}
\tau_{ij} = c_1 \mathbf{I} + c_2 \mathbf{\Bar{S}} + c_3 \mathbf{\Bar{S}^2} + c_4 \mathbf{\Bar{\Omega}^2} + c_5 (\mathbf{\Bar{S}\Bar{\Omega} - \Bar{\Omega}\Bar{S}}) 
\\ + c_6 \mathbf{\Bar{\Omega}\Bar{S}\Bar{\Omega}} + c_7 (\mathbf{\Bar{S}^2\Bar{\Omega} - \Bar{\Omega}\Bar{S}^2}) + c_8 (\mathbf{\Bar{\Omega}\Bar{S}\Bar{\Omega}^2 - \Bar{\Omega}^2\Bar{S}\Bar{\Omega}}) 
\end{equation}
where $\mathbf{\Bar{S}}$ and $\mathbf{\Bar{\Omega}}$ denote the filtered strain rate or rotation rate tensor. For completeness, the invariants are listed below:

\begin{equation} \label{eqn9}
\{tr(\mathbf{\Bar{S}}), tr(\mathbf{\Bar{S}^2}), tr(\mathbf{\Bar{S}^3}), tr(\mathbf{\Bar{\Omega}^2}), tr(\mathbf{\Bar{S}\Bar{\Omega}^2}), tr(\mathbf{\Bar{S}^2\Bar{\Omega}^2}), tr(\mathbf{\Bar{S}^2\Bar{\Omega}^2\Bar{S}\Bar{\Omega}}) \}
\end{equation}

\begin{figure}
\centering
\includegraphics[width=.99\textwidth]{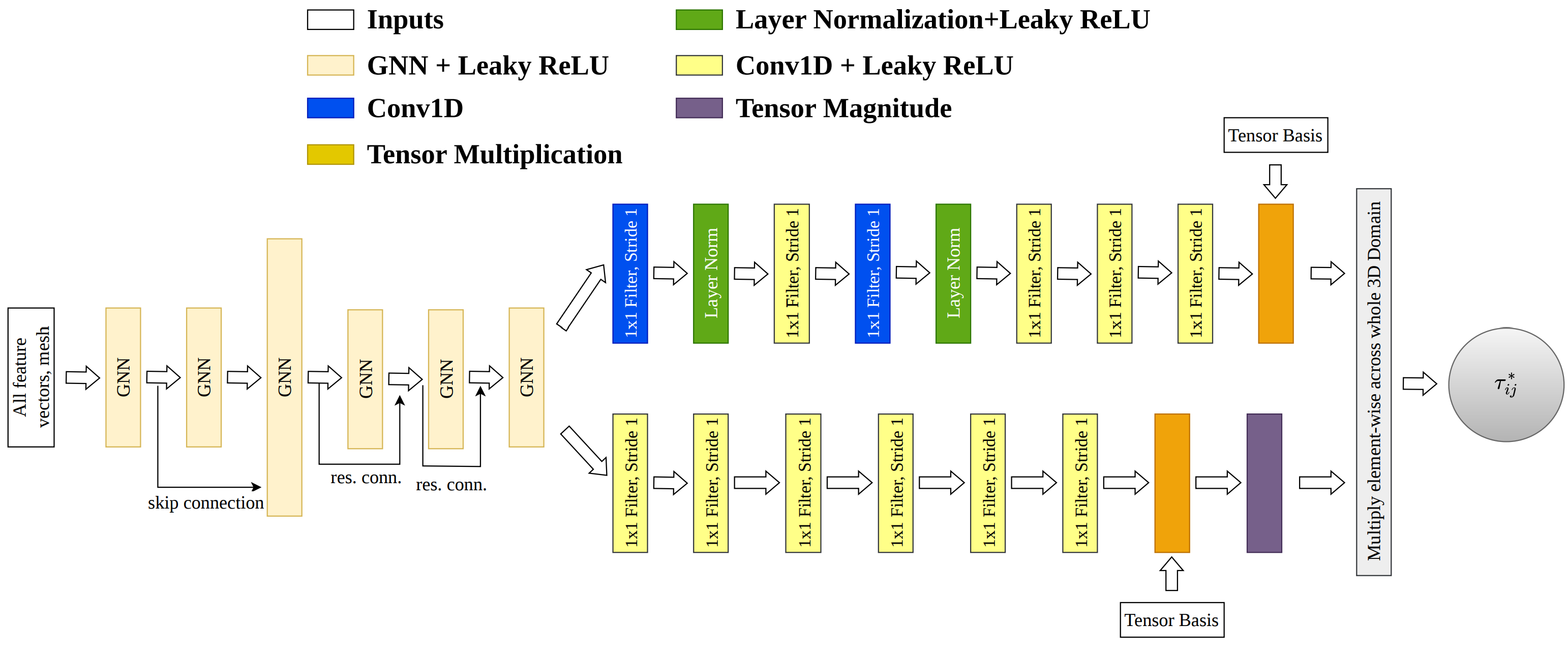}
\caption{Overall graph neural network architecture. Note that ``res. conn." denotes residual connections, and these are used in conjunction with skip connections to prevent oversmoothing. There are two layer normalization layers placed in this architecture.}
\label{fig:GNNarch}
\end{figure}

The invariants of the tensor basis expansion, which are costly to compute, are substituted with the invariants of the resolved velocity gradient tensor ($\bar{P}, \bar{Q}, \bar{R}$), as well as the magnitudes of the strain rate and rotation rate tensor as inputs instead ($|\bar{S}|, |\bar{\Omega}|$). The neural network architecture is shown in figure~\ref{fig:GNNarch}, which has many similarities to~\cite{wu2025s4nd}.

\begin{figure}
\centering
\includegraphics[width=.99\textwidth]{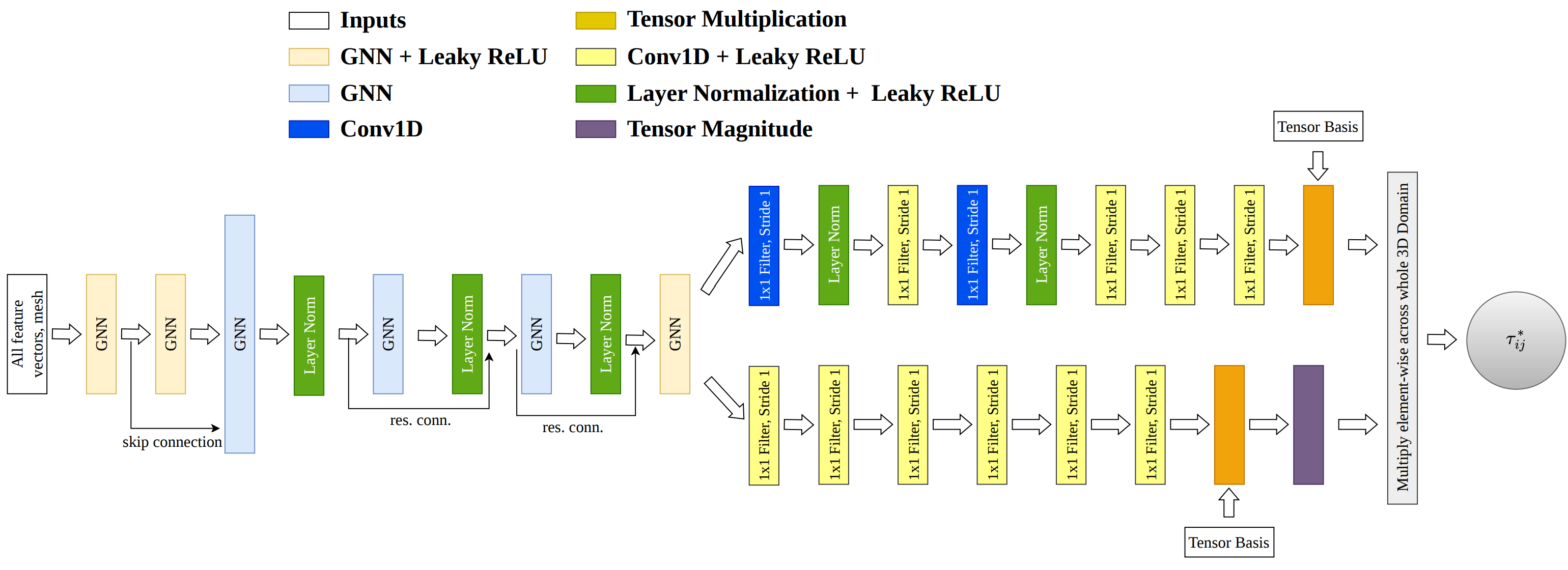}
\caption{Graph neural network variant architecture. Note that ``res. conn." denotes residual connections,  and these are used in conjunction with skip connections to prevent oversmoothing. There are five layer normalization layers placed in this architecture, three more than the original. Other than that, the architecture is exactly the same.}
\label{fig:GNNarchvar}
\end{figure}

As seen, the tensor basis provides the tensorial structure for the data-driven subgrid model, and serves as an inductive bias for the neural network so that any output of the neural network obeys the structure of $\tau_{ij}$. Meanwhile, the inputs (such as $\bar{P}, \bar{Q}, \bar{R}$) pass through layers with learnable weights and nonlinear activation functions. These are the inputs that go in from the left-most side of neural network architecture in the diagrams, and will contain features that the neural network will extract and ``learn" from. In general, the neural network is more sensitive to inputs that go through nonlinear transformations~\citep{novak2018sensitivity}. Simplifying these inputs (less higher powers of the velocity gradient tensor) would reduce their susceptibility to distribution shifts, leading to more consistent \textit{a posteriori} performance. To investigate this, neural networks are trained with two different sets of inputs, $\bar{P}, \bar{Q}, \bar{R}, |\bar{S}|, |\bar{\Omega}|$ (``complex inputs") and $\bar{P}, |\bar{S}|, |\bar{\Omega}|$ (``simple inputs"). The simpler set is hypothesized to perform more consistently \textit{a posteriori}. The neural network outputs consist of the coefficients of the tensor basis expansion, which are then multiplied by the tensor basis to obtain $\tau_{ij}$.

The neural network model inputs and outputs should be dimensionless. A global normalization (max-min normalization) is used for the input fields:
\begin{equation} \label{eqn:maxmin}
x_{norm} = \frac{x-x_{min}}{x_{max}-x_{min}}*2-1
\end{equation}
while a local normalization is adopted for the tensor basis and the subgrid stress tensor:
\begin{equation} \label{eqn:tbnnlocal}
W^{norm}_{ij} = \frac{W_{ij}}{|\bar{V}_{ij}|^x}, \tau_{ij}^* = \frac{\tau_{ij}}{\Delta_{LES}^2|\bar{V}_{ij}|^2}
\end{equation}
where $W_{ij}$ is the tensor to be normalized, $\bar{V}_{ij}$ denotes the velocity gradient tensor, $|{\bar{V}_{ij}}|$ its magnitude, $x$ the exponent to ensure dimensionlessness, $\Delta_{LES}$ corresponds to the LES grid spacing, and repeated indices do not imply summation. 

To ensure that the results are consistent across slight perturbations to the neural network architecture, a variant architecture is introduced with additional layer normalization (ALN) operations between select GNN layers, which is shown in figure~\ref{fig:GNNarchvar}. Input normalization is also varied, using local normalization. Training follows~\citep{Wu_PrF} with a composite loss function that minimizes the root mean squared error between the predicted and actual subgrid stress tensor (denoted as non-bolded RMSE when only reporting the value), and the RMSE between the predicted and actual energy dissipation, $\epsilon = - \tau_{ij} \Bar{S}_{ij}$ (denoted as non-bolded DRMSE when only reporting the value): 
\begin{equation} \label{eqn12}
\mathcal{L} = \mathbf{RMSE}(\tau_{ij,pred}-\tau_{ij})/\mathbf{RMS}(\tau_{ij})+ \mathbf{RMSE}(\epsilon_{pred}-\epsilon)/\mathbf{RMS}(\epsilon)
\end{equation}

where $\mathbf{RMSE}$ is the root-mean squared error operation and $\mathbf{RMS}$ is the root-mean square operation. 

\subsection{Training Data}
\subsubsection{Training Configuration}
The training data is obtained from querying the Johns Hopkins Turbulence Database (JHTDB) for forced HIT and channel flow data~\citep{JHTB1, JHTB2}. In all cases, the data is split into a 80-10-10 split, where 80\% of the training data is used for training, 10\% of the data is used as a validation set, and 10\% of the data is used as the hold-out test set. For HIT, 5 snapshots are queried, and for channel flow, 4 snapshots are queried. Each snapshot is queried at equal time intervals that are statistically de-correlated from each other. The DNS HIT resolution queried is $1024^3$, with a Taylor Reynolds number ($Re_{\lambda}$) of 433. The channel flow resolution queried is $2048 \times 512 \times 1536$, with a friction Reynolds number ($Re_{\tau}$) of 1000.

In all cases, the training data is filtered using two different filter widths, 16 $\Delta_{DNS}$ and 32 $\Delta_{DNS}$, as filtering on two different filter widths shows more generalizability to other filter widths, as seen in~\cite{park}. Furthermore, two different training data domain sizes (such as $16^3$ and $32^3$) are also used to sensitize the neural network to different domain sizes, following~\cite{Wu_PrF}, where detailed training data processing procedures can be found. This means that the filtered DNS fields will be broken up into ``subdomains" where each subdomain is a training data snapshot. If the Kolmogorov scale is defined as $\eta$, then the total training set for each filter consists of the following, which is consistent with the training set used in~\cite{Wu_PrF}:
\begin{itemize}
    \item 320 realizations of filter width $\bar{\Delta} = 16\Delta_{DNS}$ ($\bar{\Delta}/\eta \approx 35$), domain size $16^3$ HIT
    \item 320 realizations of filter width $\bar{\Delta} = 32\Delta_{DNS}$ ($\bar{\Delta}/\eta \approx 70$), domain size $8^3$ HIT
    \item 40 realizations of filter width $\bar{\Delta} = 16\Delta_{DNS}$ ($\bar{\Delta}/\eta \approx 35$), domain size $32^3$ HIT
    \item 40 realizations of filter width $\bar{\Delta} = 32\Delta_{DNS}$ ($\bar{\Delta}/\eta \approx 70$), domain size $16^3$ HIT
    \item 320 realizations of filter width $\bar{\Delta} = 16\Delta_{DNS}$, domain size $16^3$ channel flow
    \item 320 realizations of filter width $\bar{\Delta} = 32\Delta_{DNS}$, domain size $8^3$ channel flow
    \item 40 realizations of filter width $\bar{\Delta} = 16\Delta_{DNS}$, domain size $32^3$ channel flow
    \item 40 realizations of filter width $\bar{\Delta} = 32\Delta_{DNS}$, domain size $16^3$ channel flow
\end{itemize}
For each filter, one can generate the above set of training data using the filtered DNS fields. The final training set consists of combining all the entries in the above list for each filter into a large training set. To demonstrate applicability to channel flow, select neural networks are applied to both HIT and channel flow without retraining. The same applies to when two different grid resolutions are used, as some neural networks are applied to a coarser grid spacing to test applicability of the results to a different grid resolution. Even if the neural network trained is only applied to HIT at one specific resolution (such as $\Delta_{LES} = 16 \Delta_{DNS}$), they are still trained on the full set of training data with various domain sizes, filter widths, and two different flows for consistency.

\subsubsection{Effect of Filtering on Ground Truth}
Before training the neural networks, an analysis of the ground truth used for supervised learning is presented. Here, the impact of filtering is investigated for the ground truth labels, where both spatial distributions and probability density functions (PDFs) are shown. For the spatial distributions shown, a random slice out of the 3D field was selected for visualization, while the PDF is plotted for the full 3D field. The ground truth values are shown for the box, BTF, and DSCFs.

\begin{figure}
\centering
     \begin{subfigure}[hbt!]{.95\textwidth}
         \centering
         \includegraphics[width=\textwidth]{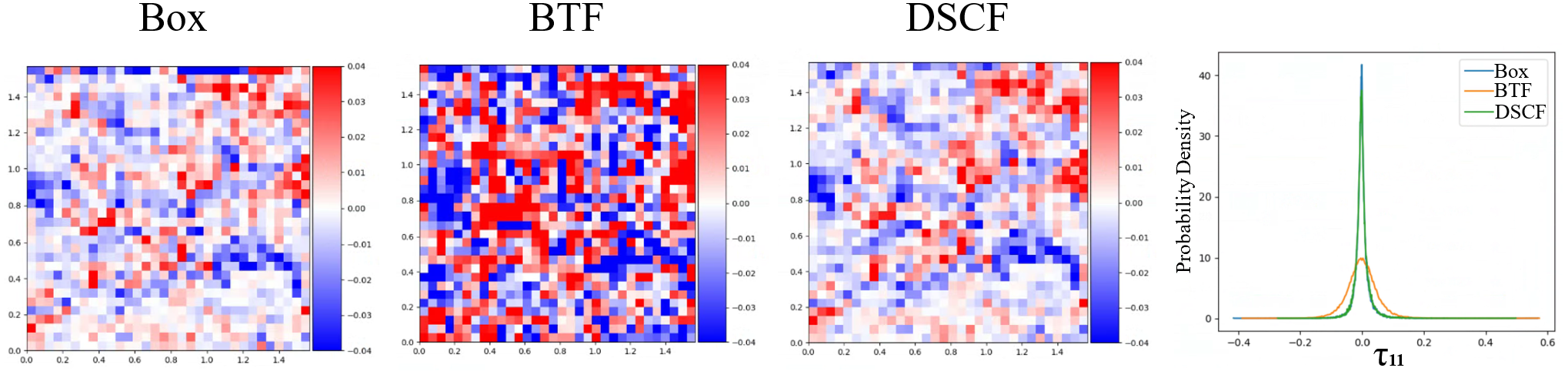}
         \caption{$\tau_{11}$ spatial distribution and PDF. The x and y axis correspond to the coordinates in the physical domain for the spatial distribution visualizations.}
         \label{fig:tau11filt}
     \end{subfigure}
     \vspace{1em}
     \begin{subfigure}[hbt!]{.95\textwidth}
         \centering
         \includegraphics[width=\textwidth]{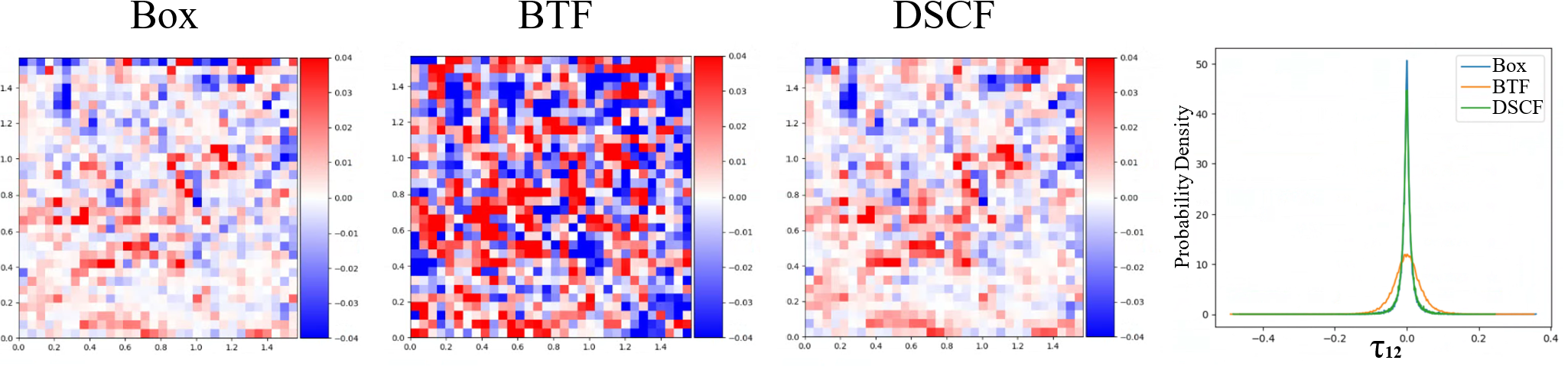}
         \caption{$\tau_{12}$ spatial distribution and PDF. The x and y axis correspond to the coordinates in the physical domain for the spatial distribution visualizations.}
         \label{fig:tau12filt}
     \end{subfigure}
     \vspace{1em}
     \begin{subfigure}[hbt!]{.95\textwidth}
         \centering
         \includegraphics[width=\textwidth]{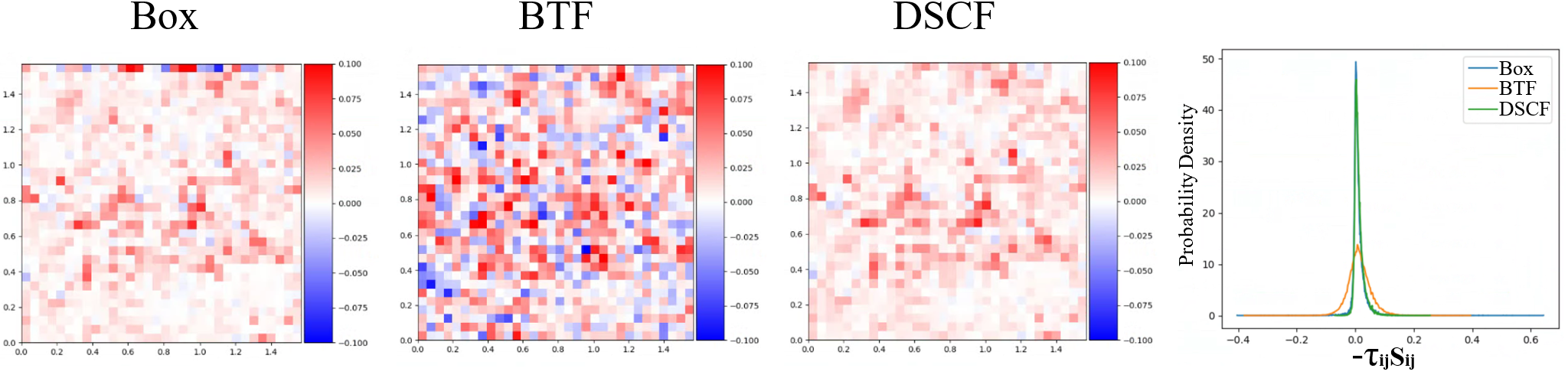}
         \caption{$\epsilon = -\tau_{ij}\bar{S}_{ij}$ spatial distribution and PDF. The x and y axis correspond to the coordinates in the physical domain for the spatial distribution visualizations.}
         \label{fig:dissfilt}
     \end{subfigure}
     \caption{Effect of various filters on the ground truth subgrid stress quantities. All colorbars are scaled the same per row.}
\end{figure}

As seen in figure~\ref{fig:tau11filt} and figure~\ref{fig:tau12filt} , the act of applying two-thirds dealiasing completely changes the subgrid stress field, which is seen visually in the increasing intensity of the subgrid stress and the additional oscillations in the overall field. Meanwhile, the box and DSCF fields look more similar. This is further seen in the PDF, where the act of two-thirds dealiasing creates far more variance in the PDF as compared to the box filter or the DSCF. The phenomenon happens for both the normal and the shear components of the subgrid stress tensor. This suggests that when considering the subgrid stress, a Fourier-Spectral code with two-thirds dealiasing applied will likely produce a completely different environment than a 6th order compact code, and the differences can be quite large. Therefore, these two numerical methods are sufficiently different when considering the subgrid stress.

The same effects are seen in figure~\ref{fig:dissfilt} in the interscale energy transfer field ($\epsilon = -\tau_{ij}\bar{S}_{ij}$), where the box filter and DSCF produce more similar fields, while the BTF produces far more oscillatory fields. This is reflected in the PDF, where the BTF once again produces far more variance in the PDF as compared to the other filters. Overall, for all the PDFs, the box filter has the largest peak, followed by the DSCF, and then there is quite a large gap between those two filters and the BTF. This also reveals that if one were to train on only one filter (oftentimes a box filter), if applied to a Fourier-Spectral code with two-thirds dealiasing, there will automatically be a distribution shift as the neural network is trained to produce subgrid stress distributions similar to the training data, which is filtered with a box filter. As such, training on two or more filters allows the neural network to learn a wider range of plausible subgrid stress distributions it may see in the flow solver, reducing the distribution shift.

\subsection{All Neural Network Configurations}
A summary of all the neural networks trained are seen in table~\ref{tab:tabletrain}. Neural networks are trained with different training data (one filter versus augmented training data), different inputs (complex versus simple), and slightly different architectures (more or less layer normalization). The training spans a wide range of filters, from purely training on only the box filter, only the BTF, only the DSCF, to combinations of two filters and also three filters. Neural networks trained on only the box filter are applied to both the Fourier-Spectral with two-thirds dealiasing solver and the 6th order compact solver. The neural network trained only on the BTF filter is applied to the Fourier-spectral solver to give a baseline on how a neural network trained with only BTF would do. Similarly, the neural network trained with only the DSCF is applied to the 6th order compact solver to give a baseline. Meanwhile, the neural networks trained with two filters are applied to different flow solvers depending on which combination of filters it were trained on. For example, a neural network trained on the box filter and BTF is applied to the Fourier-Spectral code with two-thirds dealiasing (and not to the 6th order compact code which does not have two-thirds dealiasing). Similarly, a neural network trained on the box filter and DSCF is applied to the 6th order compact code. The neural network trained with three filters (box, BTF, and DSCF) is applied to both solvers without retraining to show how the data augmentation approach can work on different flow solvers without retraining. In all cases, the neural networks used have roughly 24000 trainable parameters.

\begin{table}
\caption{\label{tab:tabletrain} Neural network configurations, ``S" stands for the spectral solver with two-thirds dealiasing (PadeOps), and ``C" stands for the 6th order compact solver (PadeLibs). Normalization and neural network structure are also given.}
\centering
\resizebox{\textwidth}{!}{
\begin{tabular}{l c c c c c}  
\hline
Number of Filters & Inputs & Neural Network Structure & Flow Solver & Normalization & Neural Network Name \\
\hline
One filter (box) & $\bar{P}, \bar{Q}, \bar{R}, |\bar{S}|, |\bar{\Omega}|$ & Original & S, C & Global & NN-Box-Complex-Original-G \\
One filter (BTF) & $\bar{P}, \bar{Q}, \bar{R}, |\bar{S}|, |\bar{\Omega}|$ & Original & S & Global & NN-BTF-Complex-Original-G \\
One filter (DSCF) & $\bar{P}, \bar{Q}, \bar{R}, |\bar{S}|, |\bar{\Omega}|$ & Original & C & Global & NN-DSCF-Complex-Original-G \\
Two filter (box and BTF) & $\bar{P}, \bar{Q}, \bar{R}, |\bar{S}|, |\bar{\Omega}|$ & Original & S & Global & NN-BoxBTF-Complex-Original-G \\
Two filter (box and DSCF) & $\bar{P}, \bar{Q}, \bar{R}, |\bar{S}|, |\bar{\Omega}|$ & Original & C & Global & NN-BoxDSCF-Complex-Original-G \\
Three filter (box, BTF, DSCF) & $\bar{P}, \bar{Q}, \bar{R}, |\bar{S}|, |\bar{\Omega}|$ & Original & S, C & Global & NN-BoxBTFDSCF-Complex-Original-G \\
One filter (box) & $\bar{P}, \bar{Q}, \bar{R}, |\bar{S}|, |\bar{\Omega}|$ & Additional LN & S, C & Global & NN-Box-Complex-ALN-G \\
Two filter (box and BTF) & $\bar{P}, \bar{Q}, \bar{R}, |\bar{S}|, |\bar{\Omega}|$ & Additional LN & S & Global & NN-BoxBTF-Complex-ALN-G \\
Two filter (box and DSCF) & $\bar{P}, \bar{Q}, \bar{R}, |\bar{S}|, |\bar{\Omega}|$ & Additional LN & C & Global &NN-BoxDSCF-Complex-ALN-G \\
One filter (box) & $\bar{P}, |\bar{S}|, |\bar{\Omega}|$ & Original & S, C & Global &NN-Box-Simple-Original-G \\
Two filter (box and BTF) & $\bar{P}, |\bar{S}|, |\bar{\Omega}|$ & Original & S & Global &NN-BoxBTF-Simple-Original-G \\
Two filter (box and DSCF) & $\bar{P}, |\bar{S}|, |\bar{\Omega}|$ & Original & C & Global & NN-BoxDSCF-Simple-Original-G \\
One filter (box) & $\bar{P}, |\bar{S}|, |\bar{\Omega}|$ & Additional LN & S, C & Global & NN-Box-Simple-ALN-G \\
Two filter (box and BTF) & $\bar{P}, |\bar{S}|, |\bar{\Omega}|$ & Additional LN & S & Global & NN-BoxBTF-Simple-ALN-G \\
Two filter (box and DSCF) & $\bar{P}, |\bar{S}|, |\bar{\Omega}|$ & Additional LN & C & Global & NN-BoxDSCF-Simple-ALN-G \\
One filter (box) & $\bar{P}, |\bar{S}|, |\bar{\Omega}|$ & Original & S, C & Local & NN-Box-Simple-Original-L \\
Two filter (box and BTF) & $\bar{P}, |\bar{S}|, |\bar{\Omega}|$ & Additional LN & S & Local & NN-BoxBTF-Simple-ALN-L \\
Two filter (box and DSCF) & $\bar{P}, |\bar{S}|, |\bar{\Omega}|$ & Additional LN & C & Local & NN-BoxDSCF-Simple-ALN-L \\
\hline
\end{tabular}
}
\end{table}

\begin{table}
\caption{\label{tab:nnperf}
\textit{A priori} evaluation on the hold out test set, each value is given as mean (standard deviation). For the RMSE and DRMSE, lower values are better, while higher values are better for the correlation.}
\centering
\renewcommand{\arraystretch}{1.15}
\setlength{\tabcolsep}{6pt}
\resizebox{0.75\textwidth}{!}{
\begin{tabular}{lccc}
\hline
Neural Network & RMSE & DRMSE & Correlation \\
\hline
NN-Box-Complex-Original-G     & 0.751\,(0.017) & 0.664\,(0.026) & 0.660\,(0.020) \\
NN-BoxBTF-Complex-Original-G  & 0.772\,(0.017) & 0.693\,(0.028) & 0.636\,(0.020) \\
NN-BoxDSCF-Complex-Original-G & 0.755\,(0.017) & 0.675\,(0.028) & 0.660\,(0.020) \\
NN-BoxDSCFBTF-Complex-Original-G & 0.759\,(0.016) & 0.681\,(0.029) & 0.659\,(0.020) \\
NN-Box-Complex-ALN-G          & 0.750\,(0.017) & 0.661\,(0.026) & 0.661\,(0.020) \\
NN-BoxBTF-Complex-ALN-G       & 0.753\,(0.017) & 0.664\,(0.026) & 0.659\,(0.021) \\
NN-BoxDSCF-Complex-ALN-G      & 0.758\,(0.017) & 0.675\,(0.028) & 0.660\,(0.021) \\
NN-Box-Simple-Original-G      & 0.751\,(0.017) & 0.662\,(0.024) & 0.660\,(0.020) \\
NN-BoxBTF-Simple-Original-G   & 0.758\,(0.017) & 0.668\,(0.028) & 0.657\,(0.021) \\
NN-BoxDSCF-Simple-Original-G  & 0.757\,(0.017) & 0.672\,(0.027) & 0.661\,(0.020) \\
\hline
\end{tabular}
}
\end{table}

\section{\textit{A Priori Analysis}}

\subsection{One Filter versus Two Filter Results}
As seen in table~\ref{tab:nnperf}, where the neural networks are evaluated on the test set box filtered data, there is no \textit{a priori} performance degradation when training on two or more filters versus training on one filter since all RMSE, DRMSE, and correlation coefficient values are within 5 percent of each other. For example, neural networks trained with a max-min normalization and using the set of inputs $\bar{P}, \bar{Q}, \bar{R}, |\bar{S}|, |\bar{\Omega}|$ all have RMSE values in the 0.75-0.78 range, DRMSE values in the 0.66-0.70 range, and correlation coefficients in the 0.63-0.66 range. Also, using a complex or simple set of inputs has negligible difference on the neural network performance, suggesting that just using $|\bar{S}|, |\bar{\Omega}|$ is sufficient ($\bar{P} = 0$ for incompressible flow, and is manually masked to be zero to prevent the neural network from overfitting to noise. It is kept so that the model can generalize to compressibility in future work). 

\section{\textit{A Posteriori Analysis}}
The neural networks are evaluated on two different flow solvers running forced HIT on a $64 \times 64 \times 64$ grid (corresponding to 16$\Delta_{DNS}$) and a $Re_{\lambda} = 433$. A quantitative metric is also given, sum spectral error ($SSE = \sum_{k} |(log(E_{DNS_k})-log(E_{LES_k}))|, k_{PadeLibs} \in [0, 35],  k_{PadeOps} \in [0, 24]$), where $E_{DNS_k}$ denotes the filtered DNS spectra, $E(k)$ denotes the LES energy spectra, with differences in ``k" since PadeOps has two-thirds dealiasing. While SSE is a good first-order metric, it must be interpreted in a physical context. SSE is computed using the logarithm of the energy spectra values, rendering it artificially sensitive to deviations in the higher-wavenumber, dissipative range. This can manifest in SSE oftentimes assigning a larger SSE value to dissipative spectra even if it is physically desirable, while not putting enough emphasis on high-wavenumber energy buildup. Therefore, visual inspection of the spectral decay or pile-up is still essential. Other metrics that are investigated for HIT include plotting the PDF of the production term of the enstrophy equation, $\bar{\omega}_i \bar{S}_{ij} \bar{\omega}_j$, as well as analyzing the intermittency by plotting longitudinal flatness as a function of spatial spacing, where longitudinal flatness is defined as:

\begin{equation}
    F(r) = \frac{\langle (\delta u)^4 \rangle}{\langle (\delta u)^2 \rangle^2}
\end{equation}

and $\delta u$ is defined as $\delta u = \bar{u}(x+r) - \bar{u}(x)$, where r is the spatial spacing.

\subsection{Fourier-Spectral with Two-Thirds Dealiasing (PadeOps)}
The neural networks are integrated into the Fourier-Spectral code with two-thirds dealiasing (PadeOps), and after running to statistical stationarity, data is collected and plotted. In addition, two traditional models are run for reference. The Static Smagorinsky model is used with a coefficient set to $C = 0.18$, while the Sigma model is also run with a coefficient set to $C=1.3$. These are all within the ranges commonly used for these two models~\citep{nicoud2011using, smagorinsky1963general, meyers2006model}. The Sigma model was chosen as the coefficient range works for many different flows and does not require clipping for channel flow, whereas models such as the Dynamic Smagorinsky model may need clipping~\citep{meneveau1996lagrangian}. Table~\ref{tab:fouriernns} shows all the neural networks trained and integrated into the Fourier-Spectral solver, as well as the SSE for easy reference. 

\begin{table}
\caption{\label{tab:fouriernns}
\textit{A posteriori} evaluation for the Fourier-Spectral code for all cases tested.}
\centering
\renewcommand{\arraystretch}{1.15}
\setlength{\tabcolsep}{4pt} 
\small 
\begin{tabular}{lc c lc}
\toprule
Neural Network & SSE & \hspace{0.5cm} & Neural Network & SSE \\
\midrule
NN-Box-Complex-Original-G      & 18.87 & & NN-Box-Simple-Original-G      & 4.34 \\
NN-BTF-Complex-Original-G      & 4.32  & & NN-BoxBTF-Simple-Original-G   & 3.68 \\
NN-BoxBTF-Complex-Original-G   & 4.0   & & NN-Box-Simple-ALN-G           & 4.27 \\
NN-BoxDSCFBTF-Comp-Orig-G      & 4.58  & & NN-BoxBTF-Simple-ALN-G        & 4.16 \\
NN-Box-Complex-ALN-G           & 5.56  & & NN-Box-Simple-Original-L      & 4.59 \\
NN-BoxBTF-Complex-ALN-G        & 3.93  & & NN-BoxBTF-Simple-Original-L   & 4.98  \\

Static Smagorinsky        & 16.6  & & Sigma   & 13.34  \\
\bottomrule
\end{tabular}
\end{table}

\subsubsection{Effect of Multiple filters}
\begin{figure}
\centering
     \begin{subfigure}{0.49\textwidth}
         \centering
         \includegraphics[width=\textwidth]{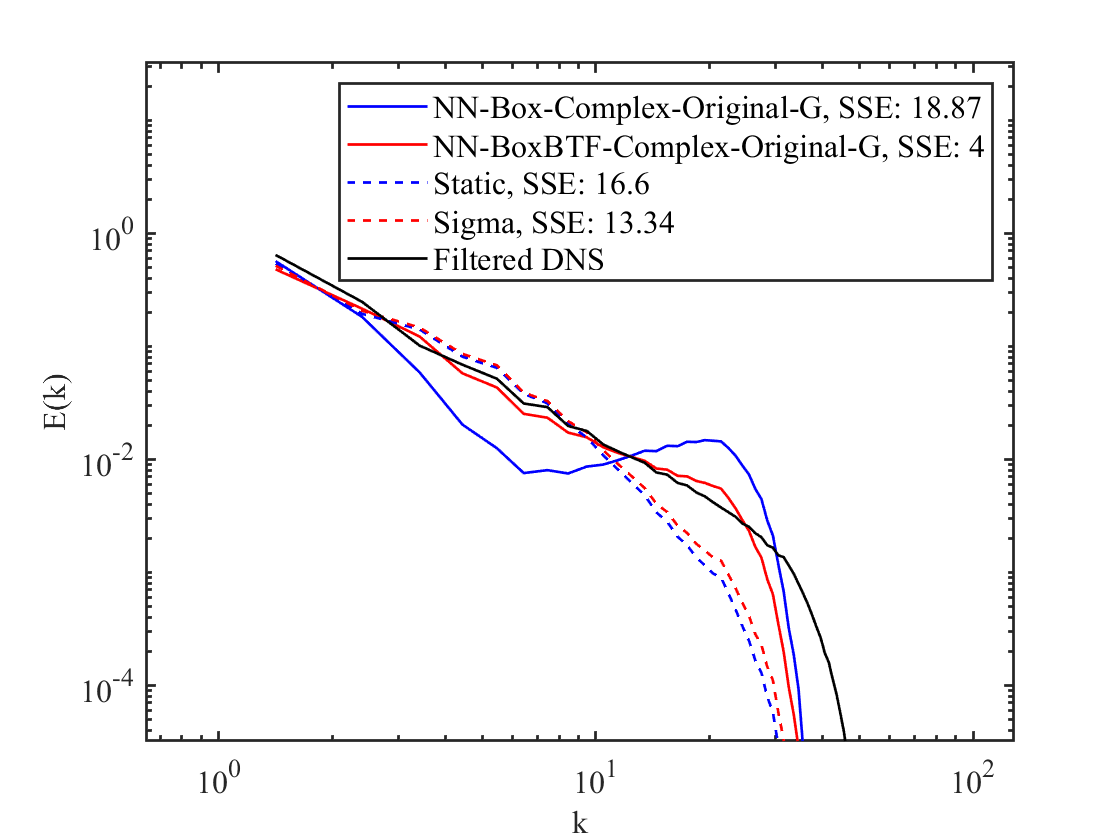}
         \caption{}
         \label{fig:CompareSpectra}
     \end{subfigure}
     \hspace{1em}
     \begin{subfigure}{0.425\textwidth}
         \centering
         \includegraphics[width=\textwidth]{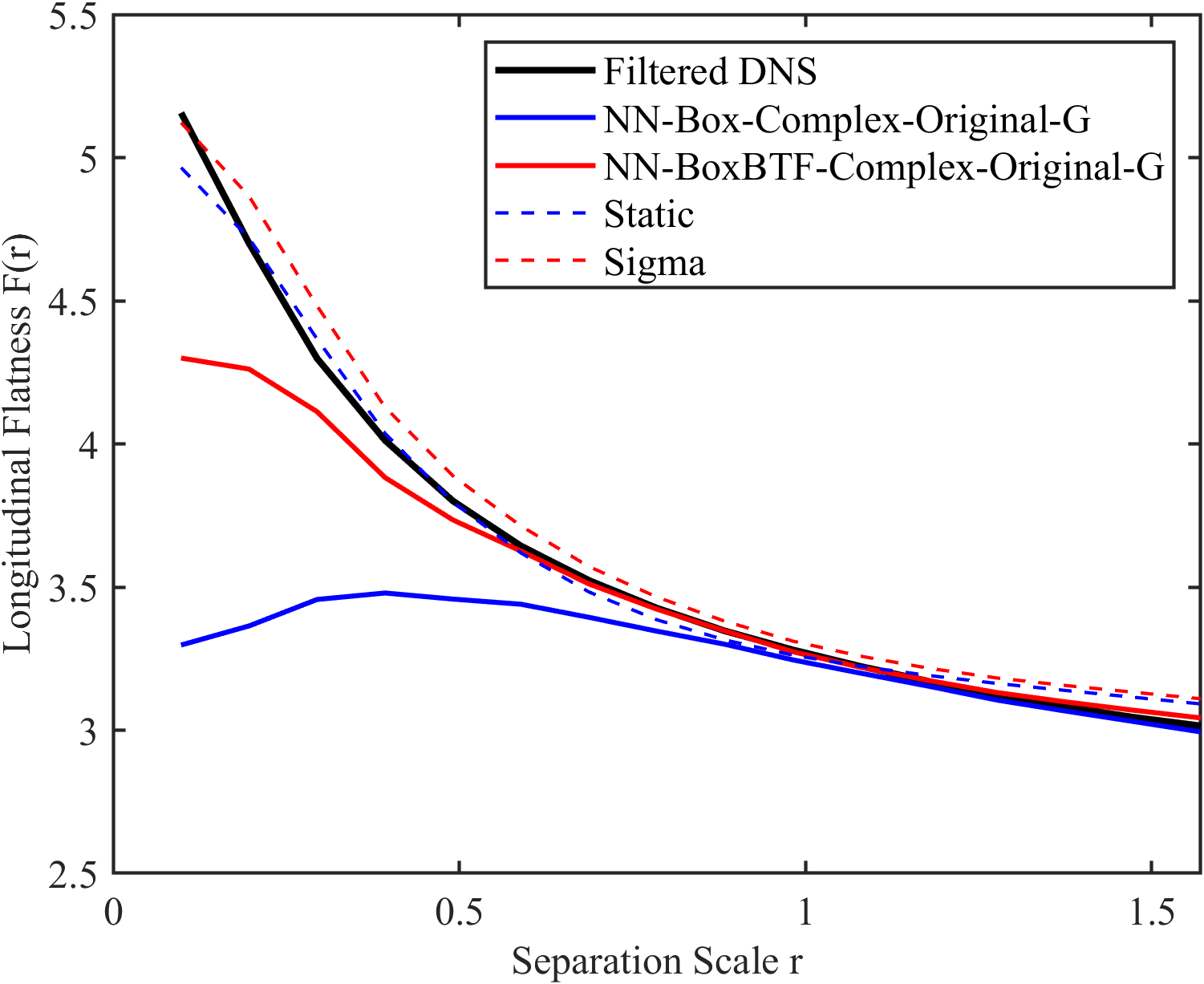}
         \caption{}
         \label{fig:CompareInter}
     \end{subfigure}
     \hspace{1em}
     \begin{subfigure}{0.49\textwidth}
         \centering
         \includegraphics[width=\textwidth]{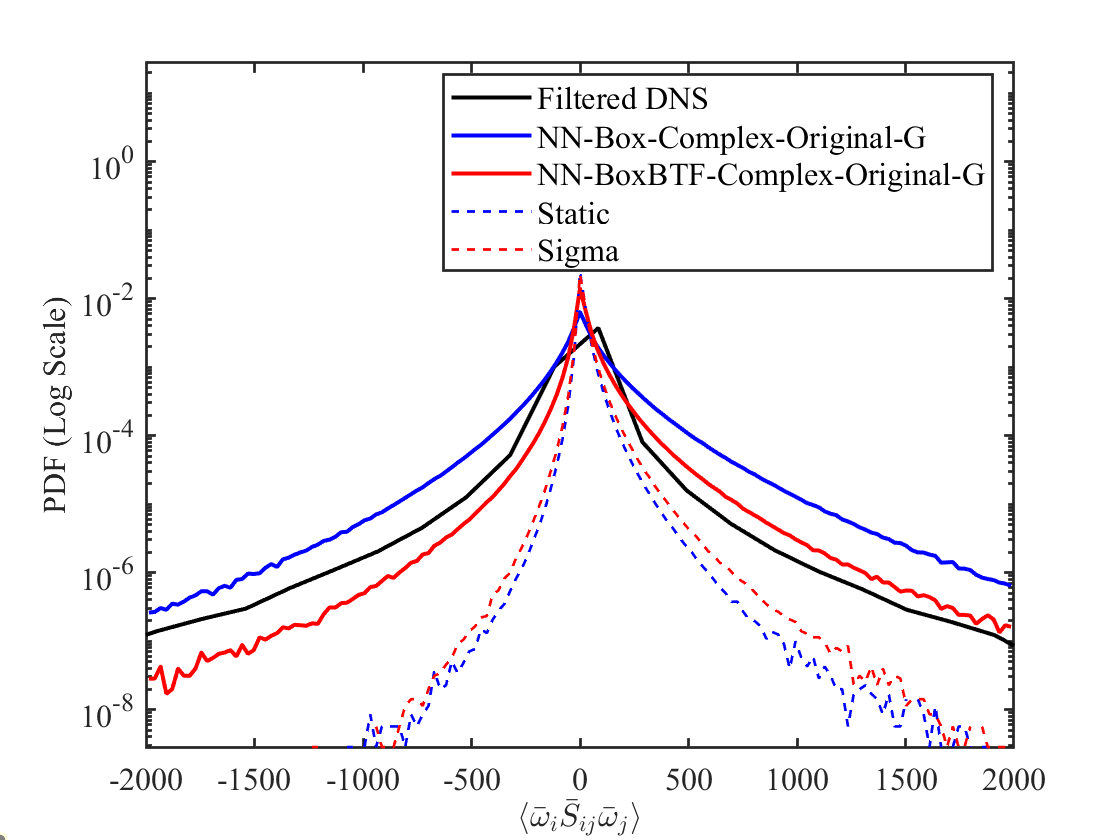}
         \caption{}
         \label{fig:CompareVortex}
     \end{subfigure}
     \caption{\textit{A posteriori} one and two filter comparisons with traditional models for the Fourier-spectral code with two-thirds dealiasing. Spectra, longitudinal flatness, and the PDF of $\overline{\omega_i}\overline{S_{ij}}\overline{\omega_j}$ are plotted. The GNN architecture used corresponds to the original architecture with the global normalization.}
\end{figure}

First, the effect of training on two filters as compared to one filter is analyzed, followed by the analysis of a three filter approach using the ``complex" set of inputs. Figures~\ref{fig:CompareSpectra} to~\ref{fig:CompareVortex} show the results of the neural network trained only with a box filter and the neural network trained with two filters (box and BTF) compared to the Static Smagorinsky and Sigma models.

From the spectra, it is seen that the Static Smagorinsky and Sigma models are over-dissipative, despite using the smallest coefficient in the recommended range for the Sigma model. This highlights that in the coarse-grid regime, traditional models may be slightly less accurate as large amounts of turbulent kinetic energy are unable to be resolved and the subgrid stress models must model larger fractions of the unresolved physics. Meanwhile, the neural network trained on one filter is under-dissipative, causing an unrealistic buildup of turbulent kinetic energy at the higher wavenumbers. The neural network trained on two filters is most able to accurately represent the spectra. The spectra fall-off for all of the LES results are expected when compared to filtered DNS due to the two-thirds dealiasing operation at the higher wavenumbers. When analyzing the longitudinal flatness, which characterizes the intermittency at different spatial scales, one can see that the two filter case is more accurate than the one filter case. However, in both these cases the neural network models underpredict the filtered DNS values, while the traditional models are more accurate in representing these values. This is expected, as over-dissipative models tend to reduce the variance (denominator of the longitudinal flatness calculation), resulting in higher values of ``rare events" compared to the variance. However, an intermittent flow that doesn't drive turbulence for HIT would not be as physically accurate. 

One of the primary drivers of 3D forced HIT is vortex stretching, which can be analyzed by plotting the production term of the resolved enstrophy equation. From figure~\ref{fig:CompareVortex}, the neural network models are revealed to drive enstrophy production far more accurately than the traditional models. The traditional models ``fall off" faster than the filtered DNS values, while the neural network models are more accurate, suggesting that they are driving the energy cascade in a more physical manner. The two-filter case is also more accurate than the one filter case, which can be seen in the positive tail of the enstrophy production. The one filter case, which over-predicts the enstrophy production, when combined with the low longitudinal flatness values at small separation scales, suggests that there is grid scale noise that is not being adequately dissipated, which is confirmed in the high-wavenumber buildup in the spectra. Overall, when combining all this analysis, the two filter case is far more physically accurate than the one-filter case, and in general the two-filter neural network case is more physically accurate than the traditional models in representing turbulent physics. This suggests that filtering plays a crucial role in the accuracy of neural network subgrid stress models that are trained outside of the LES solver. Oftentimes, for the enstrophy production, the PDF tails are lower when compared to filtered DNS (less variance). This can partially be attributed to the spectral fall-off induced by the two-thirds dealiasing scheme, as enstrophy production is dominated by the highest resolved wavenumbers. Thus, the two-thirds dealiasing scheme will cause slightly smoothed velocity gradients, which will lead to smaller peak magnitudes of enstrophy production and a tail that ``falls off" faster in the PDF.

\begin{figure}
\centering
     \begin{subfigure}{0.48\textwidth}
         \centering
         \includegraphics[width=\textwidth]{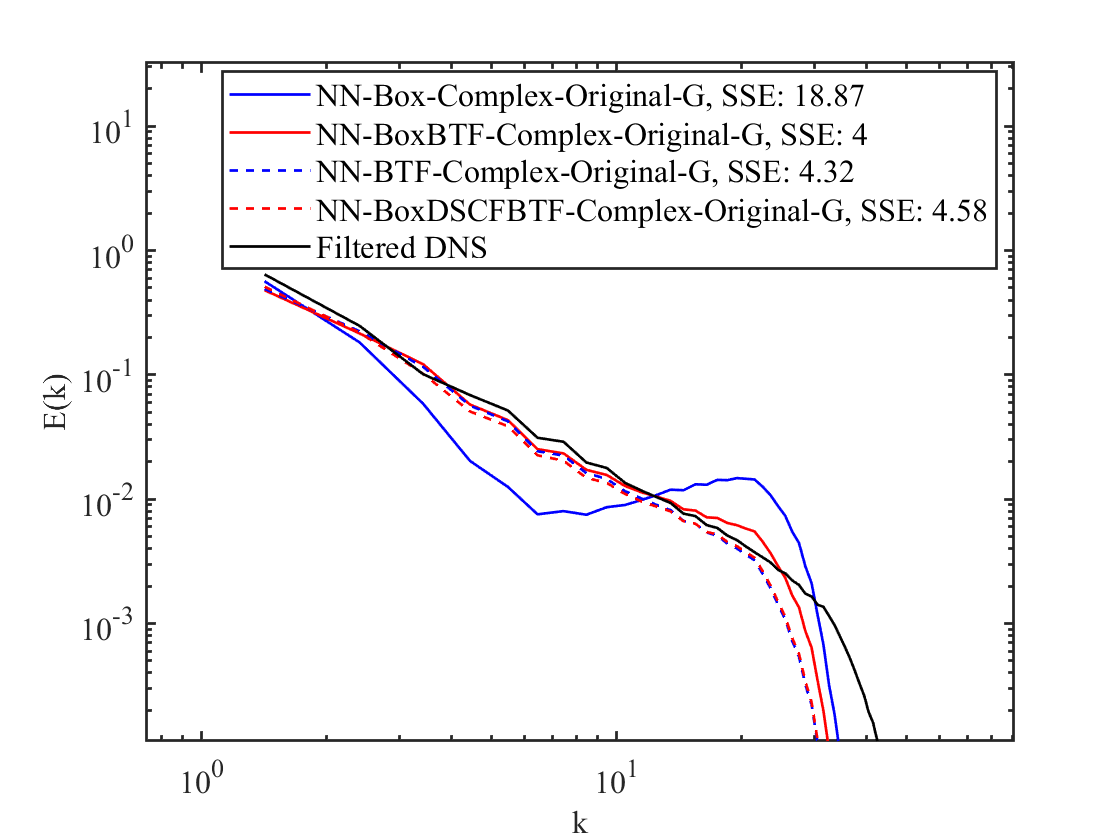}
         \caption{}
         \label{fig:FiltSpectra}
     \end{subfigure}
     \hspace{1em}
     \begin{subfigure}{0.48\textwidth}
         \centering
         \includegraphics[width=\textwidth]{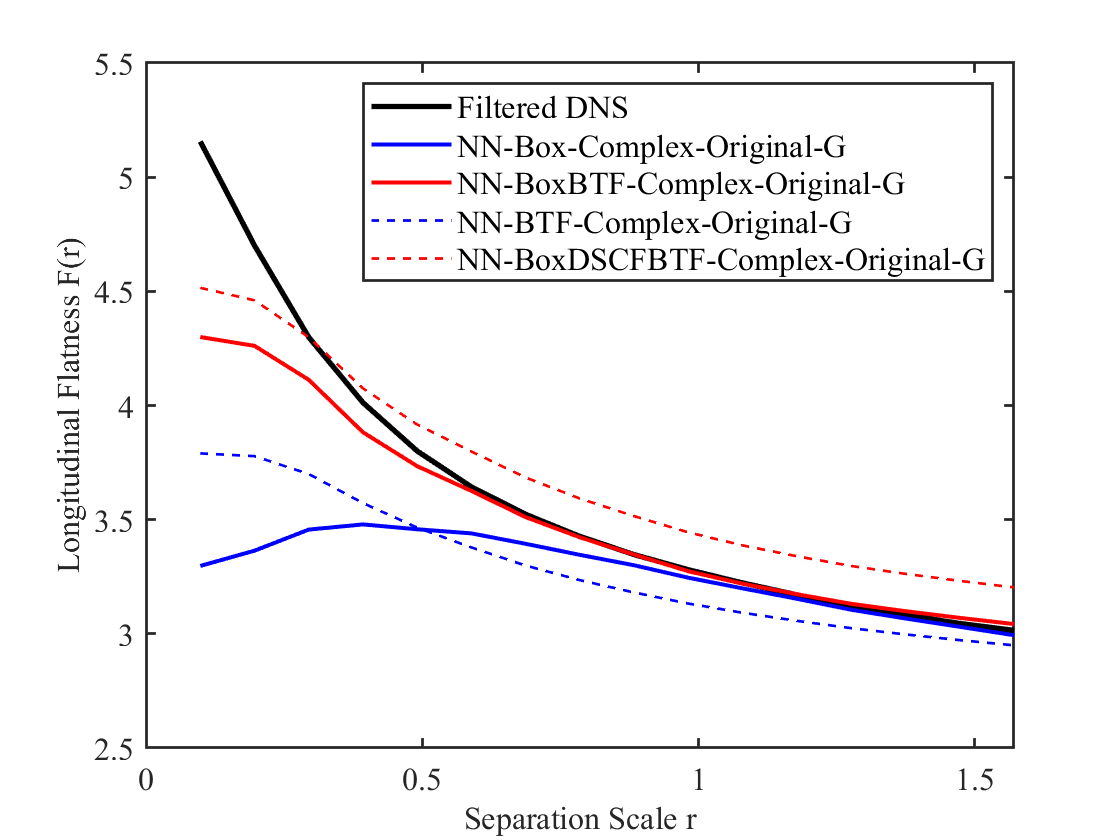}
         \caption{}
         \label{fig:FiltInter}
     \end{subfigure}
     \hspace{1em}
     \begin{subfigure}{0.49\textwidth}
         \centering
         \includegraphics[width=\textwidth]{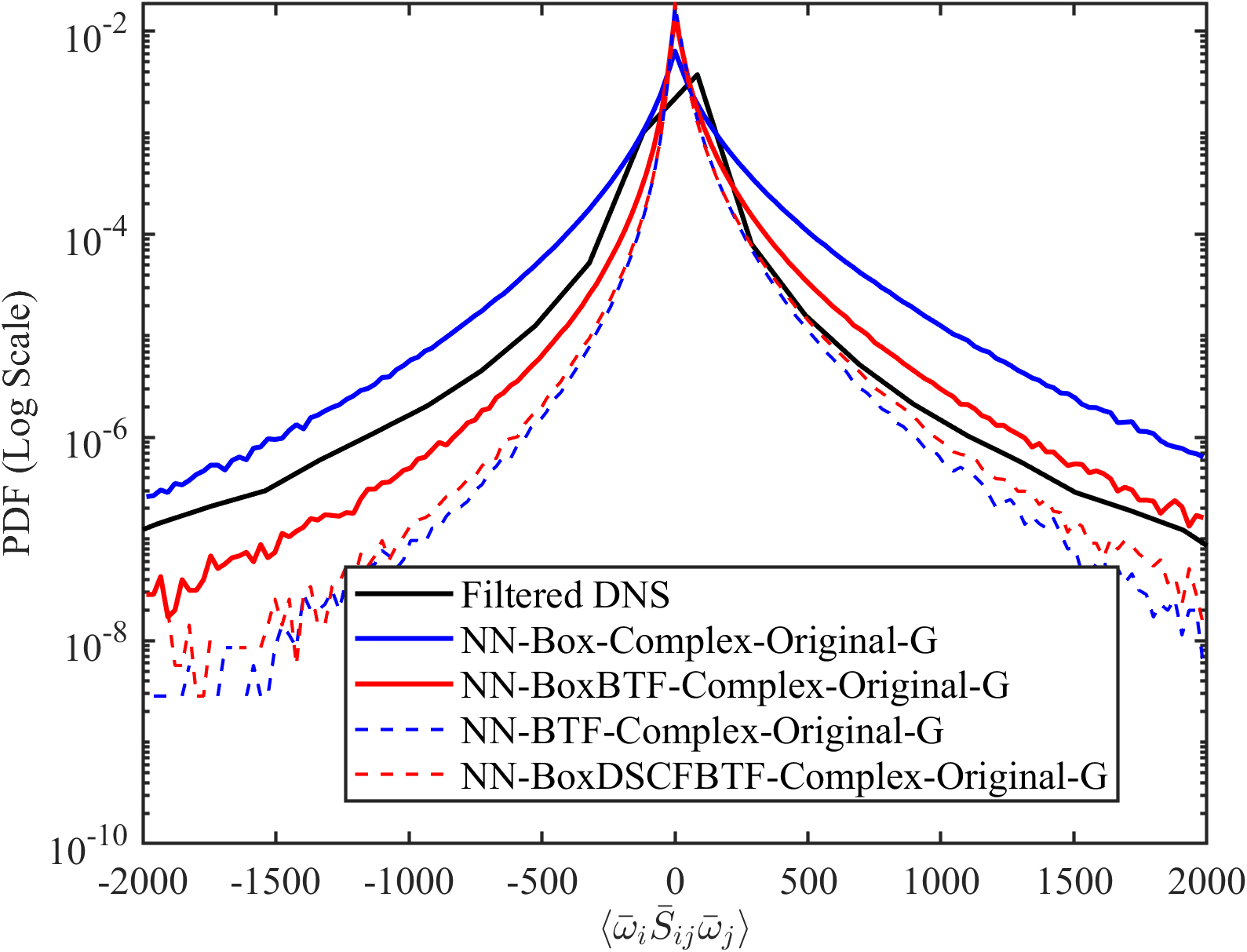}
         \caption{}
         \label{fig:FiltVortex}
     \end{subfigure}
     \caption{\textit{A posteriori} multiple filter comparisons for the Fourier-spectral code with two-thirds dealiasing. Spectra, longitudinal flatness, and the PDF of $\overline{\omega_i}\overline{S_{ij}}\overline{\omega_j}$ are plotted. The GNN architecture used corresponds to the original architecture with the global normalization, but trained with a variety of filter combinations ranging from one to three different filter shapes.}
\end{figure}

Next, the box-filter trained neural network and the two filter neural network are compared with a neural network trained on all three filters (box, BTF, DSCF) and a neural network trained on only the BTF filter. This will highlight the neural network performance when trained only on BTF, as well as the neural networks trained on all three filters, which can be applied without retraining to the 6th order compact LES solver too. The total comparison is shown in figures~\ref{fig:FiltSpectra} to~\ref{fig:FiltVortex}.

The spectra indicate that the neural network trained with only the BTF filter is more accurate than the neural network trained with only the box filter, which is to be expected as BTF accounts for two-thirds dealiasing while the box filter does not. Meanwhile, the three filter neural network also does well in producing flows with physically accurate spectra. As seen in the SSE from table~\ref{tab:fouriernns} and figure~\ref{fig:FiltSpectra}, the SSE is the largest for the box filter case, while the other neural networks trained have more reasonable SSE. However, when taking a look at the longitudinal flatness, the benefits of training on multiple filters are still clear. For the neural network trained on just the BTF, at small separation scales, the overall intermittency of the flow is still lower as compared to both the two and three filter cases. In fact, the neural networks trained on just one filter (whether it is box or BTF) are the worst, while the multiple filter cases are better in capturing the intermittency. Overall, all cases have similar behavior at larger separation scales, although the three filter case overpredicts the longitudinal flatness, but is also the best at predicting the flow intermittency at the smallest separation scales. When analyzing the production of enstrophy, the BTF-trained neural network and the three filter trained neural network show similar behavior, although the three filter trained neural network is slightly more accurate (closer to filtered DNS). For this case, the two filter (box and BTF) trained neural network is still the most accurate. In general, when considering the spectra, intermittency, and the production of enstrophy, the two cases that are closest to filtered DNS are the two and three filter trained neural networks. This suggests that it is beneficial for the neural network to be trained on multiple filter shapes, as opposed to one filter shape. By doing so, the neural network is less likely to ``overfit" to a specific filter shape and can become a more robust model for \textit{a posteriori} LES applications, where the implied filter is unknown. The implied filter width for LES is a product of both the grid and numerical method, and thus the implied SGS distribution is entirely dependent on the implied filter. By training on multiple different filter shapes, the output distribution shift for $\tau_{ij}$ can be better reduced.

\subsubsection{Neural Network Perturbation}
\begin{figure}
\centering
     \begin{subfigure}{0.44\textwidth}
         \centering
         \includegraphics[width=\textwidth]{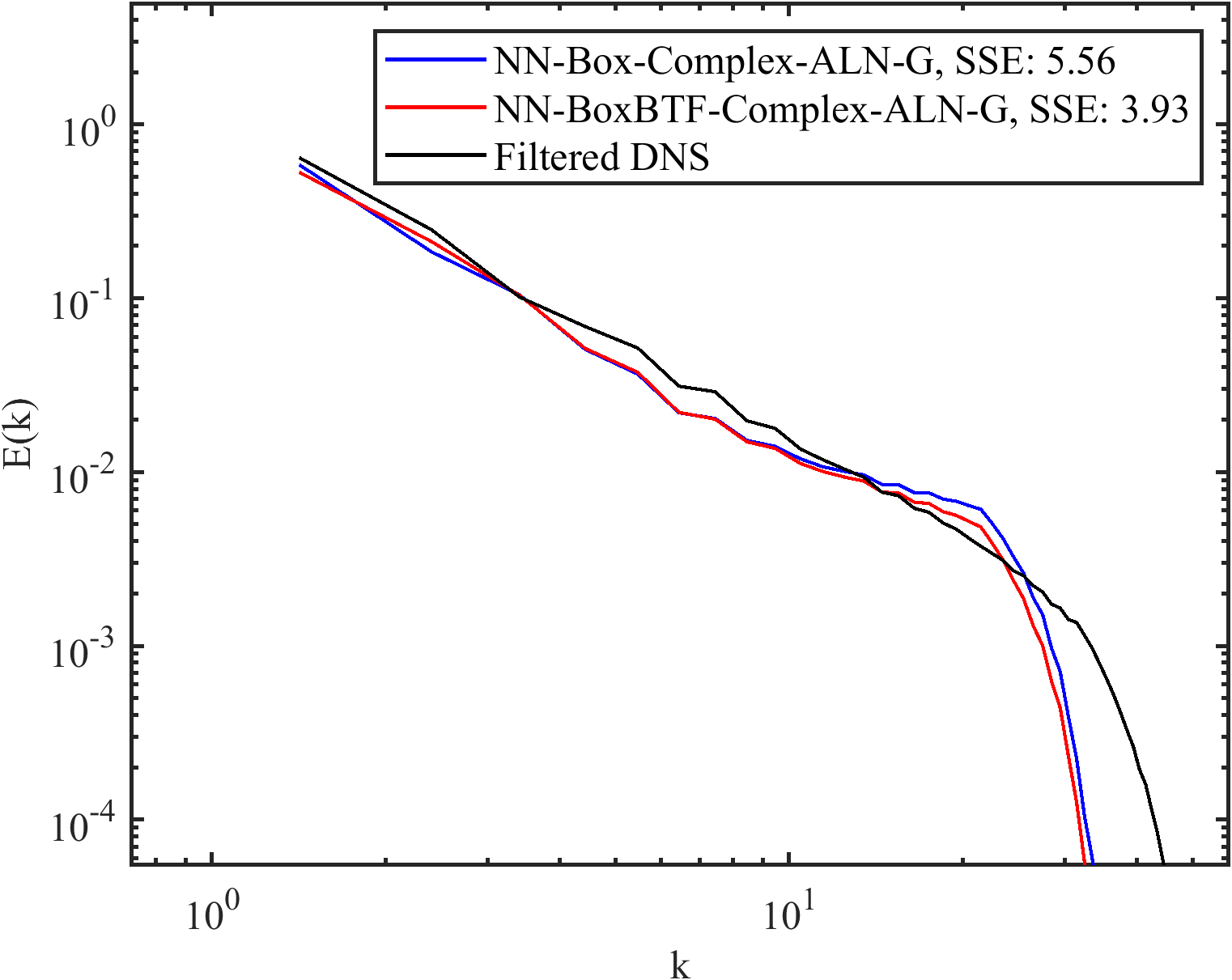}
         \caption{}
         \label{fig:ALNSpectra}
     \end{subfigure}
     \hspace{1em}
     \begin{subfigure}{0.43\textwidth}
         \centering
         \includegraphics[width=\textwidth]{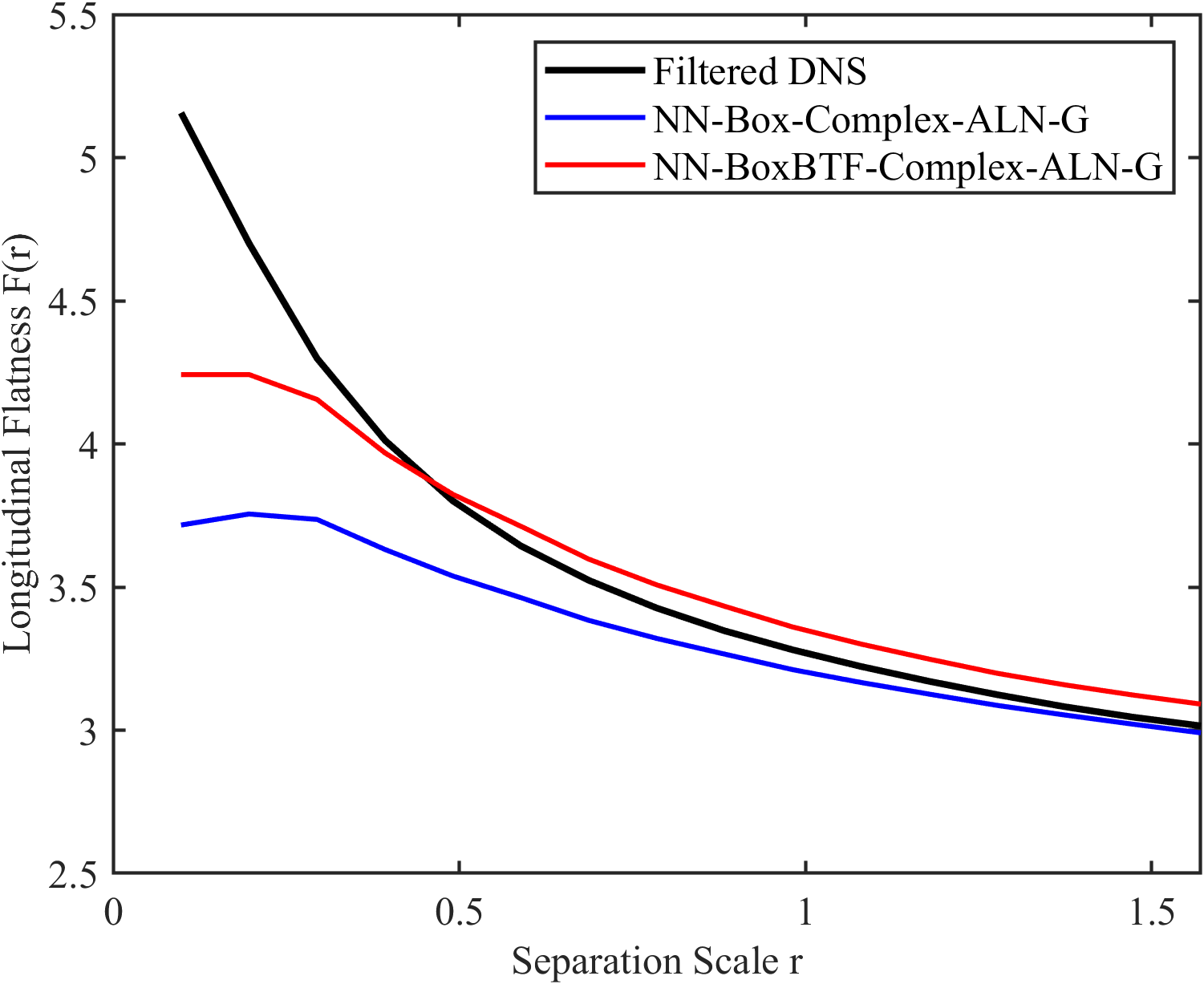}
         \caption{}
         \label{fig:ALNInter}
     \end{subfigure}
     \hspace{1em}
     \begin{subfigure}{0.49\textwidth}
         \centering
         \includegraphics[width=\textwidth]{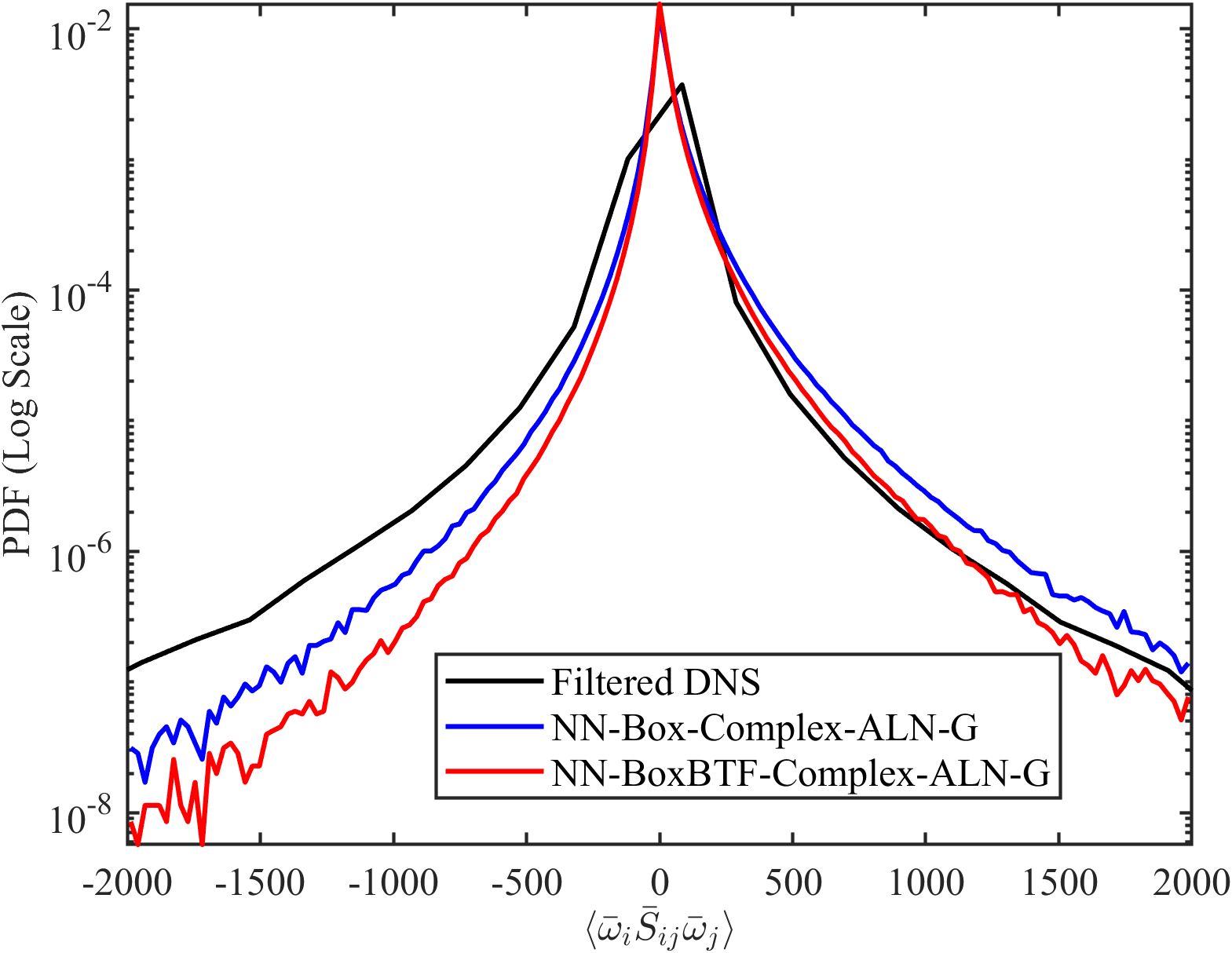}
         \caption{}
         \label{fig:ALNVortex}
     \end{subfigure}
     \caption{\textit{A posteriori} ALN comparisons for the Fourier-spectral code with two-thirds dealiasing. Spectra, longitudinal flatness, and the PDF of $\overline{\omega_i}\overline{S_{ij}}\overline{\omega_j}$ are plotted. The GNN architecture used corresponds to the variant ALN architecture with the global normalization.}
\end{figure}

The box-filtered and two filter training data is also used to train a neural network variant with additional layer normalization to evaluate if the accuracy increase due to the two filter training is present across neural network perturbations. These results are shown in figures~\ref{fig:ALNSpectra} to~\ref{fig:ALNVortex}. 

From the spectra, the one-filter trained neural network is now significantly better as compared to the previous case, suggesting that the results are very sensitive to the neural network architecture. Meanwhile, the two filter results are more robust across a wider range of neural network architectures, as the shape of the spectra doesn't change much. As seen, there is still an improvement in the spectra when training on two filters as compared to one filter, although the effect is less drastic for this neural network architecture perturbation. However, the longitudinal flatness still suggests that the two filter case is better at representing the overall ``tails" of the flow, as it shows better small scale intermittency. Meanwhile, both neural network models are accurate in enstrophy production, with the one filter case producing a more accurate PDF at the negative tail but the two filter case producing a more accurate PDF at the positive tail. Overall, the two filter case is still more physically accurate as compared to the one filter case, and when comparing with the previous case, the two filter case is far more robust to architecture perturbations than the one filter case.

\subsubsection{Using Simpler Inputs}
\begin{figure}
\centering
     \begin{subfigure}{0.49\textwidth}
         \centering
         \includegraphics[width=\textwidth]{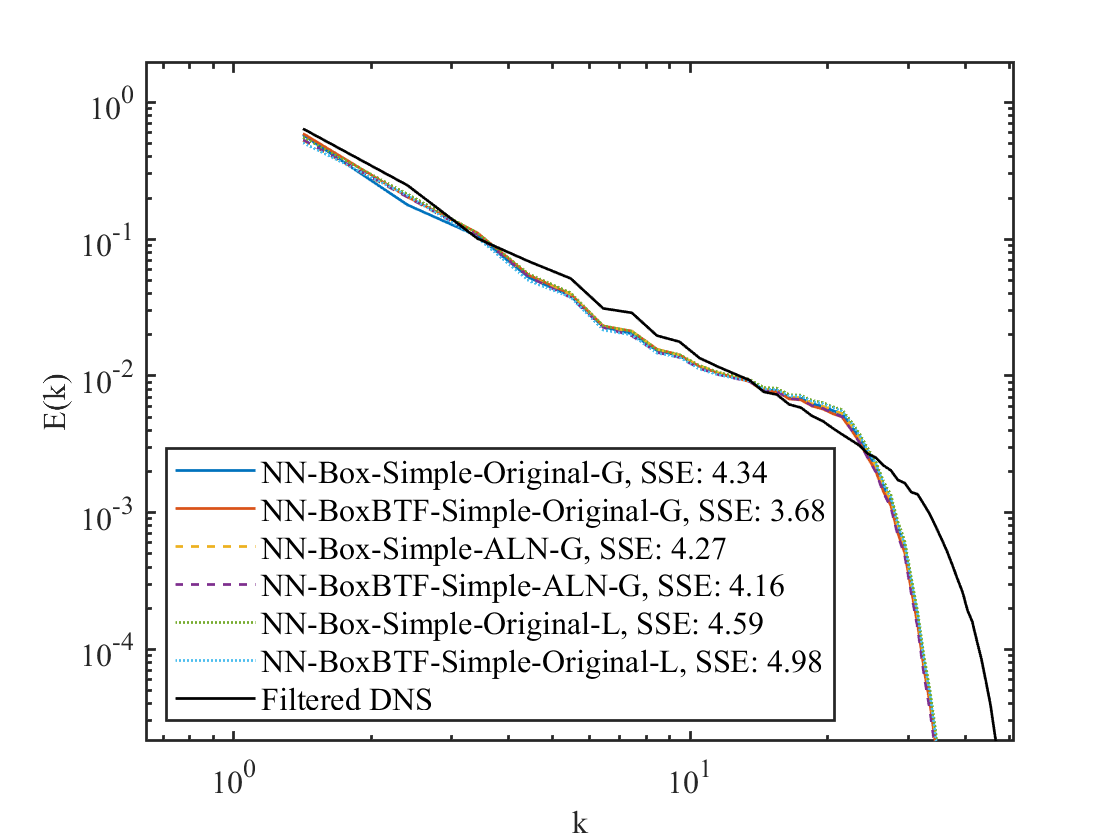}
         \caption{}
         \label{fig:SimpleSpectra}
     \end{subfigure}
     \hspace{1em}
     \begin{subfigure}{0.425\textwidth}
         \centering
         \includegraphics[width=\textwidth]{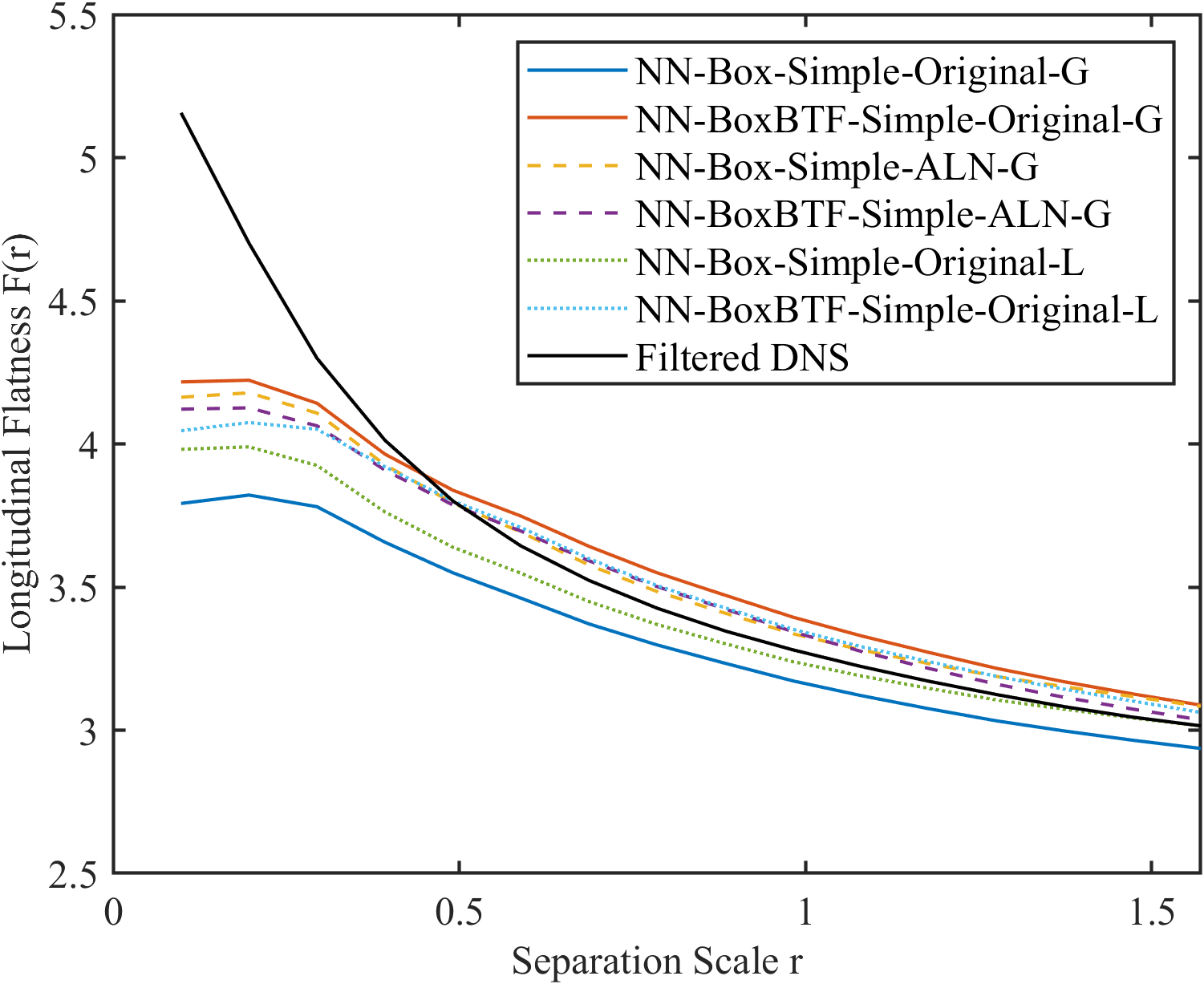}
         \caption{}
         \label{fig:SimpleInter}
     \end{subfigure}
     \hspace{1em}
     \begin{subfigure}{0.49\textwidth}
         \centering
         \includegraphics[width=\textwidth]{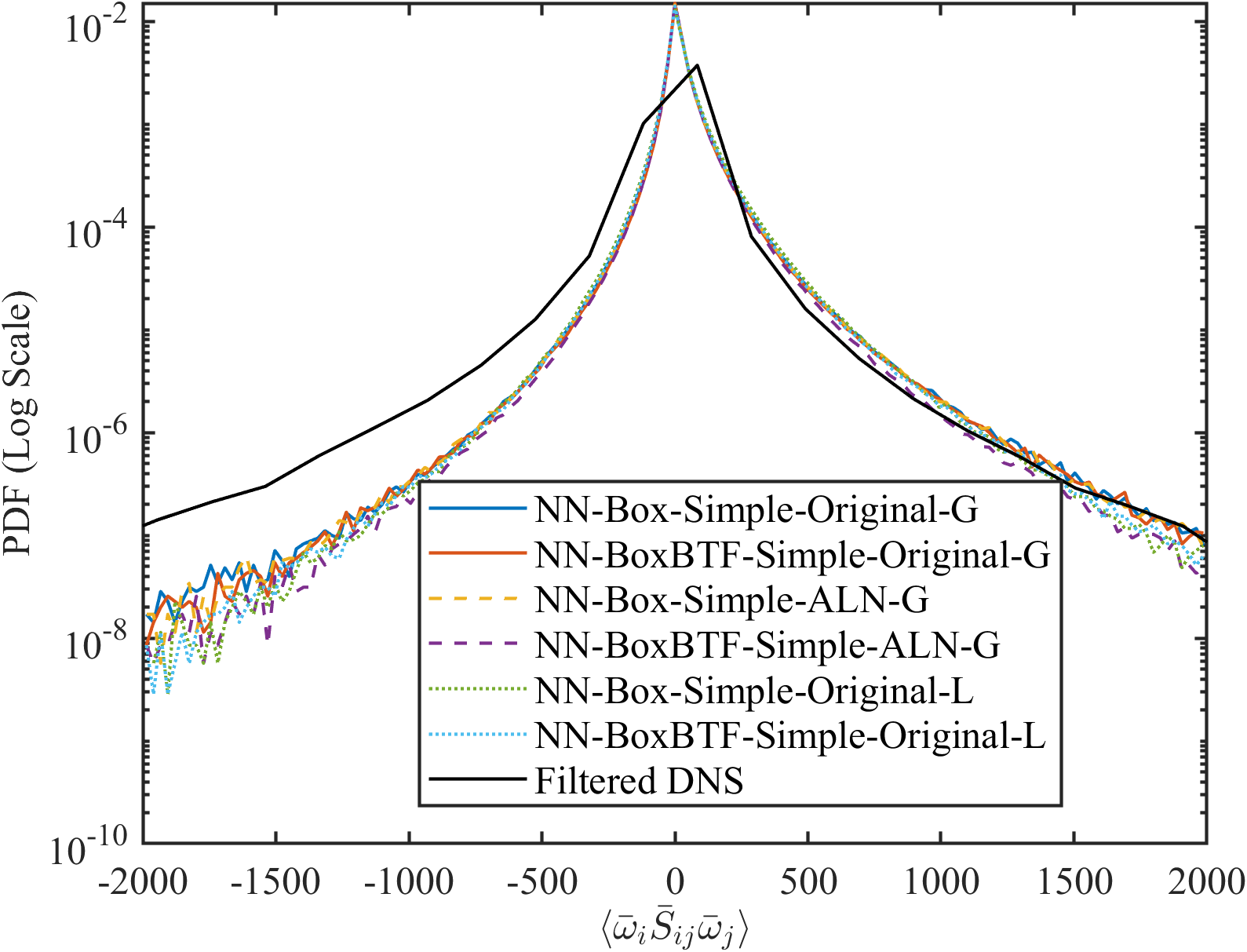}
         \caption{}
         \label{fig:SimpleVortex}
     \end{subfigure}
     \caption{\textit{A posteriori} simple input comparisons for the Fourier-spectral code with two-thirds dealiasing. Spectra, longitudinal flatness, and the PDF of $\overline{\omega_i}\overline{S_{ij}}\overline{\omega_j}$ are plotted. Both the original and ALN GNN architectures are tested with ``simple" inputs using both local and global normalization.}
\end{figure}

Now, neural networks are trained on the simpler inputs to evaluate the distribution shift associated with neural network inputs. The hypothesis is that simpler inputs suffer less from a distribution shift, and so there should be less variation in these results. The neural networks trained involve both the variant architecture and the original architecture, as well as a different input normalization to show that the results are robust to a wide range of conditions. The results are shown in figures~\ref{fig:SimpleSpectra} to~\ref{fig:SimpleVortex}.

Across all the different neural network models, the spectra are all relatively consistent, which validates the idea that simple inputs result in less distribution shift as compared to more complex inputs. From the SSE from table~\ref{tab:fouriernns} and figure~\ref{fig:SimpleSpectra}, the neural networks trained with simple inputs have relatively similar SSE values across a wide range of training conditions. When evaluating the longitudinal flatness at small separation scales, the two filter trained neural networks in general, predict a higher (closer to filtered DNS) value as compared to the neural networks trained on one filter. The enstrophy production PDFs show relatively similar performance across all the various models. This suggests that training on two filters will often do no worse than training on one filter, and can still obtain some performance improvements in select situations, highlighting the utility of training on two filters. This is seen in two different neural network architectures and two different input normalizations. Overall, using simpler inputs does reduce the distribution shift seen \textit{a priori} versus \textit{a posteriori}, and training on one versus two filters has less of a difference, but can still improve some qualities of the flow. Ultimately, this suggests that the input distribution shift has a large effect on the neural network SGS models \textit{a posteriori}, as simpler inputs (which suffer less from distribution shift) produce neural network models that have less \textit{a posteriori} variance across models. When the input distribution shift is reduced, training on two filters can still help increase robustness \textit{a posteriori} by reducing the output distribution shift, although this becomes a more secondary effect.

\subsection{6th Order Compact (PadeLibs)}
The neural networks are now integrated into a 6th order compact code, PadeLibs, to investigate if the results hold for a flow solver with a different numerical method. Similarly, forced HIT is run with the same parameters as the Fourier-Spectral code, and the spectra, longitudinal flatness, and enstrophy production PDFs are plotted. Table~\ref{tab:padenns} shows all the neural networks trained and integrated into this flow solver, as well as SSE for easy reference.

\begin{table}
\caption{\label{tab:padenns}
\textit{A posteriori} evaluation for the 6th order compact code for all cases tested.}
\centering
\renewcommand{\arraystretch}{1.15}
\setlength{\tabcolsep}{4pt} 
\small 
\begin{tabular}{lc c lc}
\toprule
Neural Network & SSE & \hspace{0.5cm} & Neural Network & SSE \\
\midrule
NN-Box-Complex-Original-G      & 32.58 & & NN-Box-Simple-Original-G      & 7.56 \\
NN-DSCF-Complex-Original-G      & 35.33  & & NN-BoxDSCF-Simple-Original-G   & 10.97 \\
NN-BoxDSCF-Complex-Original-G   & 11.93   & & NN-Box-Simple-ALN-G           & 9.50 \\
NN-BoxDSCFBTF-Comp-Orig-G      & 14.22  & & NN-BoxDSCF-Simple-ALN-G        & 12.22 \\
NN-Box-Complex-ALN-G           & 9.50  & & NN-Box-Simple-Original-L      & 12.56 \\
NN-BoxDSCF-Complex-ALN-G        & 5.59  & & NN-BoxDSCF-Simple-Original-L   & 15.84  \\
\bottomrule
\end{tabular}
\end{table}

\subsubsection{Effect of Multiple filters}
\begin{figure}
\centering
     \begin{subfigure}{0.43\textwidth}
         \centering
         \includegraphics[width=\textwidth]{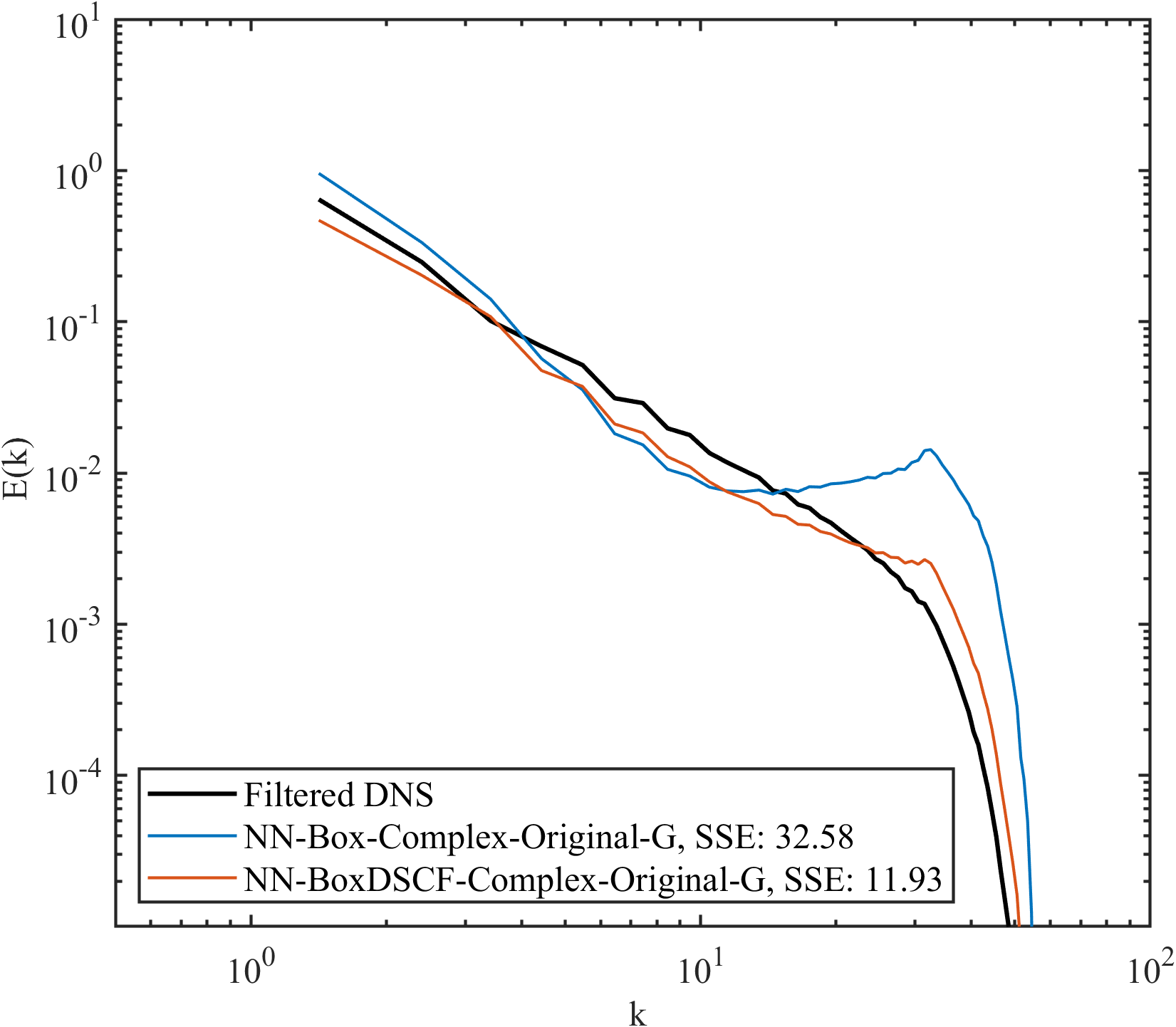}
         \caption{}
         \label{fig:PlibCompareSpectra}
     \end{subfigure}
     \hspace{1em}
     \begin{subfigure}{0.46\textwidth}
         \centering
         \includegraphics[width=\textwidth]{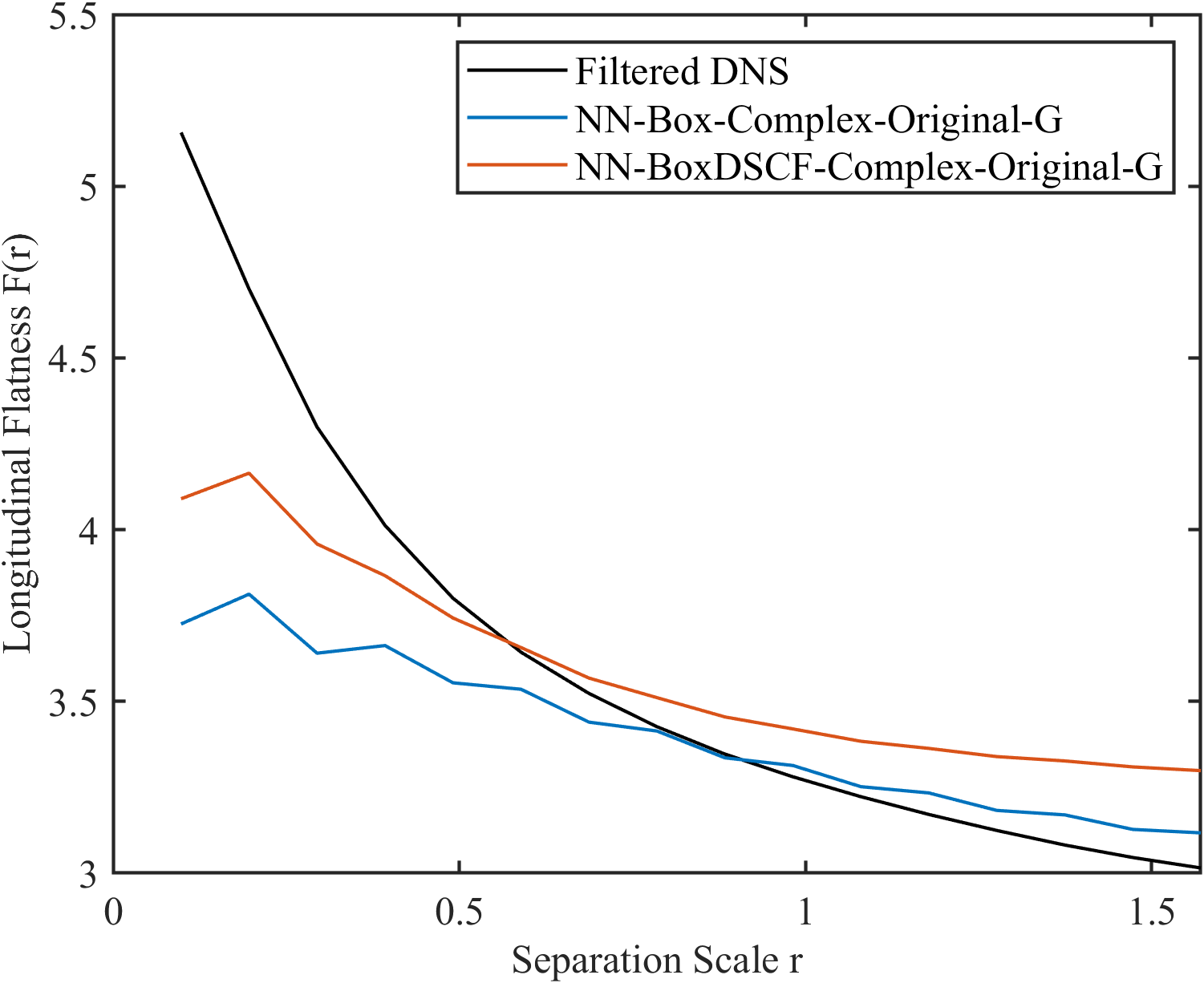}
         \caption{}
         \label{fig:PlibCompareInter}
     \end{subfigure}
     \hspace{1em}
     \begin{subfigure}{0.45\textwidth}
         \centering
         \includegraphics[width=\textwidth]{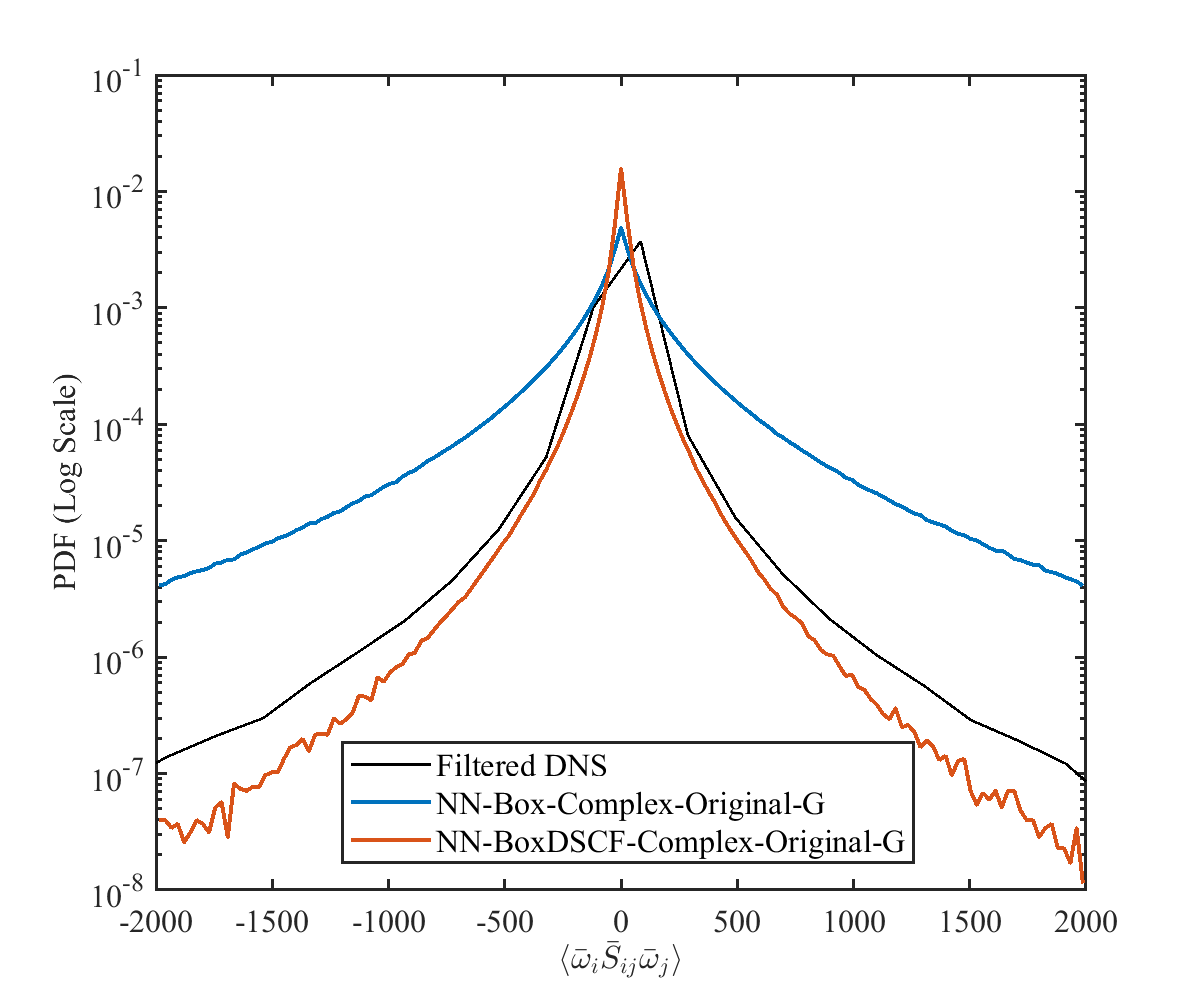}
         \caption{}
         \label{fig:PlibCompareVortex}
     \end{subfigure}
     \caption{\textit{A posteriori} one and two filter comparisons for the 6th order compact code. Spectra, longitudinal flatness, and the PDF of $\overline{\omega_i}\overline{S_{ij}}\overline{\omega_j}$ are plotted. The GNN architecture used corresponds to the original architecture with the global normalization.}
\end{figure}

First, the neural network trained on only the box filter is analyzed against the two filter case. In this case, the two filter training data corresponds to data prepared with both the DSCF and the box filter. These results are shown in figures~\ref{fig:PlibCompareSpectra} to~\ref{fig:PlibCompareVortex}.

The spectra indicate that the neural network trained on two filters is more physically accurate than the neural network trained on one filter, which suffers from an extreme buildup of high wavenumber energy. This is consistent with the previous Fourier-Spectral results, showing robustness across two different high order numerical methods. In addition, the small scale longitudinal flatness also shows improvements when training on two filters instead of one filter, which is also consistent with the previous results on the Fourier-Spectral code. For the large scale longitudinal flatness, both cases flatten out at higher values as compared to the filtered DNS profile. The enstrophy production PDF plot shows that the neural network trained on one filter is significantly over-predicting the enstrophy production, and combined with the smaller longitudinal flatness value at small separation scales (suggests high variance compared to the intermittent events), highlights that there is spurious grid-scale energy that is not being dissipated by the model. Overall, the results indicate that training on two filters is far better than training on only a box filter. This suggests that for a solver with a different high-order numerical method, the idea of training on two filters to reduce the output distribution shift of $\tau_{ij}$ holds, and produces more robust models \textit{a posteriori}.

\begin{figure}
\centering
     \begin{subfigure}{0.43\textwidth}
         \centering
         \includegraphics[width=\textwidth]{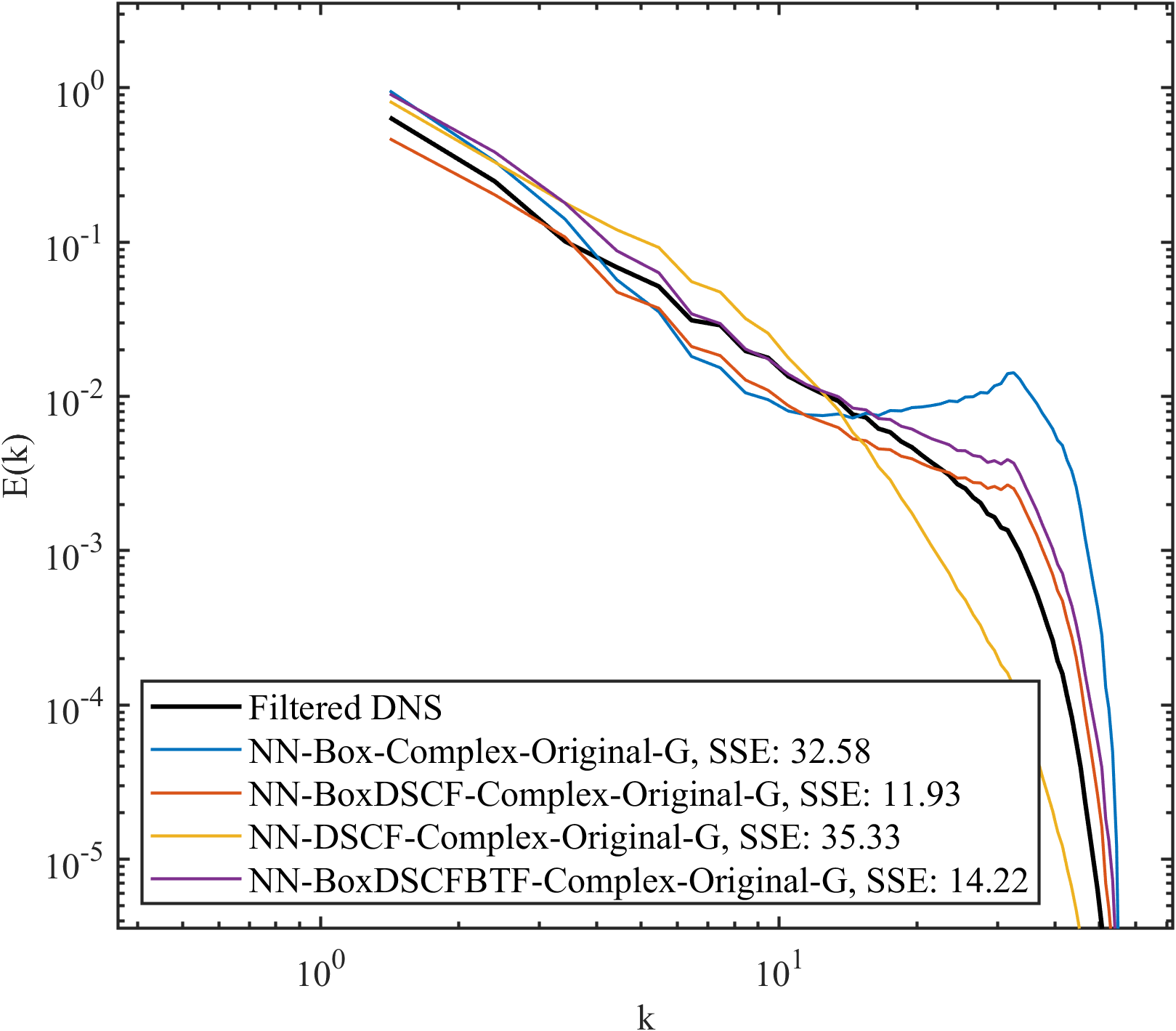}
         \caption{}
         \label{fig:PlibFiltSpectra}
     \end{subfigure}
     \hspace{1em}
     \begin{subfigure}{0.46\textwidth}
         \centering
         \includegraphics[width=\textwidth]{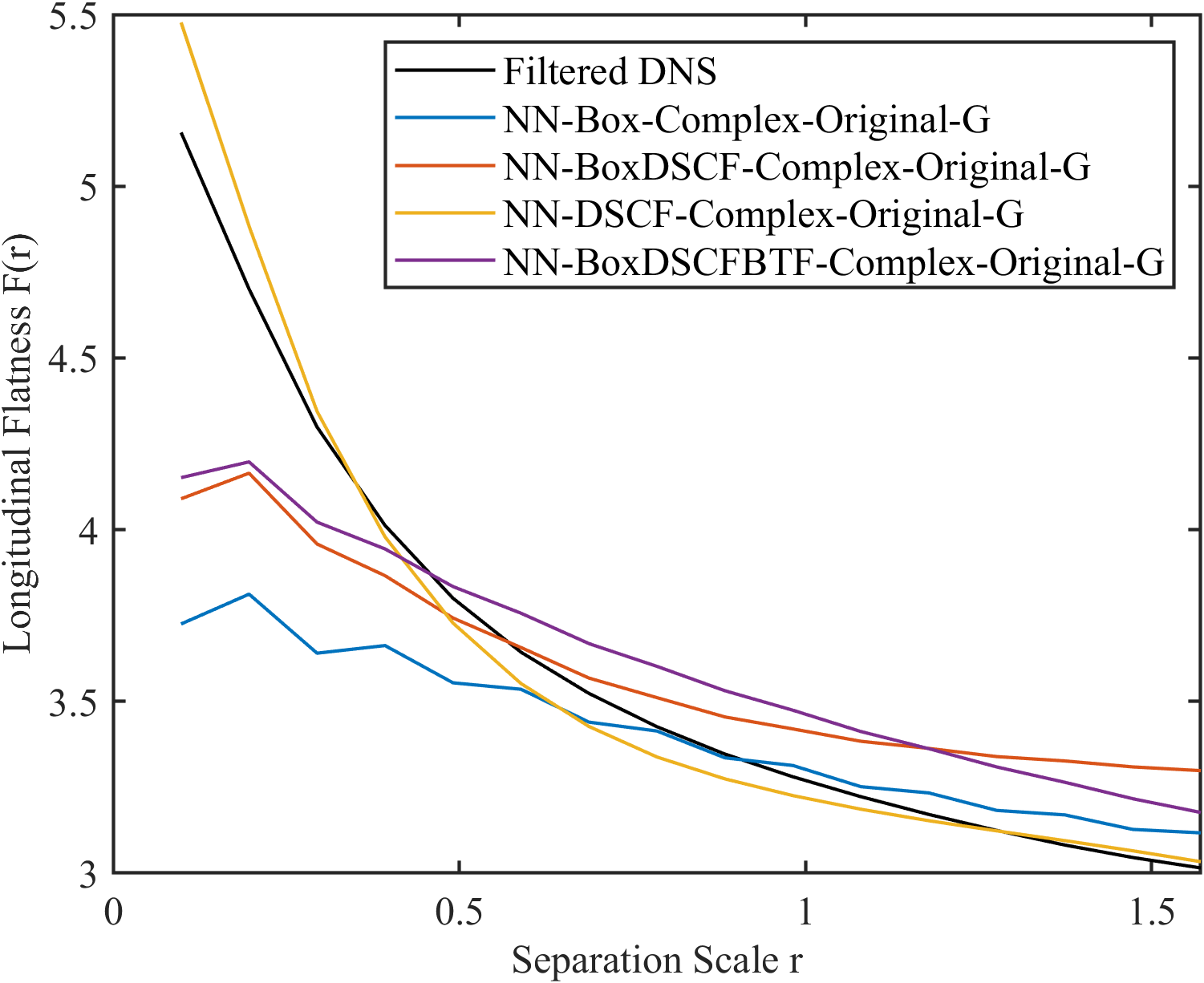}
         \caption{}
         \label{fig:PlibFiltInter}
     \end{subfigure}
     \hspace{1em}
     \begin{subfigure}{0.45\textwidth}
         \centering
         \includegraphics[width=\textwidth]{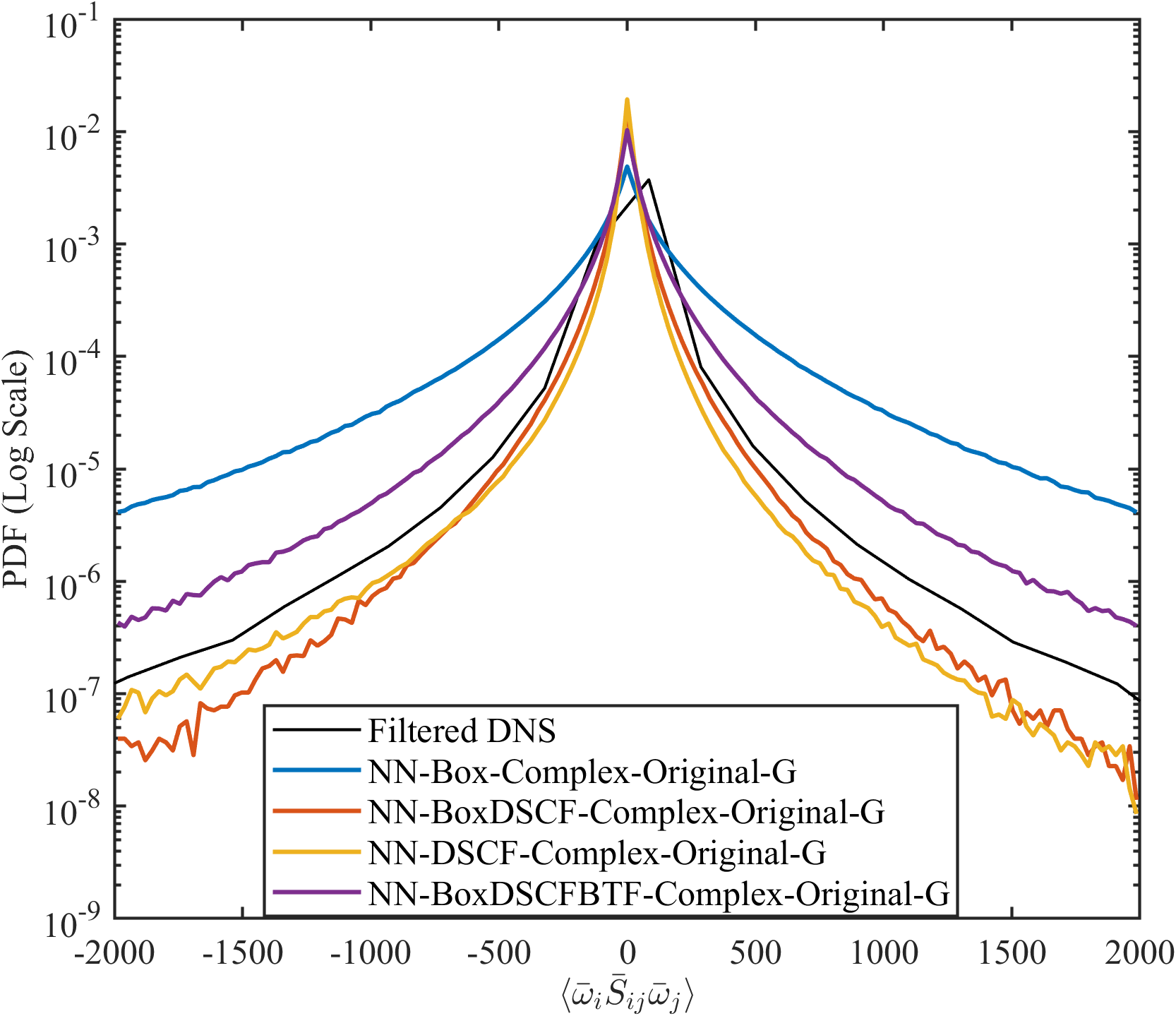}
         \caption{}
         \label{fig:PlibFiltVortex}
     \end{subfigure}
     \caption{\textit{A posteriori} multiple filter comparisons for the 6th order compact code. Spectra, longitudinal flatness, and the PDF of $\overline{\omega_i}\overline{S_{ij}}\overline{\omega_j}$ are plotted. The GNN architecture used corresponds to the original architecture with the global normalization, but trained with a variety of filter combinations ranging from one to three different filter shapes.}
\end{figure}

Next, these results are contrasted with a neural network trained on only the DSCF, as well as the same three-filter neural network that was used in the Fourier-Spectral solver. These results are shown in figures~\ref{fig:PlibFiltSpectra} to~\ref{fig:PlibFiltVortex}. 

The neural network trained with just the DSCF is over-dissipative, which highlights the utility of training on two or more filters. In this case, while the three filter case is slightly more under-dissipative as compared to the two-filter case, it is still far better than either of the one-filter cases. This can also be seen in table~\ref{tab:padenns} and figure~\ref{fig:PlibFiltSpectra}, where the SSE values for the two filter and three filter cases are lower than either of the one filter cases. The DSCF-trained neural network produces a longitudinal flatness curve that is more similar to the dissipative traditional models. This is expected, as the DSCF-trained neural network is over-dissipative too, and in this case actually over-predicts the small scale longitudinal flatness. When combined with the over-dissipative nature of DSCF, this suggests that the overall ``background" variance of the DSCF-trained neural network simulation is lower, which produces a large longitudinal flatness. Meanwhile, the three filter case in general has a more accurate longitudinal flatness than the single box filter case and the two filter case, even at the smaller separation scales. When evaluating the enstrophy production PDFs, the neural networks trained with two or three filters outperform the box filter trained neural network. The DSCF trained neural network produces an enstrophy production PDF that is also reasonable. When compared with the two filter case, it is better at predicting the negative tail but is worse at predicting the positive tail, although both underpredict the PDF. However, the DSCF achieves this prediction by sacrificing the spectra, as the rounded shape shows little indication of an inertial range, suggesting that it is unable to capture the overall physics. Meanwhile, the three filter case over-predicts the PDF at both the positive and negative tails, a potential consequence of training on the oscillatory BTF when the flow solver doesn't apply two-thirds dealiasing. In general, the two and three filter cases retain more of an inertial range in the energy spectrum, while also doing well in predicting the enstrophy production PDFs. Overall, this suggests that training on two or three filters results in models that are more physically accurate than training on one filter, and that the three filter case produces accurate results across both flow solvers, although it may not be the most optimal model as it was trained on a huge variety of filters, not all of which are applicable to the flow solver (such as training on BTF when the flow solver doesn't use two-thirds dealiasing). Despite this, three filter training can be applied to solvers with different numerical methods without re-training, and outperforms training on one filter. Overall, this suggests that training on multiple filters to reduce the neural network output distribution shift of $\tau_{ij}$ while ensuring that the neural network model doesn't ``overfit" to the specific characteristics of one specific filter shape is an effective way of training more robust models \textit{a posteriori}.

\subsubsection{Neural Network Perturbation}
\begin{figure}
\centering
     \begin{subfigure}{0.43\textwidth}
         \centering
         \includegraphics[width=\textwidth]{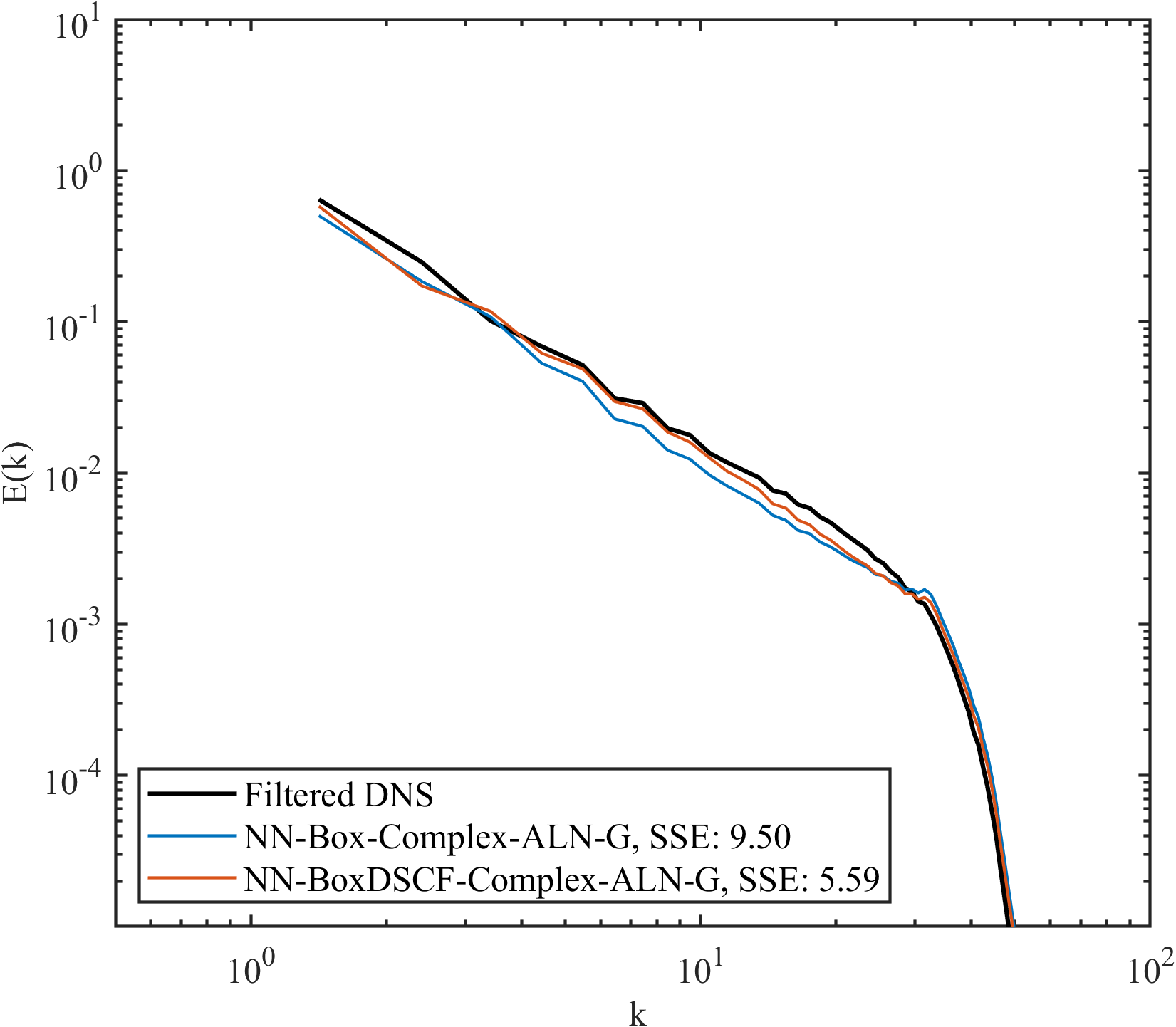}
         \caption{}
         \label{fig:PlibALNSpectra}
     \end{subfigure}
     \hspace{1em}
     \begin{subfigure}{0.46\textwidth}
         \centering
         \includegraphics[width=\textwidth]{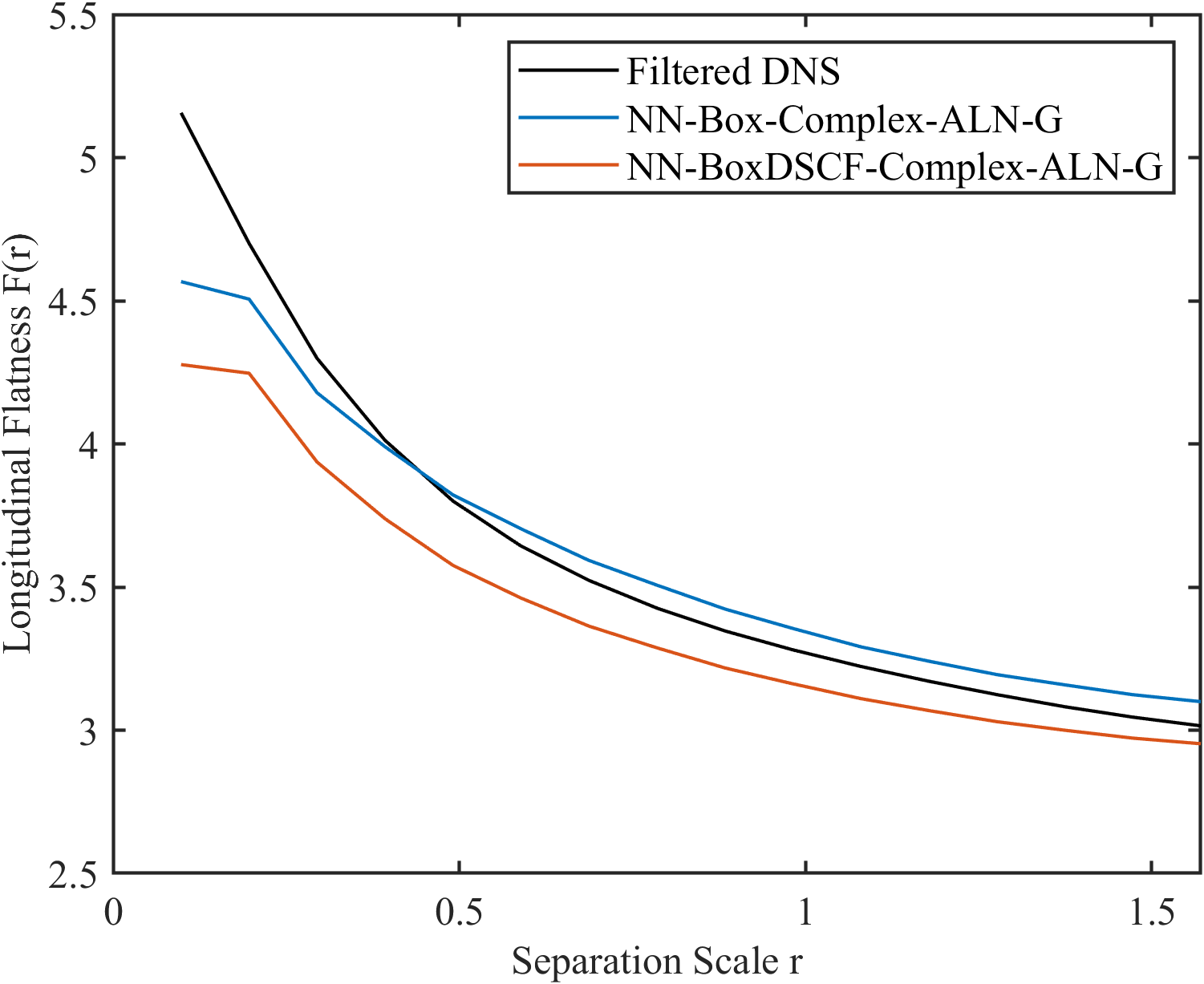}
         \caption{}
         \label{fig:PlibALNInter}
     \end{subfigure}
     \hspace{1em}
     \begin{subfigure}{0.45\textwidth}
         \centering
         \includegraphics[width=\textwidth]{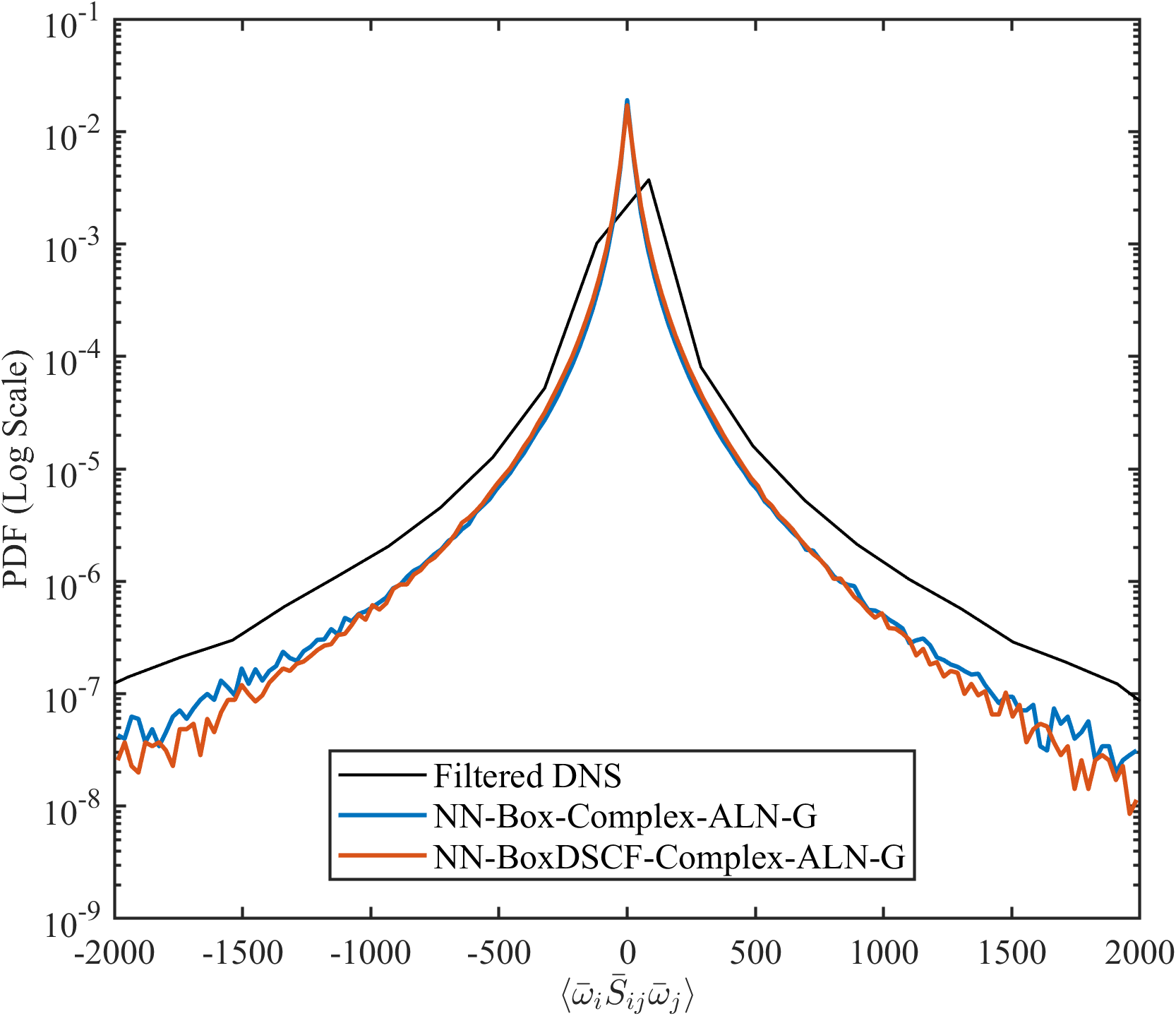}
         \caption{}
         \label{fig:PlibALNVortex}
     \end{subfigure}
     \caption{\textit{A posteriori} ALN comparisons for the 6th order compact code. Spectra, longitudinal flatness, and the PDF of $\overline{\omega_i}\overline{S_{ij}}\overline{\omega_j}$ are plotted. The GNN architecture used corresponds to the ALN architecture with the global normalization.}
\end{figure}

Similar to the Fourier-Spectral code, the box filter training data and the two filter training data (box filter and DSCF), are used to train neural networks with the variant architecture. These results are shown in figures~\ref{fig:PlibALNSpectra} to~\ref{fig:PlibALNVortex}. 

When evaluating the spectra, the same trend from the Fourier-Spectral code also holds, where the neural network trained with one filter has a wide range of results depending on the neural network architecture, while the neural network trained with two filters has less variance in the results depending on neural network architectures. In this case, both neural networks trained predict the spectra accurately. The longitudinal flatness, especially at smaller separation scales, also shows less variance across the various neural network architectures when trained on multiple filters as compared to one filter. This is seen as the longitudinal flatness at small separation scales for the neural network trained on only the box filter varies significantly depending on whether or not the ALN architecture is used (comparing figures~\ref{fig:PlibFiltInter} and~\ref{fig:PlibALNInter}). Meanwhile, the enstrophy production PDF shows little difference between the model trained with one or two filters. This highlights that once again, the general trend of training on two filters is that it often does no worse than training on one filter, and oftentimes increases the robustness of neural network models \textit{a posteriori}. This trend holds across two different neural network architectures.

\subsubsection{Using Simpler Inputs}
\begin{figure}
\centering
     \begin{subfigure}{0.47\textwidth}
         \centering
         \includegraphics[width=\textwidth]{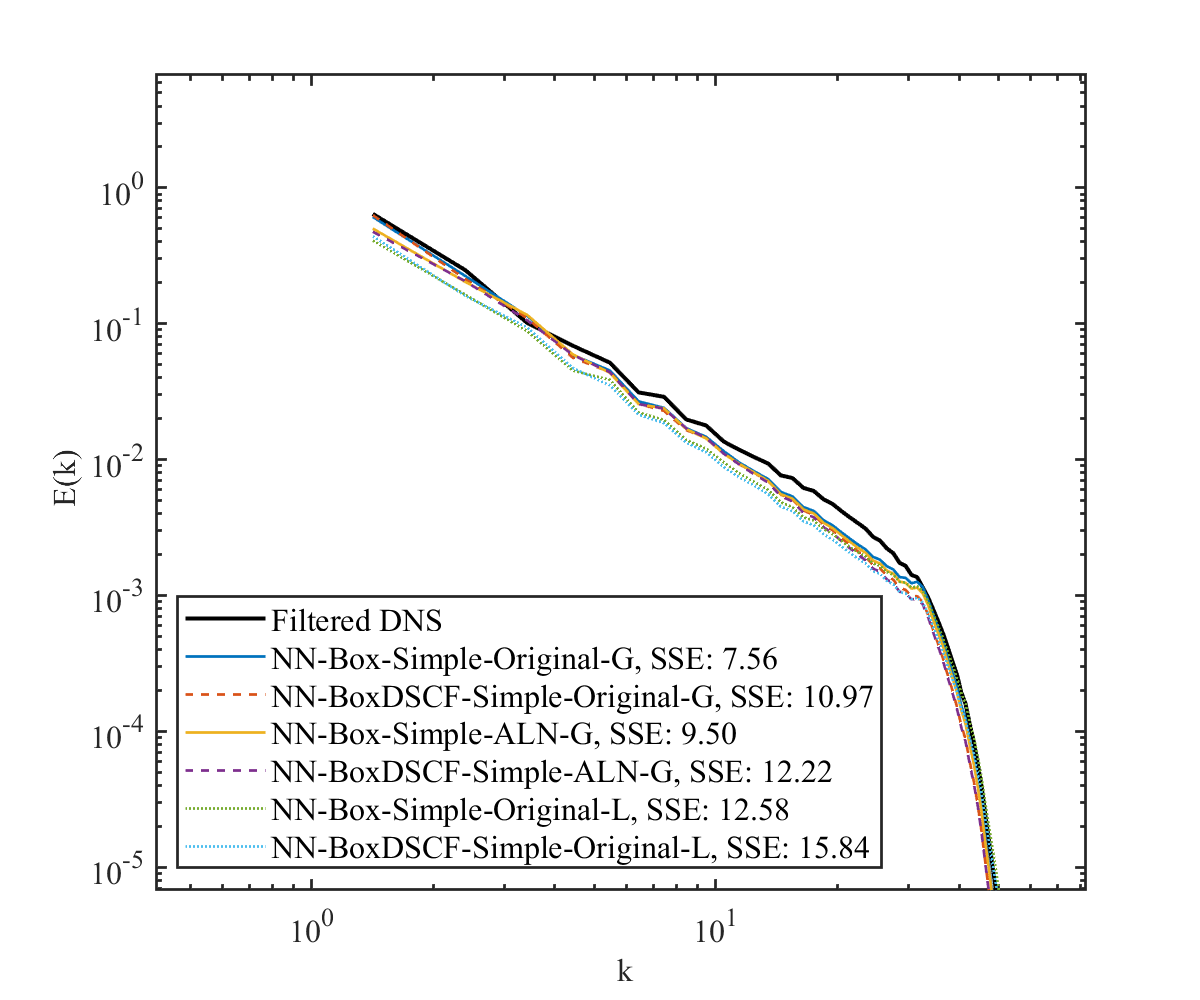}
         \caption{}
         \label{fig:PlibSimpleSpectra}
     \end{subfigure}
     \hspace{1em}
     \begin{subfigure}{0.45\textwidth}
         \centering
         \includegraphics[width=\textwidth]{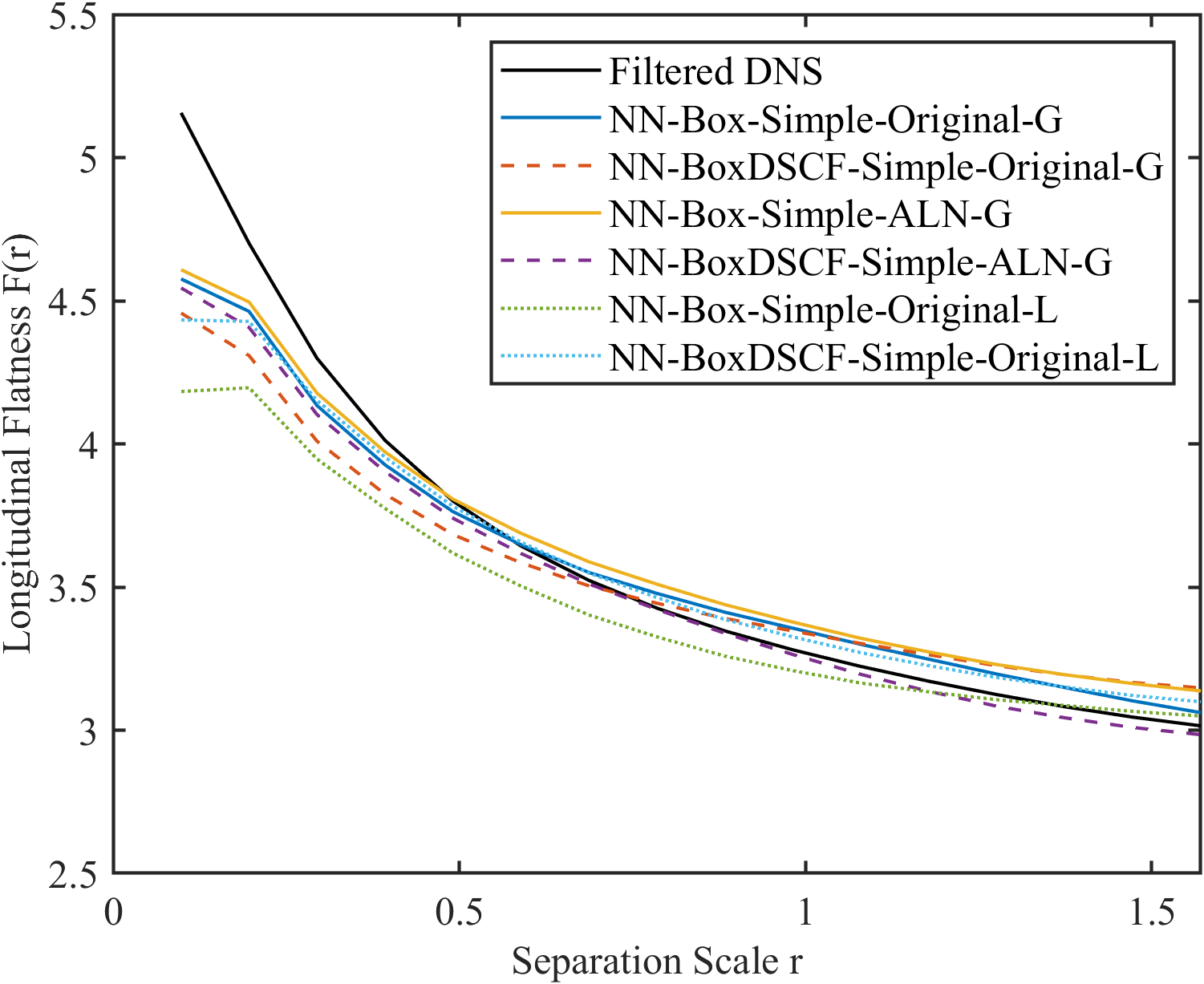}
         \caption{}
         \label{fig:PlibSimpleInter}
     \end{subfigure}
     \hspace{1em}
     \begin{subfigure}{0.49\textwidth}
         \centering
         \includegraphics[width=\textwidth]{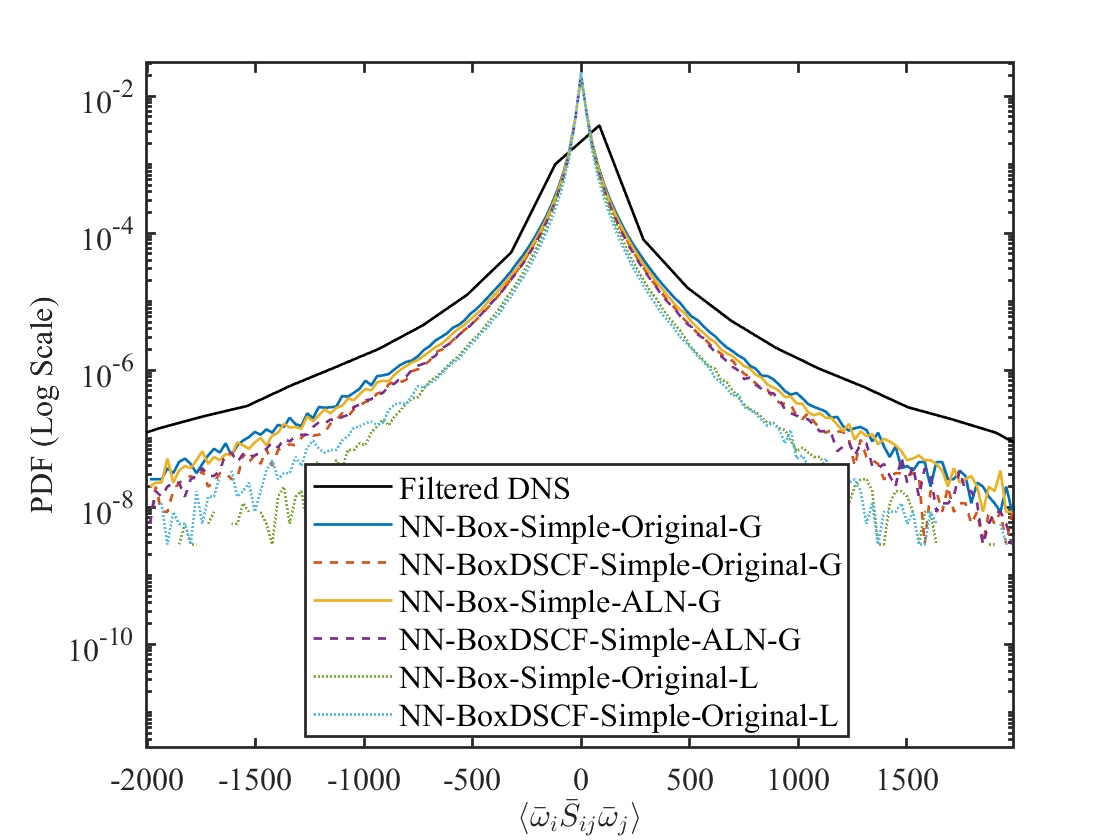}
         \caption{}
         \label{fig:PlibSimpleVortex}
     \end{subfigure}
     \caption{\textit{A posteriori} simple input comparisons for the 6th order compact code. Spectra, longitudinal flatness, and the PDF of $\overline{\omega_i}\overline{S_{ij}}\overline{\omega_j}$ are plotted. Both the original and ALN GNN architectures are used with the global and local normalization for the ``simple" inputs.}
\end{figure}

In addition, to ensure that simpler inputs do result in a smaller distribution shift for a different numerical method, neural networks are trained and evaluated with the simple set of inputs, across two different neural network architectures and also two different input normalizations. These results are shown in figures~\ref{fig:PlibSimpleSpectra} to~\ref{fig:PlibSimpleVortex}.

When evaluating the spectra, the simpler inputs reduces the input distribution shift, as the neural networks trained only on the box filter are also able to produce spectra that are reasonable. There is little difference in the spectra when comparing them to the filtered DNS curve. This is also seen in the relatively smaller variance of the SSE that is shown in table~\ref{tab:padenns} and also figure~\ref{fig:PlibSimpleSpectra}, especially when compared to the larger variance seen in the complex input cases when training on one versus two filters. A similar trend holds for the longitudinal flatness, where the models produce relatively similar results, regardless of training on one or two filters. One exception is the local normalization, where training on two filters produces better performance than training on one filter. The enstrophy production plot is also relatively similar, with all models doing well. The exception is the neural networks trained with the local normalization, where they slightly under-predict the tails more than the other models, although for the negative tail, training on two filters does help reduce the behavior slightly. Overall, using simpler inputs does reduce the distribution shift in \textit{a posteriori} simulations, and two filter training with simpler inputs does relatively similar to the one filter training. This shows that for simpler inputs, there is no significant performance degradation when training on two filters as compared to one filter.

\section{Additional \textit{A Posteriori} Evaluations}
To ensure that these trends hold for a different grid resolution and for a different flow, a coarser forced HIT is run on a $32 \times 32 \times 32$ grid (corresponds to a filter width of $\Delta_{LES} = 32 \Delta_{DNS}$), which is twice as coarse as the $64 \times 64 \times 64$ grid used in the previous HIT simulations. Furthermore, a channel flow simulation is run, matching the $Re_{\tau} = 1000$ of the DNS simulation used for training. The grid used is $128 \times 96 \times 32$, which corresponds to a filter width of $\Delta_{LES} = 16 \Delta_{DNS}$. The wall model used is not a data driven wall model. Instead, a classic wall model is used that is identical to the one described by~\cite{bou2005scale}. A RK3 time stepping scheme is employed, as well as Fourier collocation in the wall parallel directions and a 6th order staggered scheme in the wall normal direction. In all cases, the Fourier-spectral code with two-thirds dealiasing is run as the primary code. 

\subsection{Evaluation on a different grid width}
\begin{figure}
\centering
     \begin{subfigure}{0.49\textwidth}
         \centering
         \includegraphics[width=\textwidth]{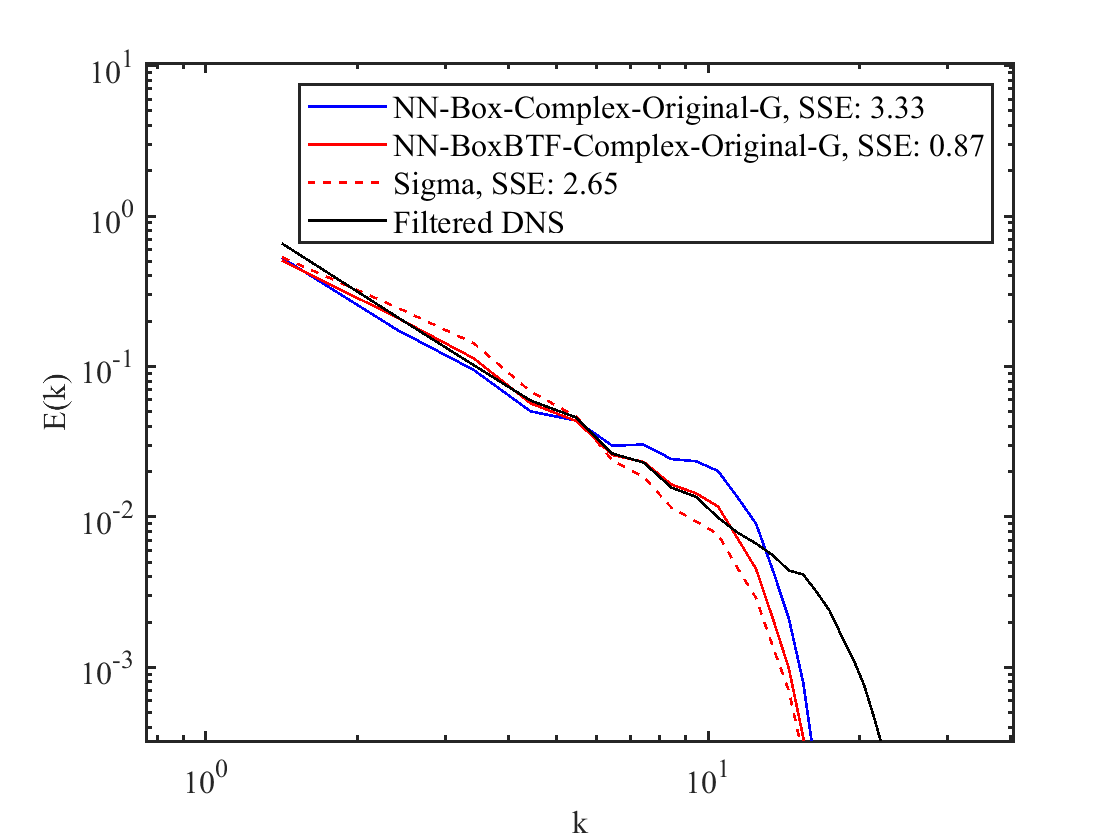}
         \caption{}
         \label{fig:Pops32CompareSpectra}
     \end{subfigure}
     \hspace{1em}
     \begin{subfigure}{0.425\textwidth}
         \centering
         \includegraphics[width=\textwidth]{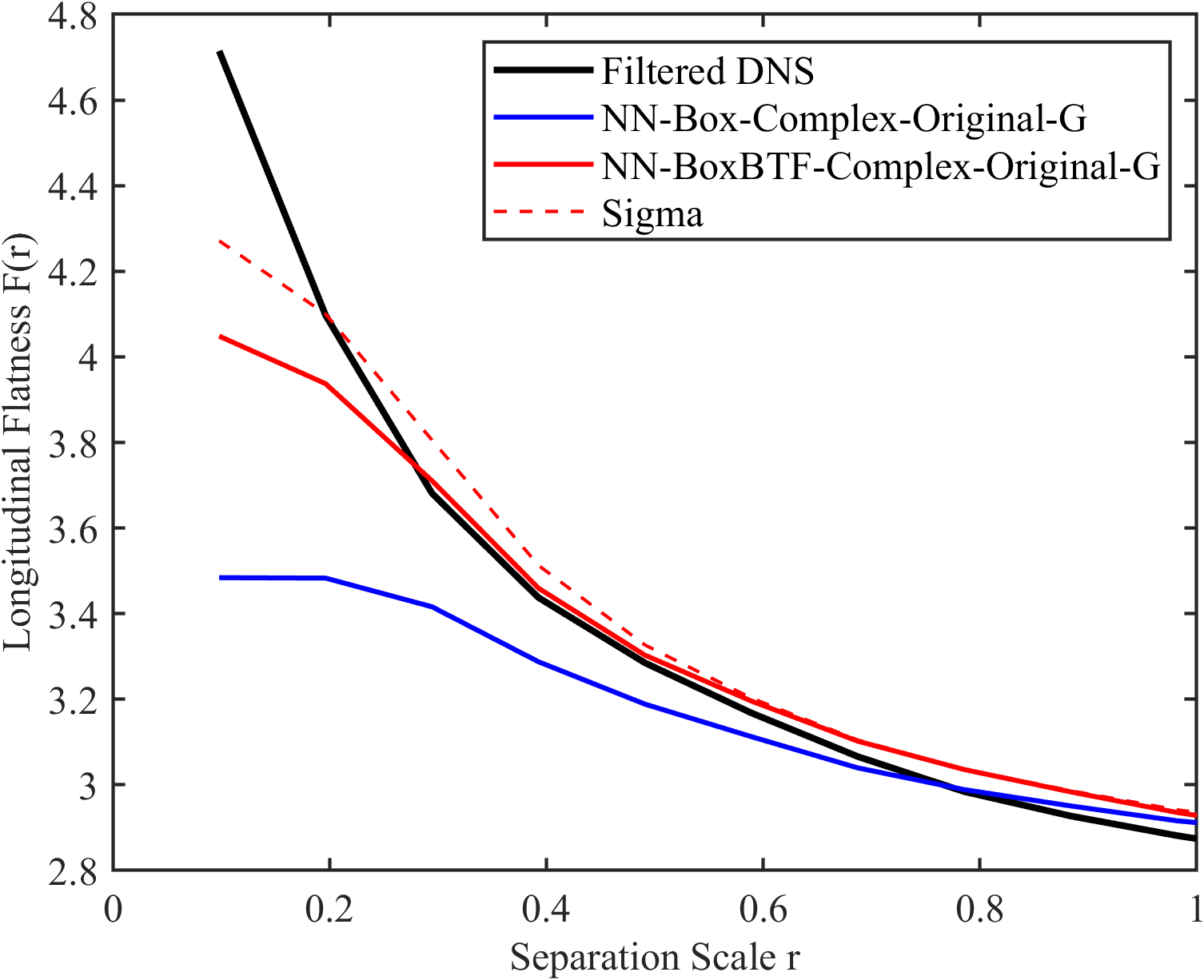}
         \caption{}
         \label{fig:Pops32CompareInter}
     \end{subfigure}
     \hspace{1em}
     \begin{subfigure}{0.49\textwidth}
         \centering
         \includegraphics[width=\textwidth]{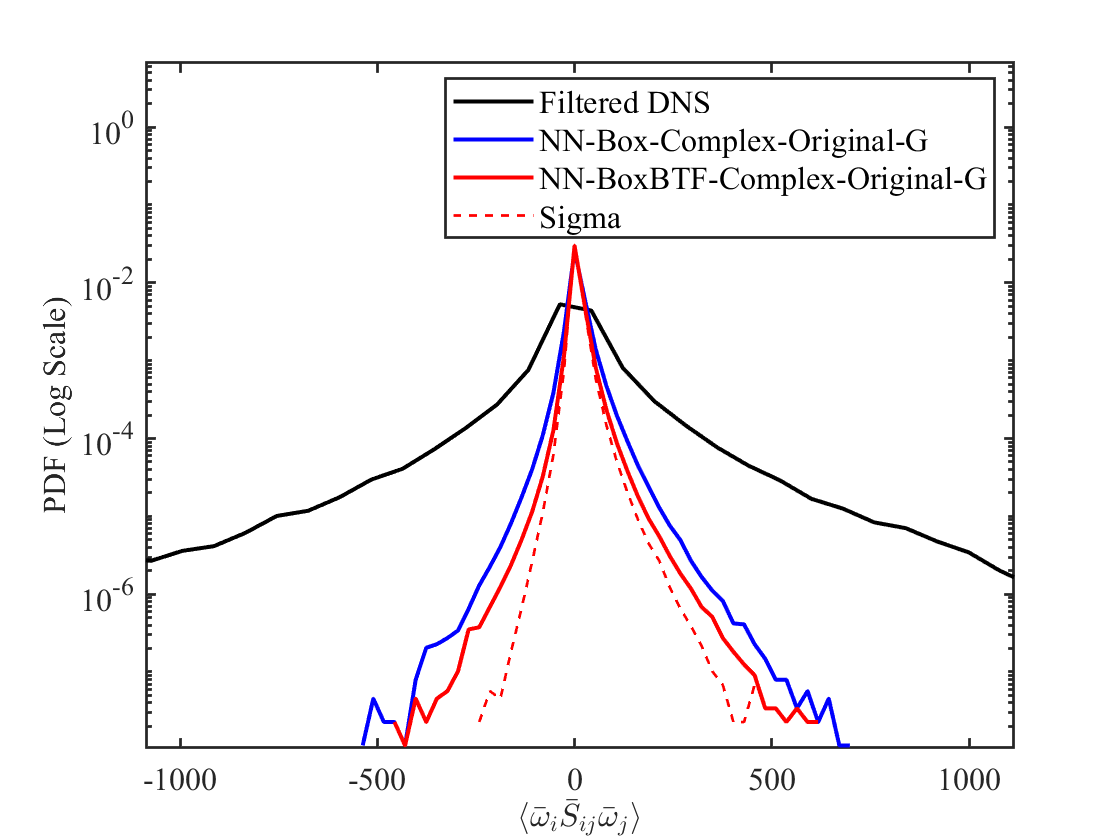}
         \caption{}
         \label{fig:Pops32CompareVortex}
     \end{subfigure}
     \caption{\textit{A posteriori} one versus two filter comparison with the Sigma model as a reference for a coarser grid spacing corresponding to $\Delta_{LES} = 32 \Delta_{DNS}$. Spectra, longitudinal flatness, and the PDF of $\overline{\omega_i}\overline{S_{ij}}\overline{\omega_j}$ are plotted. The GNN architecture used corresponds to the original architecture with the global normalization.}
\end{figure}

\begin{figure}
\centering
     \begin{subfigure}{0.49\textwidth}
         \centering
         \includegraphics[width=\textwidth]{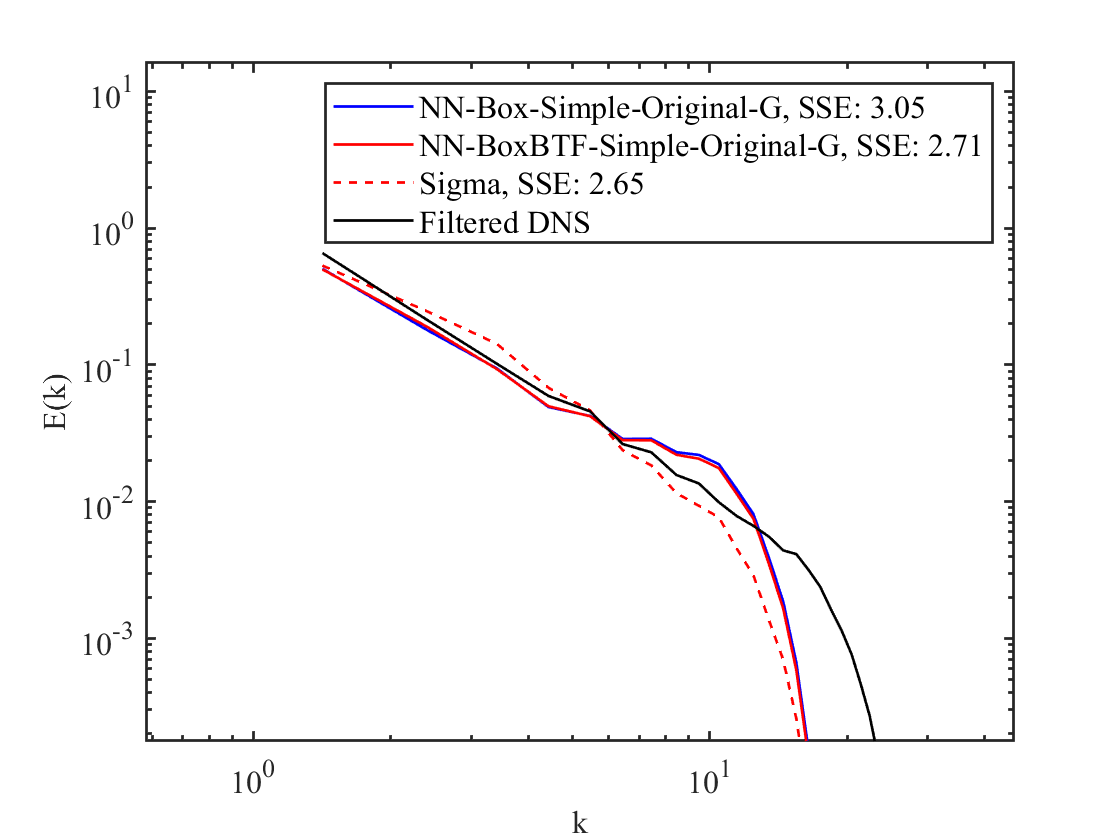}
         \caption{}
         \label{fig:Pops32SimpleSpectra}
     \end{subfigure}
     \hspace{1em}
     \begin{subfigure}{0.425\textwidth}
         \centering
         \includegraphics[width=\textwidth]{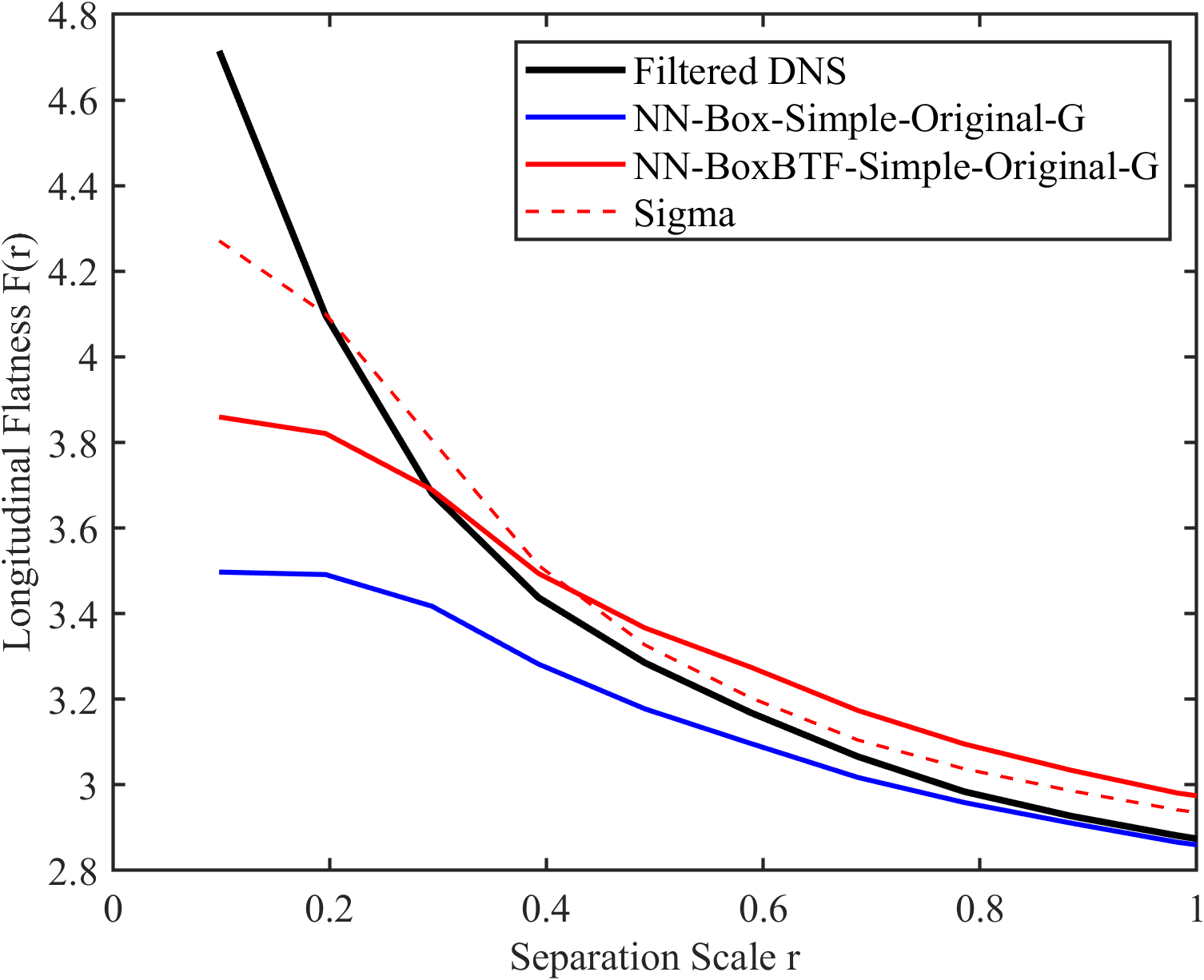}
         \caption{}
         \label{fig:Pops32SimpleInter}
     \end{subfigure}
     \hspace{1em}
     \begin{subfigure}{0.49\textwidth}
         \centering
         \includegraphics[width=\textwidth]{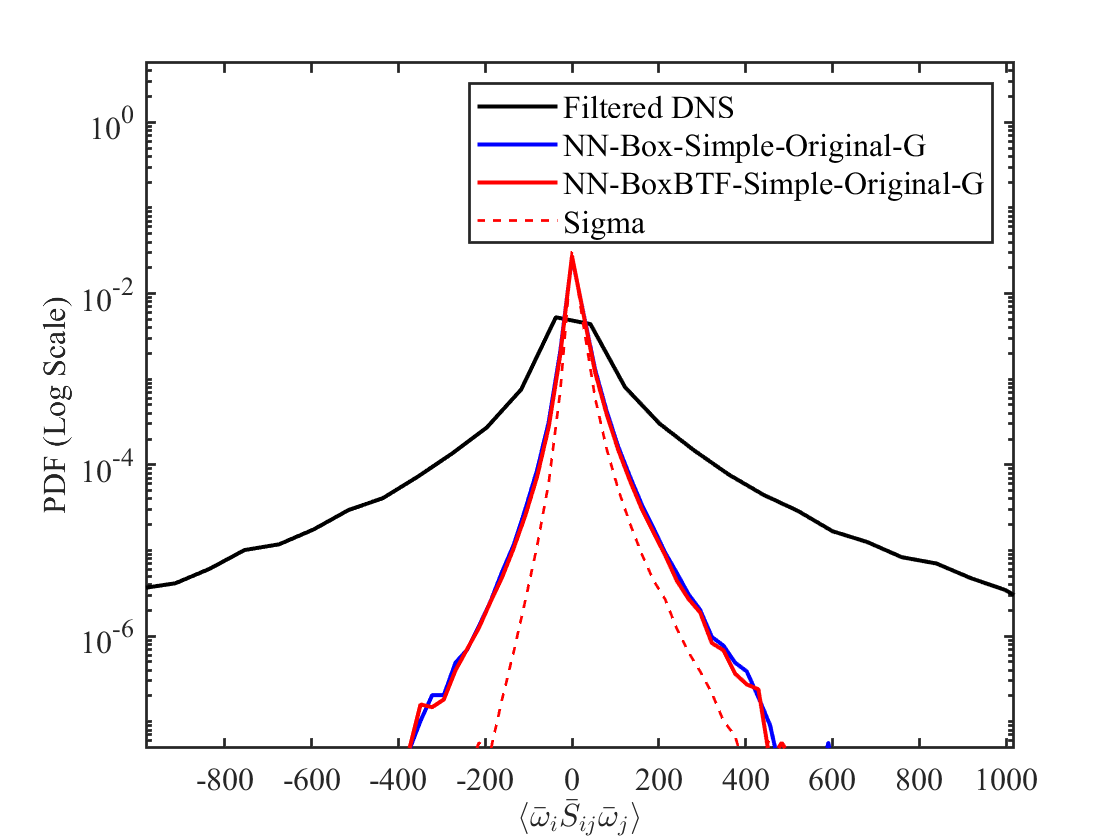}
         \caption{}
         \label{fig:Pops32SimpleVortex}
     \end{subfigure}
     \caption{\textit{A posteriori} simple input comparisons with the Sigma model as a reference on the coarser grids. Spectra, longitudinal flatness, and the PDF of $\overline{\omega_i}\overline{S_{ij}}\overline{\omega_j}$ are plotted. The GNN architecture used corresponds to the original architecture with the global normalization applied to the ``simple" inputs.}
\end{figure}

The first main trend seen was that training on two filters, in general, is better than training on one filter (especially the box filter), and this trend is more prominently seen for complex inputs. The second trend seen was that training on a simple set of inputs results in less of an input distribution shift, resulting in more consistent results across the various neural network models. In addition, the two filter cases do no worse than the one filter cases in many situations. These trends are now evaluated on a forced HIT simulation that is twice as coarse as the previous simulation. To illustrate these trends, a minimal amount of simulations were run. For the complex inputs, the box-filter trained neural network was run, as well as the box and BTF trained neural network. This is shown to highlight the performance improvement that can be gained over using a filter that is traditionally used in literature (box filter). These results are shown in figures~\ref{fig:Pops32CompareSpectra} to~\ref{fig:Pops32CompareVortex}. To illustrate the simple input set, a similar filter configuration is run, with results shown in figures~\ref{fig:Pops32SimpleSpectra} to~\ref{fig:Pops32SimpleVortex}.

From figure~\ref{fig:Pops32CompareSpectra}, when comparing the spectra of the one filter versus two filter case, the two filter case is able to improve the spectra by increasing the dissipation as compared to the one filter neural network model, which is under-dissipative. For reference, the Sigma model is also shown, and it is over-dissipative, although less so for this grid resolution, despite using the same coefficient as the previous simulations. The same trends hold for the longitudinal flatness, where the traditional models capture this better, while the two filter trained neural network significantly improves upon the one neural network case. For the enstrophy production, the neural network models do relatively similarly, but are able to outperform the traditional models such as the Sigma model. However, all models are unable to accurately capture the PDF of the enstrophy production term, likely due to the very coarse grid resolution, as in the finer resolution simulations, the two filter neural network model was able to more accurately capture the enstrophy production PDF when compared with the one filter case. Overall, for the complex set of inputs, especially when comparing the energy spectra and longitudinal flatness, training with two filters is still better than training with one filter, showing that this trend still holds even at a different resolution.

Meanwhile, when considering the simple inputs (figures~\ref{fig:Pops32SimpleSpectra} to~\ref{fig:Pops32SimpleVortex}), the neural networks do not show as much difference in the spectra when training with one or two filters, which is consistent with the previous cases. For longitudinal flatness, there is still improvement when training on two filters instead of one filter, especially at the smaller separation scales. The enstrophy production PDFs look similar for both neural network models, although they are still better than the traditional models. Overall, the same trend here holds, where for the simple inputs, there is less of a difference when training on one versus two filters, and that the two filter case often does no worse than the one filter case, and can sometimes do better in some metrics such as longitudinal flatness. This suggests that the mechanisms of reducing input and output distribution shift are not specific to one grid resolution, and are general across a range of grid resolutions. Specifically, using simpler inputs to the neural network can reduce the input side distribution shift, which the neural network is sensitive to since these are the features the neural network is ``learning" from. Meanwhile, training on multiple filters can reduce the output side distribution shift and ensure that the neural network does not ``overfit" to one specific filter shape.

\subsection{Evaluation on channel flow}
Now, these main trends are also illustrated for channel flow. Once again, the box filter is chosen as the filter used for single filter training to highlight the performance improvement that can be gained using a filter that is traditionally used in literature, and a minimal set of simulations were run to illustrate these trends. However, for channel flow, an additional set of neural networks are trained with the variant neural network architecture (ALN) to highlight that the variant neural network architecture, while being more robust, can also be enhanced with the two-filter training as compared to the one filter training. 

\subsubsection{Two Filter Complex Input Analysis}
\begin{figure}
\centering
     \begin{subfigure}{0.45\textwidth}
         \centering
         \includegraphics[width=\textwidth]{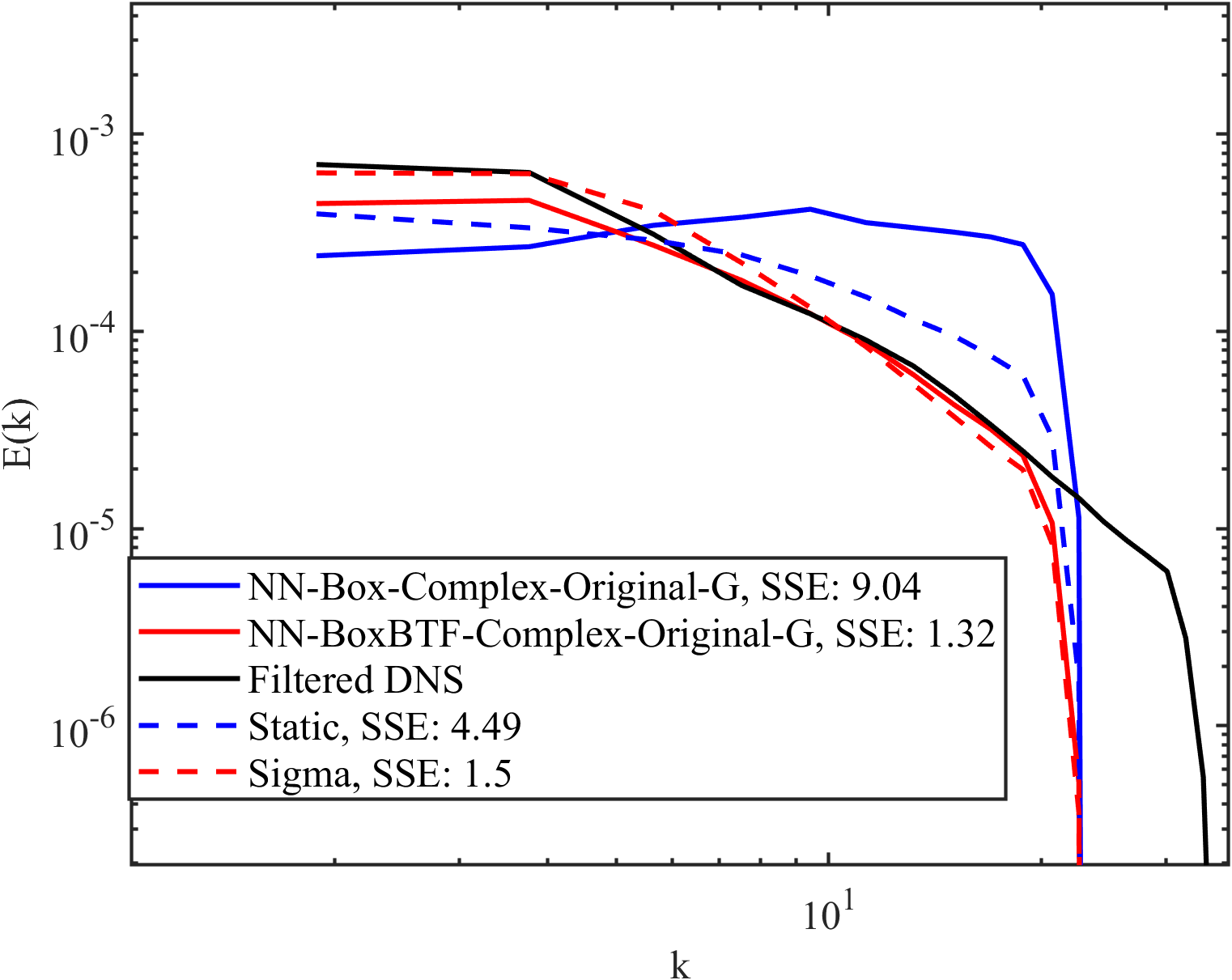}
         \caption{}
         \label{fig:Channel_Compare_Spectra}
     \end{subfigure}
     \hspace{1em}
     \begin{subfigure}{0.45\textwidth}
         \centering
         \includegraphics[width=\textwidth]{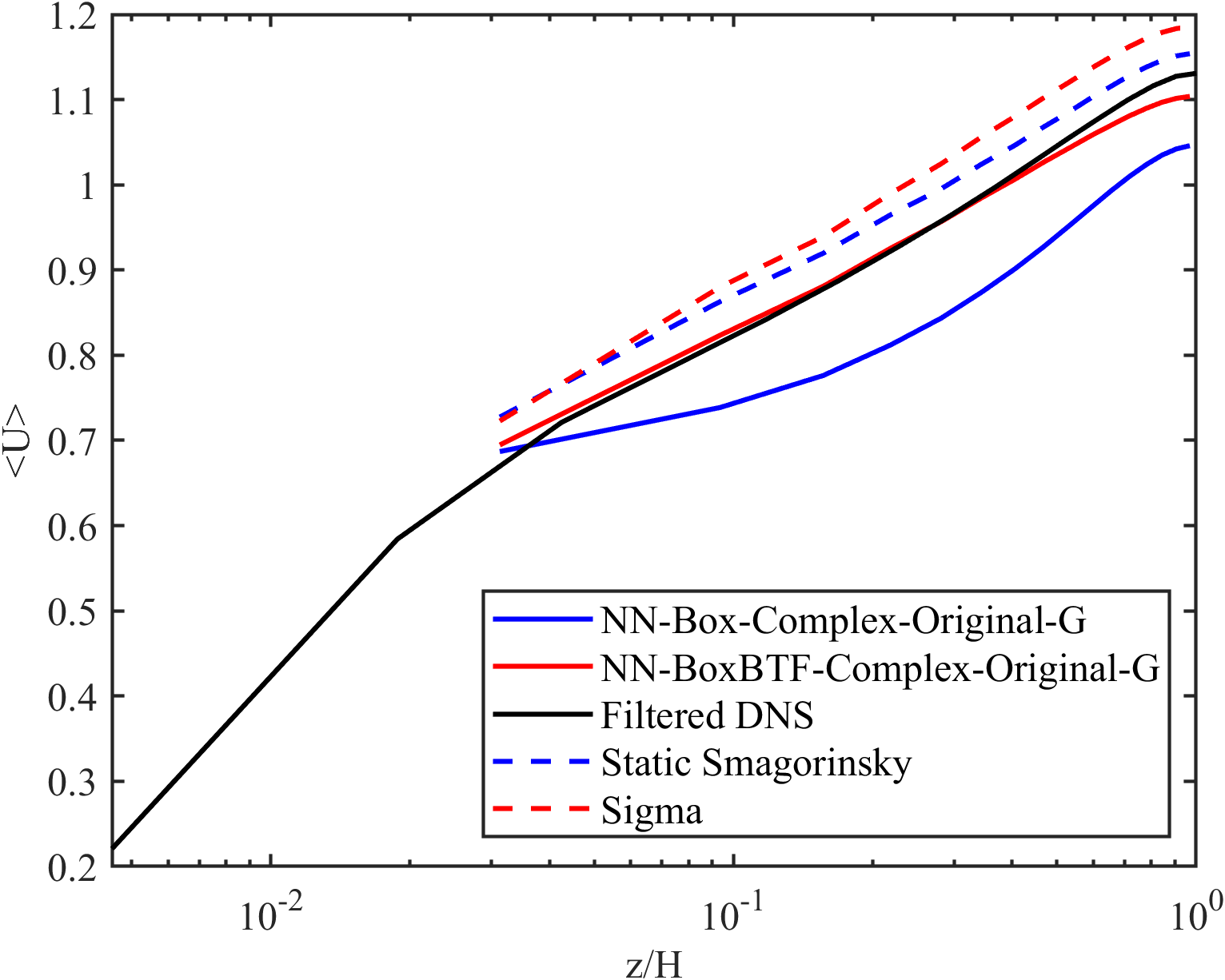}
         \caption{}
         \label{fig:Channel_Compare_Mean}
     \end{subfigure}
     \caption{\textit{A posteriori} streamwise velocity spectra and mean profiles for channel flow, comparing one versus two filter GNN models with traditional models. The GNN architecture used corresponds to the original architecture with the global normalization.}
\end{figure}

First, the box filter trained neural network is compared with the box and BTF trained neural network, along with the traditional models. The spectra and mean profiles are shown in figures~\ref{fig:Channel_Compare_Spectra} to~\ref{fig:Channel_Compare_Mean}. The spectra reported is for the streamwise velocity, at a particular wall normal location of $z/H = 0.46875$.

\begin{figure}
\centering
     \begin{subfigure}{0.38\textwidth}
         \centering
         \includegraphics[width=\textwidth]{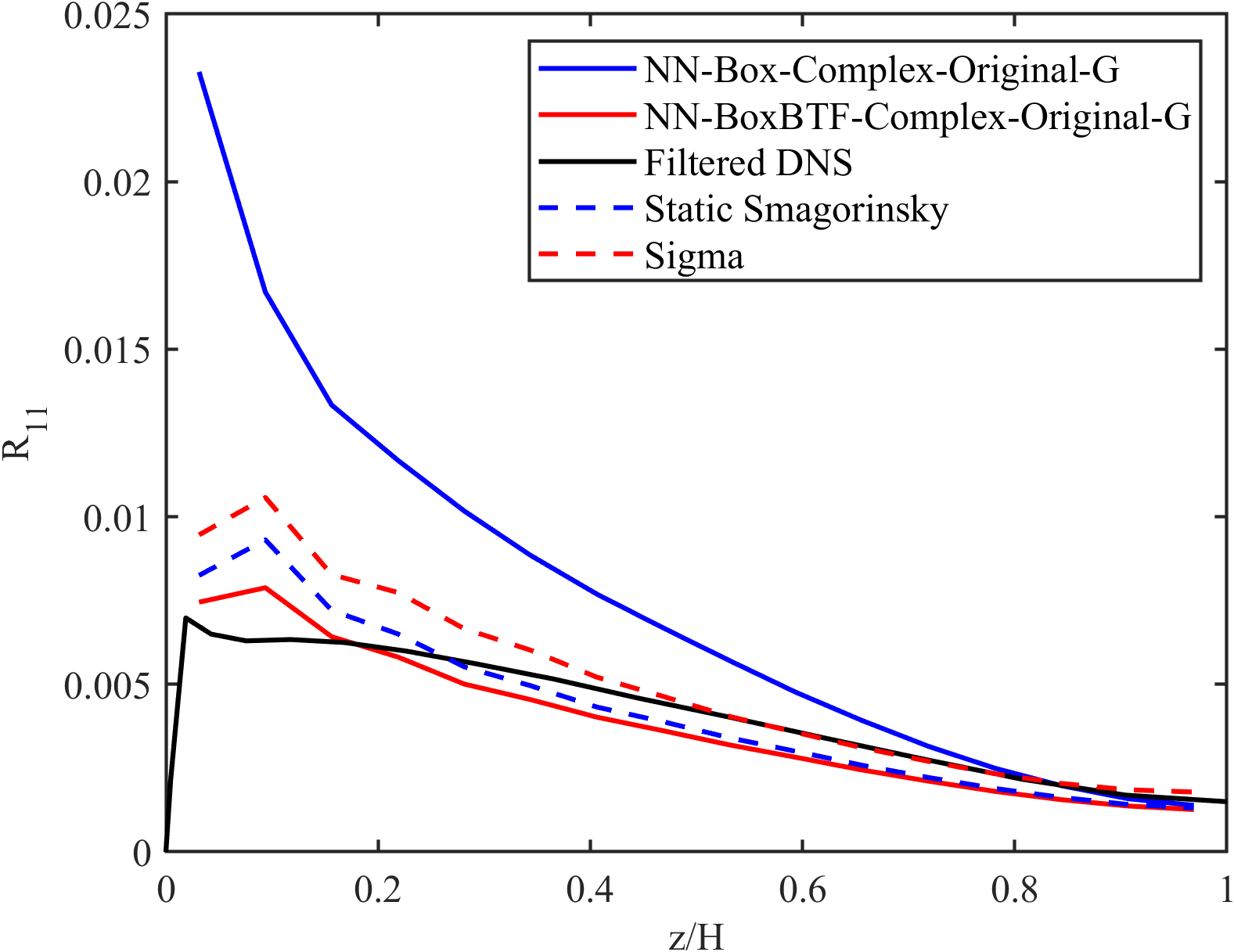}
         \caption{}
         \label{fig:Channel_Compare_R11}
     \end{subfigure}
     \hspace{1em}
     \begin{subfigure}{0.38\textwidth}
         \centering
         \includegraphics[width=\textwidth]{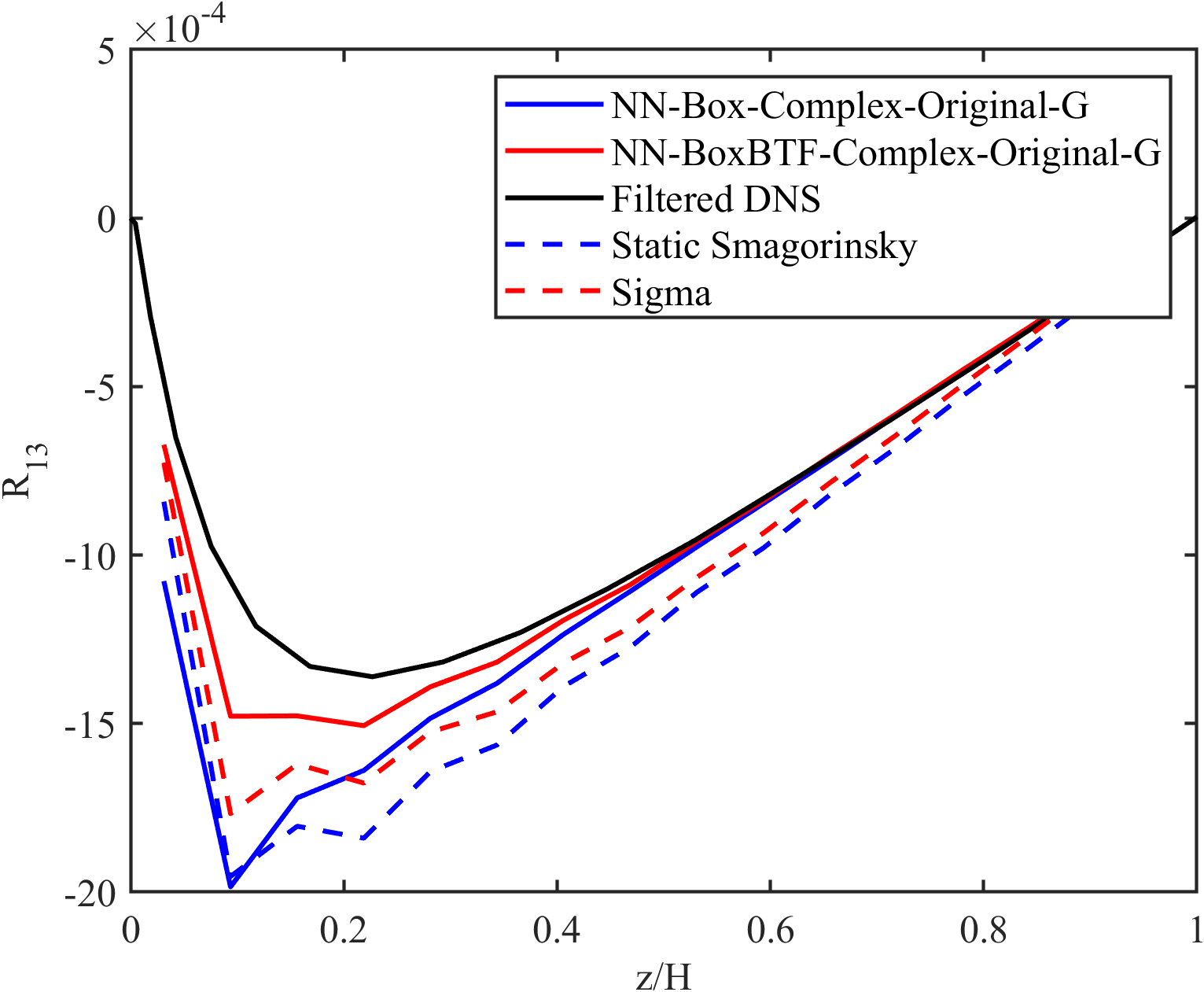}
         \caption{}
         \label{fig:Channel_Compare_R13}
     \end{subfigure}
     \hspace{1em}
     \begin{subfigure}{0.38\textwidth}
         \centering
         \includegraphics[width=\textwidth]{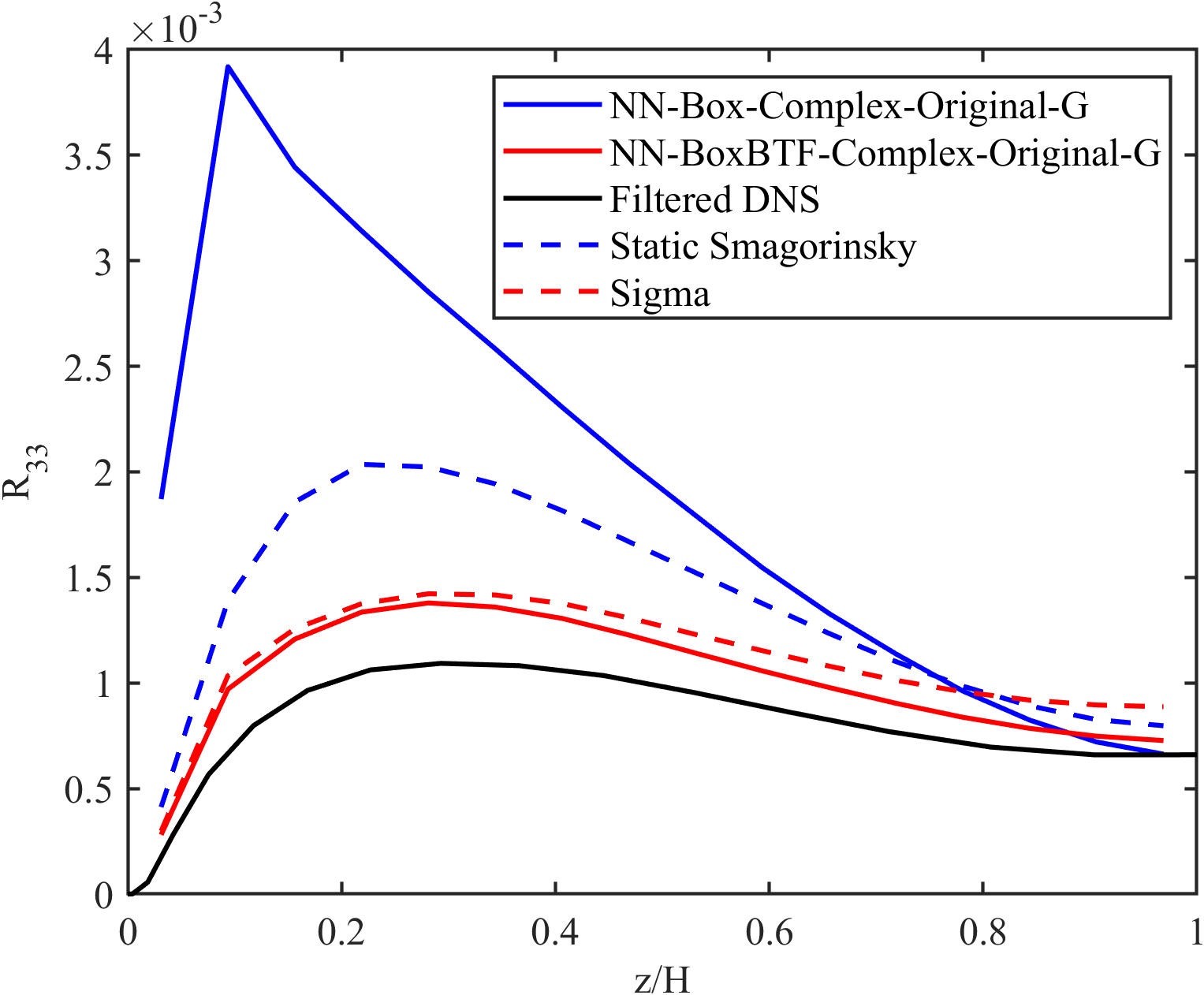}
         \caption{}
         \label{fig:Channel_Compare_R33}
     \end{subfigure}
     \caption{\textit{A posteriori} one versus two filter Reynolds stress comparisons with traditional models for channel flow. Here, $z/H = 0$ corresponds to the wall while $z/H=1$ corresponds to the channel centerline. The GNN architecture used corresponds to the original architecture with the global normalization.}
\end{figure}

In this case, the one filter neural network completely fails, creating an unphysical spectra that causes the channel flow velocity fields to be extremely choppy. In addition, the Static Smagorinsky model also has a large buildup of turbulent kinetic energy at the high wavenumbers. Meanwhile, the Sigma model and the neural network model trained on two filters are able to accurately represent the spectra. This is also seen in the mean profiles, where the single filter neural network has poor agreement with the DNS profile. Meanwhile, the two filter case lines up very well with the filtered DNS profile, even when compared with the traditional models. This provides early indications that training on two filters is also more robust than training on one filter for channel flow. The Reynolds stress profiles are shown in figures~\ref{fig:Channel_Compare_R11} to figure~\ref{fig:Channel_Compare_R33}.

The Reynolds stress profiles show a similar story, where the one filter case oftentimes performs the worst, overpredicting the Reynolds stress profiles, especially for $R_{11}$ and $R_{33}$. In addition, the Static Smagorinsky model also predicts poor Reynolds stress profiles. Meanwhile, the neural network trained on both the box filter and BTF is able to produce far more accurate Reynolds stress profiles, oftentimes producing the most accurate profiles even when compared to the Sigma model. Improvements due to the one versus two filter training is clearly seen in all the Reynolds stress profiles, where the one filter over-predictions are significantly reduced when training on two filters. Thus, for the complex set of inputs, the overall trend of training on two different filters being more accurate that was seen in forced HIT is also seen in channel flow. This suggests that the idea of reducing the output distribution shift and ensuring that the neural network sees multiple filter shapes through two filter training is more general as it shows consistent improvement across both forced HIT and channel flow for complex inputs.

\subsubsection{Two Filter Complex Input Analysis ALN}
\begin{figure}
\centering
     \begin{subfigure}{0.45\textwidth}
         \centering
         \includegraphics[width=\textwidth]{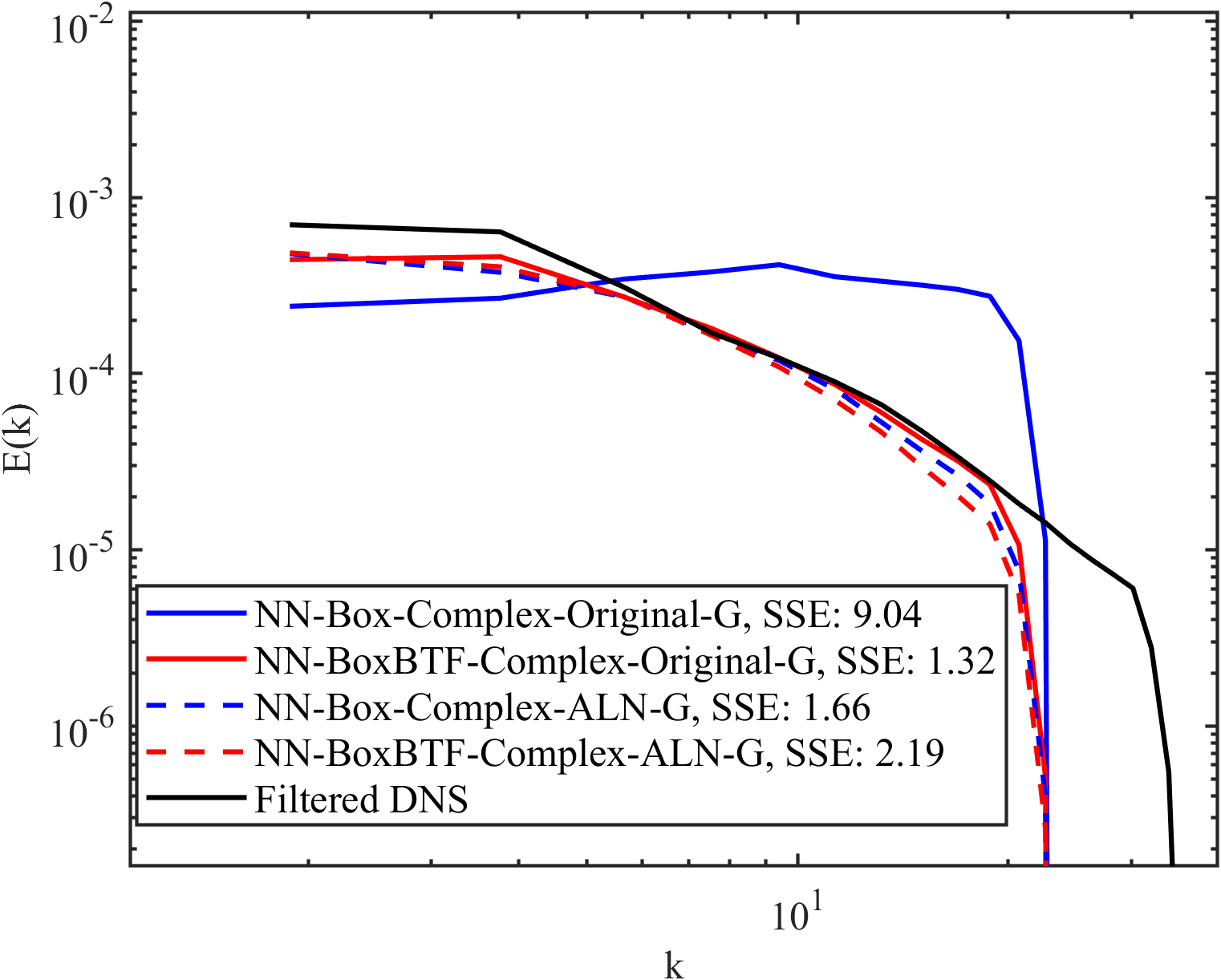}
         \caption{}
         \label{fig:Channel_Filt_Spectra}
     \end{subfigure}
     \hspace{1em}
     \begin{subfigure}{0.45\textwidth}
         \centering
         \includegraphics[width=\textwidth]{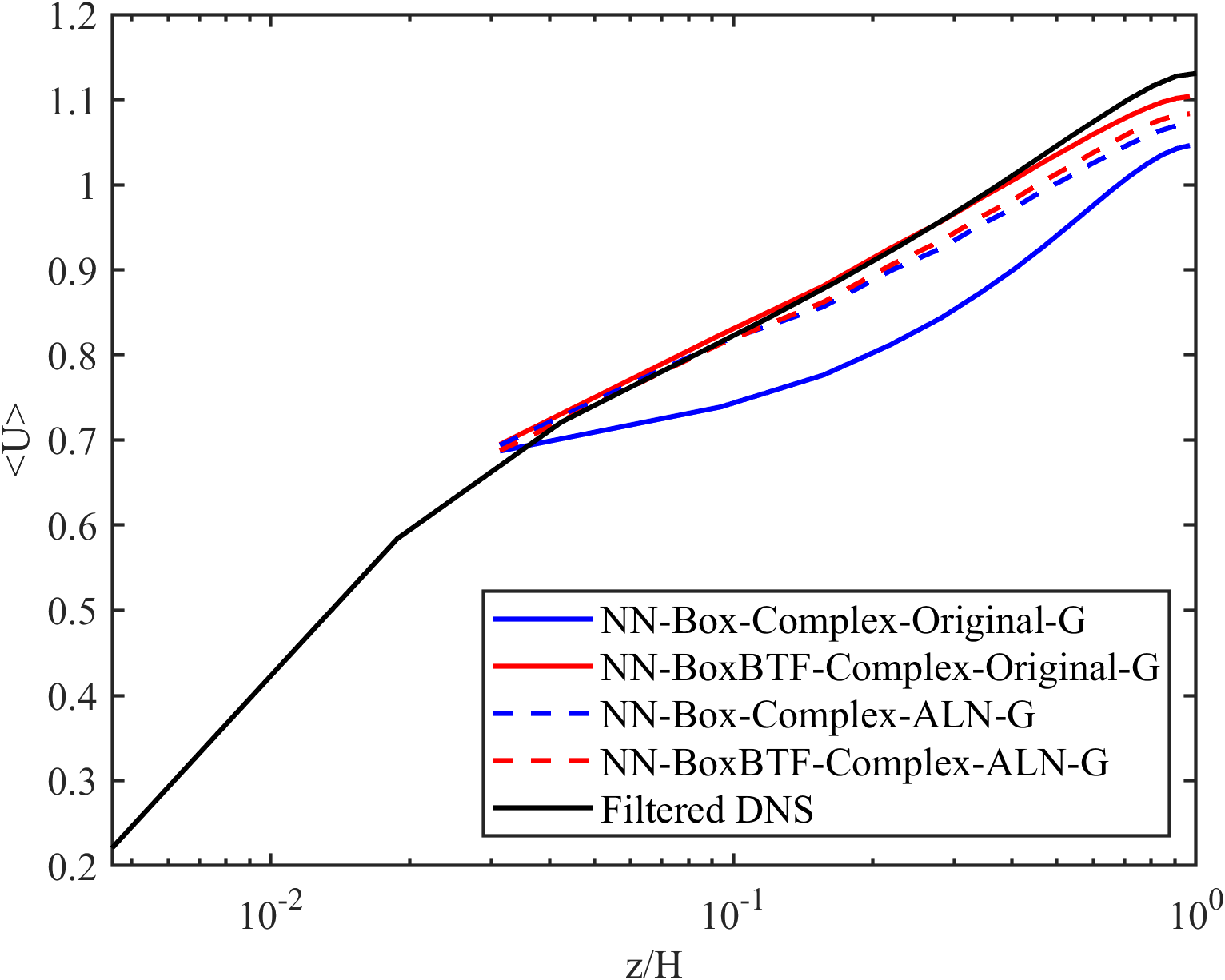}
         \caption{}
         \label{fig:Channel_Filt_Mean}
     \end{subfigure}
     \caption{\textit{A posteriori} spectra and mean profile comparison of different neural network perturbations. Both the original and ALN GNN architecture are used, with global normalization applied to the ``complex" inputs.}
\end{figure}

Now, the variant architecture (ALN) is analyzed and compared to the original architecture for channel flow. The streamwise velocity spectra and streamwise velocity mean profiles are shown in figures~\ref{fig:Channel_Filt_Spectra} and~\ref{fig:Channel_Filt_Mean}.

\begin{figure}
\centering
     \begin{subfigure}{0.38\textwidth}
         \centering
         \includegraphics[width=\textwidth]{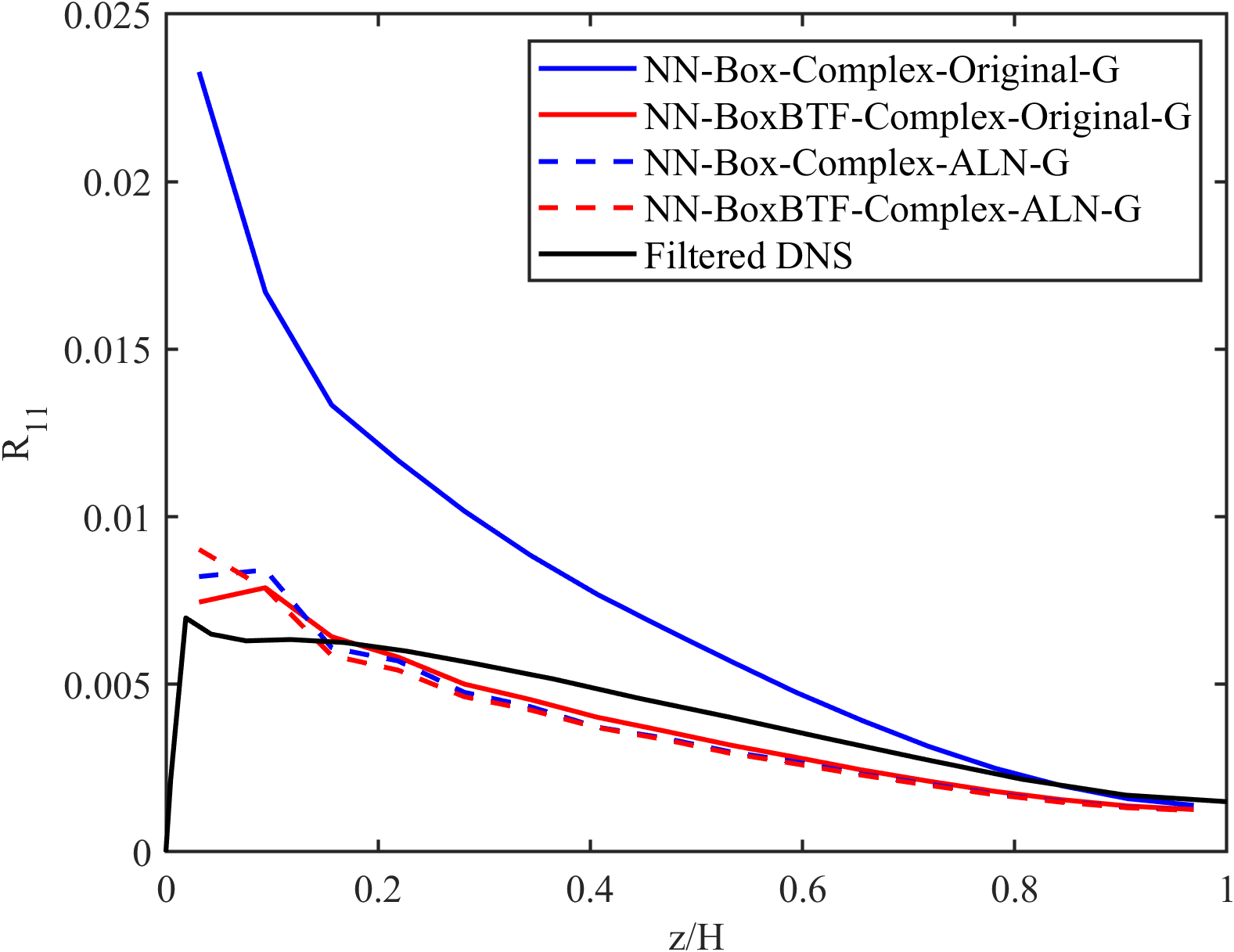}
         \caption{}
         \label{fig:Channel_Filt_R11}
     \end{subfigure}
     \hspace{1em}
     \begin{subfigure}{0.38\textwidth}
         \centering
         \includegraphics[width=\textwidth]{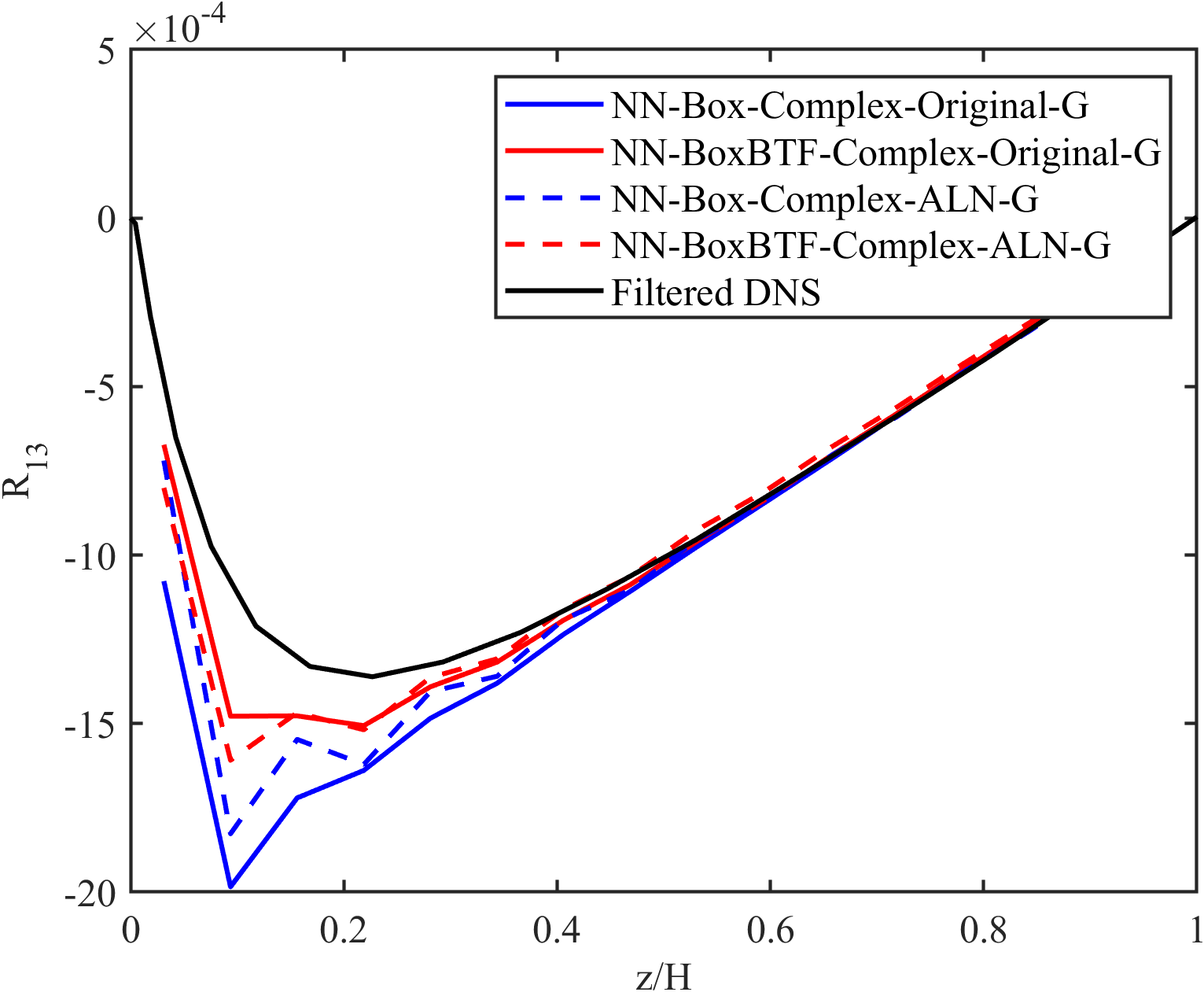}
         \caption{}
         \label{fig:Channel_Filt_R13}
     \end{subfigure}
     \hspace{1em}
     \begin{subfigure}{0.38\textwidth}
         \centering
         \includegraphics[width=\textwidth]{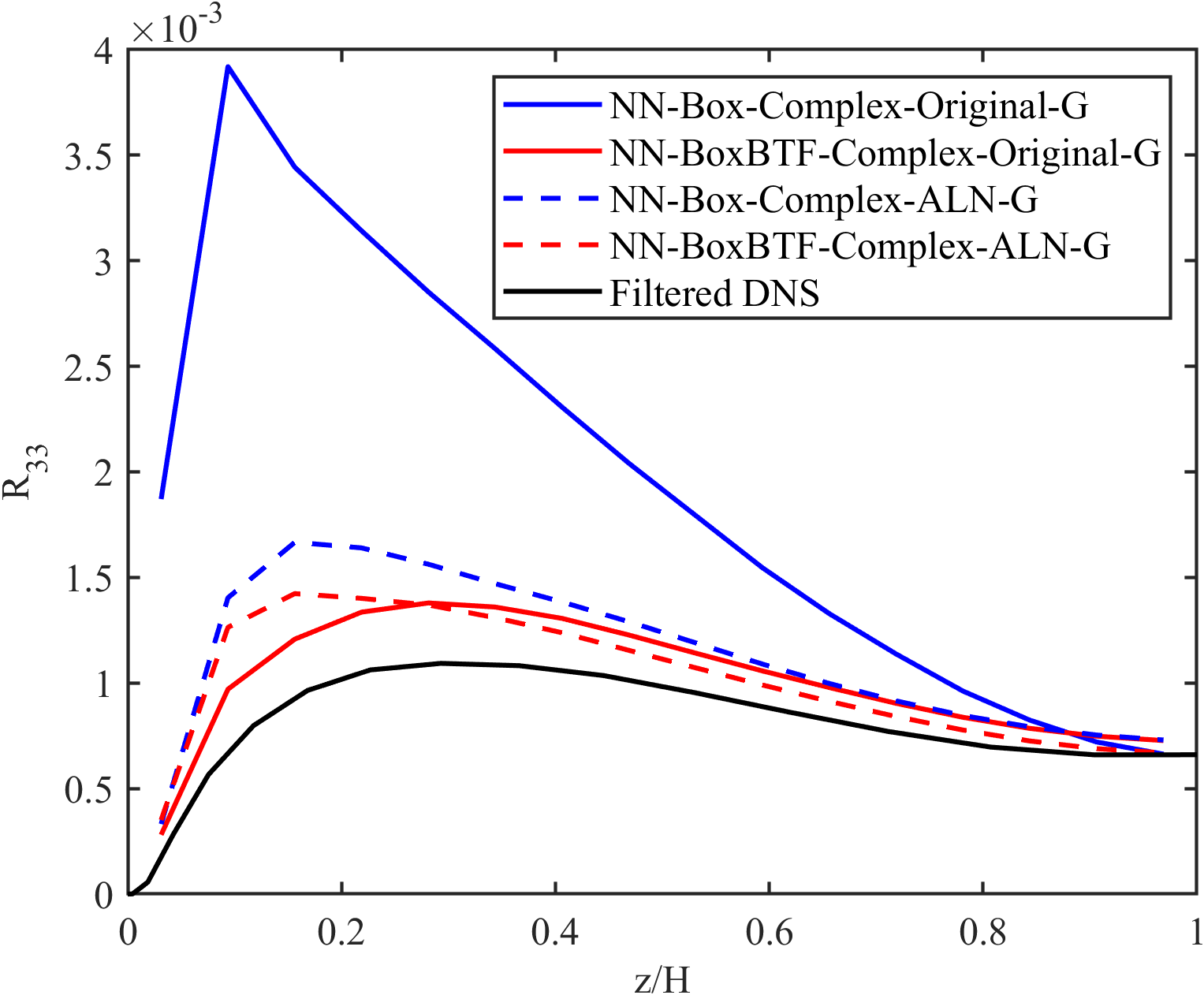}
         \caption{}
         \label{fig:Channel_Filt_R33}
     \end{subfigure}
     \caption{\textit{A posteriori} Reynolds stress comparison of different neural network perturbations. Here, $z/H = 0$ corresponds to the wall while $z/H=1$ corresponds to the channel centerline. Both the original and ALN GNN architecture are used, with global normalization applied to the ``complex" inputs.}
\end{figure}

Overall, both the one filter and two filter cases with the variant neural network architecture do well, especially when compared with the one filter case with the original architecture. This is also shown in the mean profiles, where the majority of the neural network models do well while the original neural network architecture trained with the box filter data does poorly. From both the spectra and the mean profiles, the training on two filters significantly reduces the variance in the results across neural network architectures as compared to training on one filter. This suggests that overall, the trend seen from forced HIT holds, where the neural network that is trained with one filter is far more sensitive to neural network architectures, while the two filter trained neural networks are more robust across neural network architectures. The Reynolds stress profiles are seen in figures~\ref{fig:Channel_Filt_R11} to~\ref{fig:Channel_Filt_R33}.

For $R_{11}$, there is not too much difference for the neural network models that do well. However, for $R_{13}$ and $R_{33}$, one can see that training on two filters still helps improve the model as compared to training on one filter, even for the variant neural network architecture. In this case, the red curves are more close to the filtered DNS profile as compared to the blue curves. This leads to a more nuanced conclusion for the ALN (variant) neural network architecture.

First, the ALN acts as a stabilizer that prevents the single-filter model from producing wildly inaccurate results, essentially masking its lack of robust physical knowledge that can generalize \textit{a posteriori}. In addition, the two filter training provides robustness, since the network trained on two filters learns a more general representation of the physics that is accurate regardless of whether the stabilizing ALN layers are present. This means that the two filter case is robust to architectural perturbations. Even with the ALN neural network architecture, as seen in the Reynolds stress profiles, one can still produce a more physically accurate model when training on two filters versus one. This suggests that by incorporating two filters into the training process, the neural network is able to learn a more robust representation of the subgrid scale physics to produce more accurate turbulence in the LES flow fields, regardless of architecture choice. The differences between one and two filter training may be large or small depending on the architecture choice, but there is often improvement in either architecture when training with two filters.

\subsubsection{Simple Inputs}
\begin{figure}
\centering
     \begin{subfigure}{0.45\textwidth}
         \centering
         \includegraphics[width=\textwidth]{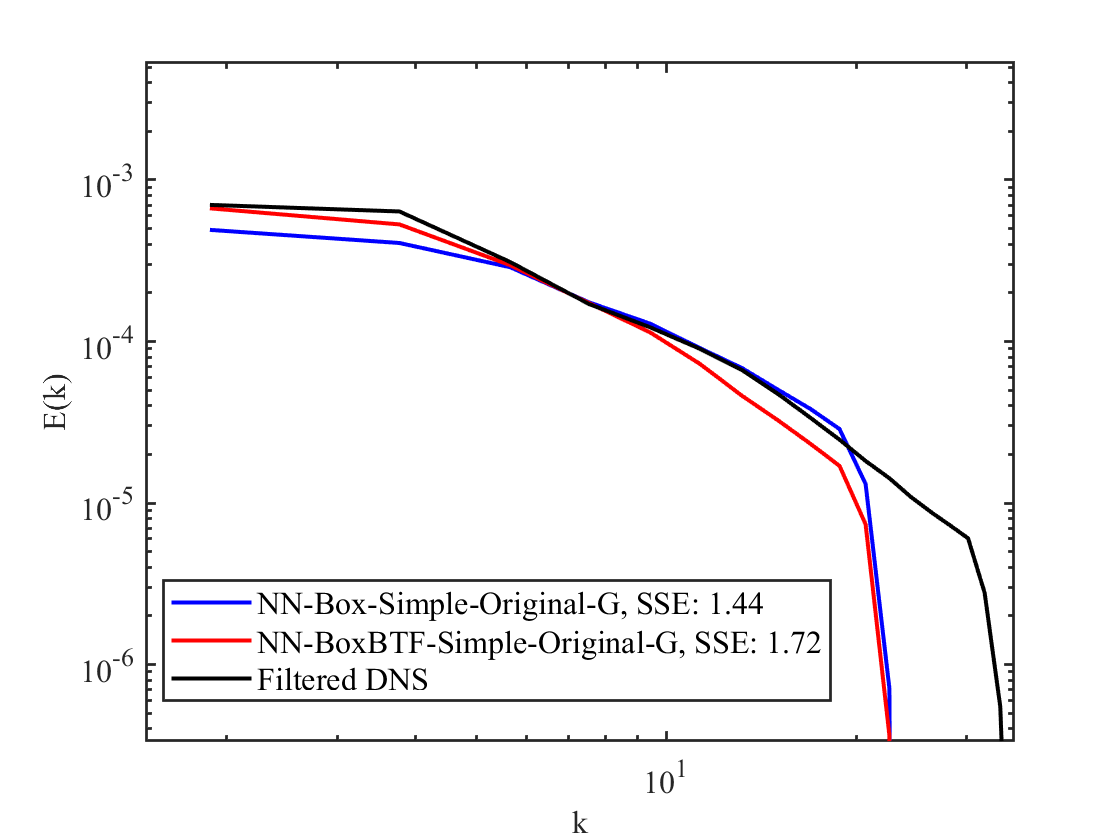}
         \caption{}
         \label{fig:Channel_Simple_Spectra}
     \end{subfigure}
     \hspace{1em}
     \begin{subfigure}{0.45\textwidth}
         \centering
         \includegraphics[width=\textwidth]{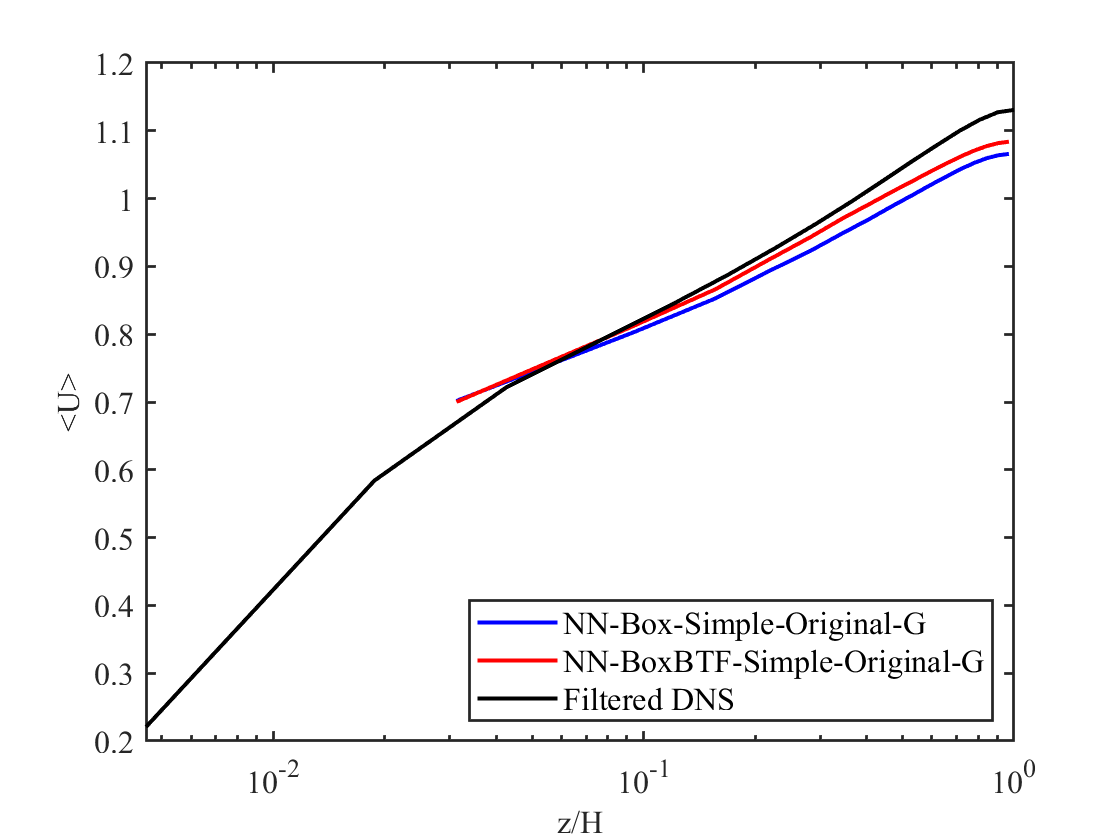}
         \caption{}
         \label{fig:Channel_Simple_Mean}
     \end{subfigure}
     \caption{\textit{A posteriori} simple inputs streamwise velocity spectra and mean profiles. The GNN original architecture is used, with global normalization applied to the ``simple" inputs.}
\end{figure}

Now, neural networks are trained with simple inputs and tested in channel flow to evaluate if there is also less input distribution shift for a more complex flow when using simpler inputs. The spectra and mean velocity profiles are shown in figures~\ref{fig:Channel_Simple_Spectra} and~\ref{fig:Channel_Simple_Mean}.

\begin{figure}
\centering
     \begin{subfigure}{0.4\textwidth}
         \centering
         \includegraphics[width=\textwidth]{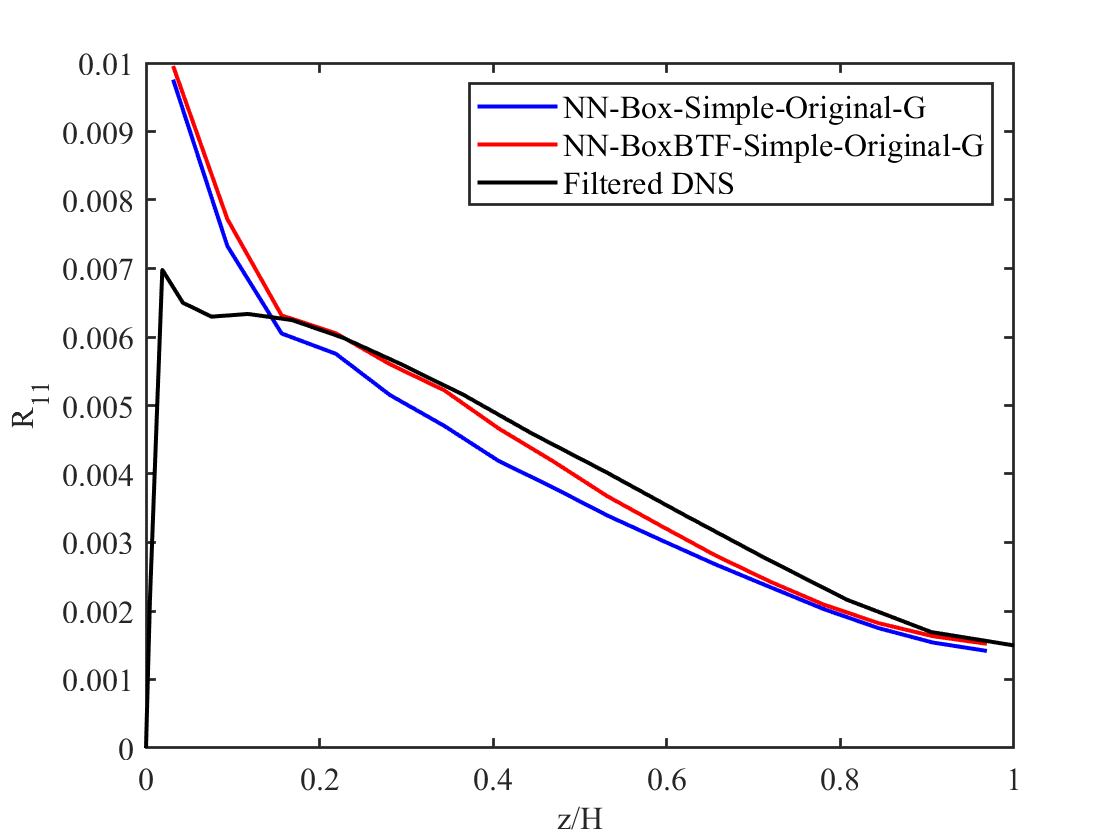}
         \caption{}
         \label{fig:Channel_Simple_R11}
     \end{subfigure}
     \hspace{1em}
     \begin{subfigure}{0.4\textwidth}
         \centering
         \includegraphics[width=\textwidth]{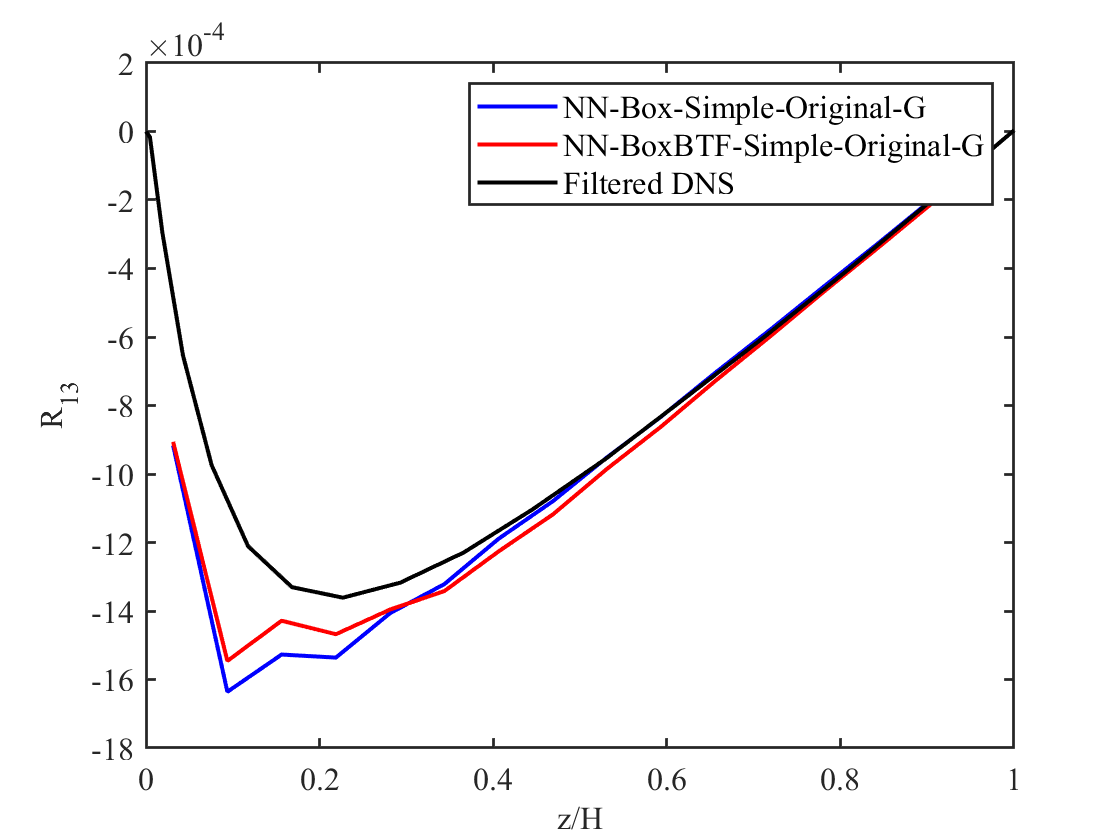}
         \caption{}
         \label{fig:Channel_Simple_R13}
     \end{subfigure}
     \hspace{1em}
     \begin{subfigure}{0.4\textwidth}
         \centering
         \includegraphics[width=\textwidth]{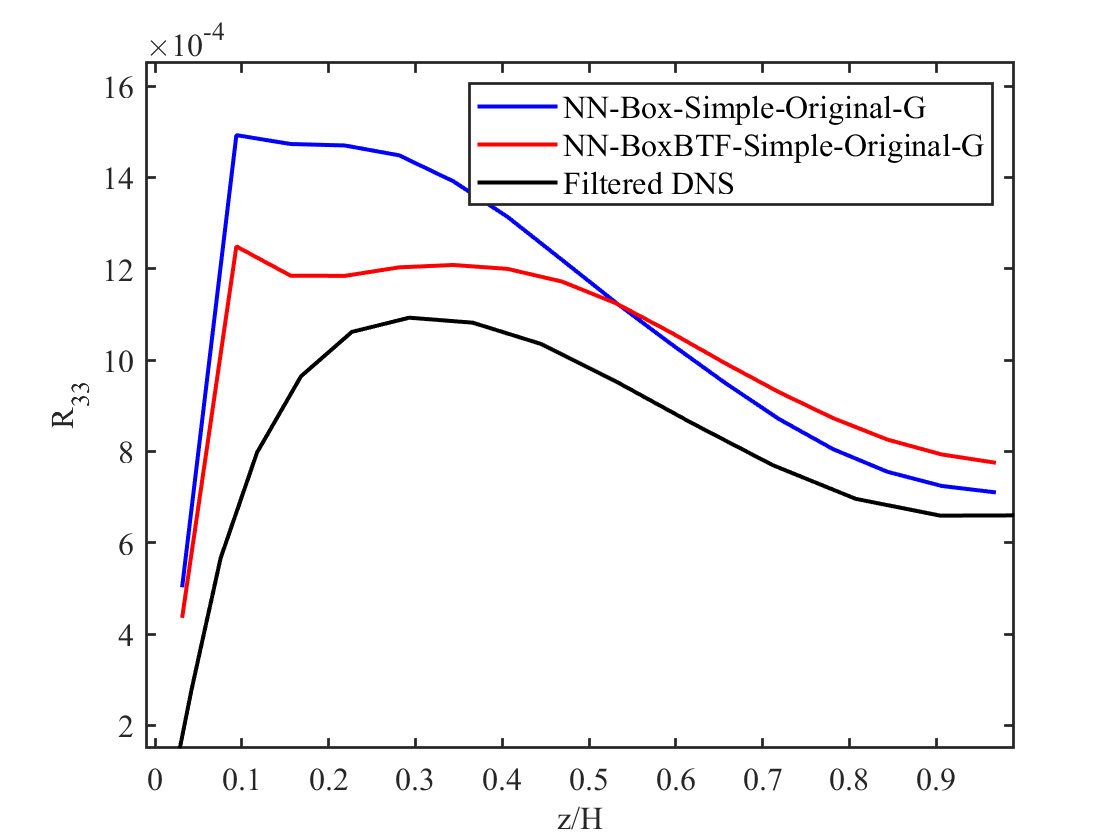}
         \caption{}
         \label{fig:Channel_Simple_R33}
     \end{subfigure}
     \caption{\textit{A posteriori} simple inputs Reynolds stress profiles for channel flow. Here, $z/H = 0$ corresponds to the wall while $z/H=1$ corresponds to the channel centerline. The GNN original architecture is used, with global normalization applied to the ``simple" inputs.}
\end{figure}

Overall, similar trends are seen for the spectra and the mean profiles that were observed in forced HIT. In these cases, there is less of a difference between the neural networks trained with one filter or two filters as there is less input distribution shift, so neural networks behave more similarly \textit{a posteriori}. A similar trend is seen in figures~\ref{fig:Channel_Simple_R11} to~\ref{fig:Channel_Simple_R33} for the Reynolds stress profiles. Despite less difference, it can be seen that in some situations, particularly $R_{33}$, training on two filters produces more physically accurate profiles as compared to training on one filter. This establishes the trend that was seen in forced HIT, that for simpler inputs, there is less of a difference between training with one or two filters, and that the two filter case often does no worse than the one filter case, and in some situations can do better.

\section{Combined \textit{A Posteriori} Analysis}
Overall, combining data augmentation and simpler inputs yield robust \textit{a posteriori} performance across solvers with different high-order numerical methods. Training with two filters allows the neural network to see more diverse input and output distributions, leading to reduced distribution shift \textit{a posteriori}. This allows the neural network, even though it is trained on generic filters not exactly mimicking the LES implicit transfer function, to perform more accurately to unseen implicit filters as it is not overfitting to the kernel artifacts associated with one specific filter (e.g. Gibbs Oscillations). This is seen as the neural network trained on all three filters (and is used in both flow solvers) performs better than the neural network trained only on the box filter (that is also applied to both flow solvers). Meanwhile, the ``simple" training set produces results that are more consistent than the more complex training set because it is able to reduce the input distribution shift, as higher order powers of the velocity gradient tensor suffer more from numerical artifacts such as aliasing, especially at the grid cutoff resolution. By relying on the lower-order terms, the ``simple" inputs allow the neural network to be less exposed to and not overfit to high-wavenumber numerical artifacts, increasing the \textit{a posteriori} robustness of the neural network model. As seen, there are two overall trends. First, using a simpler set of inputs is more robust, especially to neural network architecture perturbations. Second, the overall trend indicates that training on two filters produces results that are better than one filter training for complex inputs. For simple inputs, the overall trend indicates that two filter training is relatively similar to one filter training, as the input distribution shift is reduced due to the use of simpler inputs, although in some cases two filter training can still produce improvements. In general, applying both methods creates the most robust model. Future work can involve evaluating these approaches on lower-order methods and even more complex flows.

\section{Conclusion}
Two methods to increase the \textit{a posteriori} robustness of neural networks are proposed, data augmentation and decreasing the complexity of neural network inputs. Data augmentation involves training with multiple filters in the training data, and two filters have been proposed. BTF accounts for two-thirds dealiasing as an additional filter, whereas DSCF is non-oscillatory in physical space and uses a single mode approximation to the sharp spectral cutoff filter. \textit{A priori}, neural networks suffer no performance degradation when trained on one filter as compared to two filters, suggesting that neural network models are able to distinguish between various filters. \textit{A posteriori}, neural network models trained with two filters are more robust than neural networks trained with one filter, and this trend is confirmed from running neural network models on two high-order solvers. In addition, the neural network trained with three filters is able to be applied to both flow solvers without retraining, and outperforms the neural network trained only with the box filter. Furthermore, neural network models with less complex inputs are more consistent, reducing the distribution shift \textit{a posteriori}. Training with two different filters often does not reduce the performance of a neural network subgrid stress model either \textit{a priori} or \textit{a posteriori}. These trends hold across small neural network architecture perturbations and input normalizations, but are drawn from two high-order solvers and will need to be further evaluated on lower order solvers. In general, for higher order solvers, training with data filtered with two different filters and also using less complex inputs to neural networks increases the robustness of neural networks \textit{a posteriori}.

\backsection[Acknowledgements]{This work used CPU and GPU resources via Bridges2 at the Pittsburgh Supercomputing Center through allocation PHY230025 from the Advanced Cyberinfrastructure Coordination
Ecosystem: Services \& Support (ACCESS) program, which is supported by National Science Foundation grants
\#2138259, \#2138286, \#2138307, \#2137603, and \#2138296. DNS data is provided by the Johns Hopkins Turbulence Database. We thank the anonymous referees whose comments helped improve the paper.}

\backsection[Funding]{Andy Wu is partially supported by NASA Cooperative Agreement number 80NSSC22M0108 and Northrop Grumman, as well as the NDSEG Fellowship.}

\backsection[Declaration of interests]{\bf Declaration of Interests. The authors report no conflict of interest.}

\bibliographystyle{jfm}
\bibliography{jfm}

\end{document}